\documentclass{article}

\PassOptionsToPackage{table}{xcolor}

\usepackage{microtype}
\usepackage{graphicx}
\usepackage{booktabs} %

\usepackage{hyperref}

\usepackage[accepted]{icml2026/icml2026}

\usepackage{amsmath}
\usepackage{amssymb}
\usepackage{mathtools}
\usepackage{amsthm}

\usepackage[capitalize,noabbrev]{cleveref}

\usepackage{amsmath,amsfonts,bm}

\newcommand{\agent}{\texttt{agent}}
\newcommand{\expert}{\texttt{expert}}
\newcommand{\code}{\texttt{Code}}
\newcommand{\codebaseline}{{\code_0}}
\newcommand{\codeagent}{{\code_\agent}}
\newcommand{\codeexpert}{{\code_\expert}}
\newcommand{\workload}{\texttt{workload}}

\newcommand{\speedup}{\texttt{speedup}}
\newcommand{\adv}{\texttt{Adv}}
\newcommand{\modelcost}{\texttt{cost}}

\def\eqref#1{equation~\ref{#1}}
\def\Eqref#1{Equation~\ref{#1}}

\def\1{\bm{1}}

\DeclareMathAlphabet{\mathsfit}{\encodingdefault}{\sfdefault}{m}{sl}
\SetMathAlphabet{\mathsfit}{bold}{\encodingdefault}{\sfdefault}{bx}{n}

\usepackage[utf8]{inputenc}
\usepackage[T1]{fontenc}
\usepackage{amsfonts}
\usepackage{nicefrac}
\usepackage{subcaption}
\usepackage{url}
\usepackage{xcolor}
\usepackage{xspace}
\usepackage{colortbl}

\usepackage{multirow}
\usepackage{caption}
\usepackage{array}
\usepackage{wrapfig}
\usepackage{adjustbox}
\usepackage{enumitem}
\usepackage{tabularx}     %
\usepackage{makecell}     %
\usepackage{longtable,ragged2e}
\usepackage{graphicx,xcolor,colortbl,tabularx,booktabs,makecell,adjustbox,array,pifont}
\usepackage{listings}
\usepackage{tcolorbox}
\tcbuselibrary{listings,skins}

\usepackage{inconsolata} %

\newcommand{\pWidth}{5in}
\newcounter{pCount}
\newcommand{\pStart}{\setcounter{pCount}{1}}
\newcommand{\pRow}[1]{%
  \texttt{\fontsize{6pt}{7pt}\selectfont\color{gray}\arabic{pCount}} &
  \texttt{\fontsize{6pt}{7pt}\selectfont #1}\stepcounter{pCount}\\%
}

\newcommand{\pSection}[1]{\pRow{{\bfseries\MakeUppercase{#1}}}}

\newcommand{\pRowTable}[1]{%
  \texttt{\fontsize{6pt}{7pt}\selectfont\color{gray}\arabic{pCount}} &
  \begin{minipage}[t]{\pWidth}
    \centering
    \ttfamily\fontsize{6pt}{7pt}\selectfont
    #1
  \end{minipage}\stepcounter{pCount}\\%
}

\setlist[itemize]{leftmargin=1em}

\newcommand{\finalversion}[1]{}

\def\hide#1{}

\newcommand{\fc}{\textsc{FormulaCode}\xspace}
\newcommand{\fcv}{\textsc{FormulaCode-V}\xspace}

\newcommand{\cmark}{\ding{51}} %
\newcommand{\xmark}{\ding{55}} %

\usepackage{xfp}

\definecolor{MutedGreen}{RGB}{76,120,95}
\definecolor{MutedOrangeRed}{RGB}{176,98,78}
\definecolor{NumGray}{RGB}{0,0,0}

\newcommand{\trendnum}[1]{%
  \begingroup
  \edef\val{#1}%
  \ifdim \val pt > 0pt
    \edef\mix{\fpeval{round(40 + 82*sqrt(min(1,max(0,\val/0.0423))),0)}}%
    \textcolor{MutedGreen!\mix!NumGray}{#1}%
  \else\ifdim \val pt < 0pt
    \edef\mix{\fpeval{round(40 + 82*sqrt(min(1,max(0,abs(\val)/0.3529))),0)}}%
    \textcolor{MutedOrangeRed!\mix!NumGray}{#1}%
  \else
    \textcolor{NumGray}{#1}%
  \fi\fi
  \endgroup
}

\makeatletter
\let\fc@oldvspace\vspace
\newcommand{\fc@gobblevspace}{\@ifstar\@gobble\@gobble}
\newcommand{\disablevspace}{\renewcommand\vspace{\fc@gobblevspace}}
\newcommand{\enablevspace}{\let\vspace\fc@oldvspace}
\makeatother

\icmltitlerunning{\fc: Evaluating Agentic Optimization on Large Codebases}

\begin{document}

\disablevspace

\twocolumn[
\icmltitle{\fc: Evaluating Agentic Optimization on Large Codebases}

\icmlsetsymbol{equal}{*}

\begin{icmlauthorlist}
\icmlauthor{Atharva Sehgal}{utaustin,equal}
\icmlauthor{James Hou}{caltech,equal}
\icmlauthor{Akanksha Sarkar}{cornell}
\icmlauthor{Ishaan Mantripragada}{caltech}\\
\icmlauthor{Swarat Chaudhuri}{utaustin}
\icmlauthor{Jennifer J. Sun}{cornell}
\icmlauthor{Yisong Yue}{caltech}
\end{icmlauthorlist}

\icmlaffiliation{utaustin}{The University of Texas at Austin}
\icmlaffiliation{caltech}{California Institute of Technology}
\icmlaffiliation{cornell}{Cornell University}

\icmlcorrespondingauthor{Atharva Sehgal}{atharvas@utexas.edu}

\icmlkeywords{Code Generation, Benchmarking}

\vskip 0.3in
]

\printAffiliationsAndNotice{\icmlEqualContribution}  %

\newcommand{\fcsize}{$957$\xspace}
\newcommand{\numworkloads}{$264.6$\xspace}

\newcommand{\numFinalRepos}{$70$\xspace} %

\newcommand{\numTotalPRs}{$105074$\xspace} %

\newcommand{\numGithubRepos}{$2.8 \times 10^6$\xspace}
\newcommand{\numCodeFiles}{$2 \times 10^9$\xspace}

\newcommand{\numReposDiscovered}{$766$\xspace} %

\newcommand{\minStars}{100}

\newcommand{\numReposAttributeFiltering}{$127$\xspace} %
\newcommand{\numReposPerfFiltering}{$101$\xspace}
\newcommand{\numPRsScraped}{26717} %

\newcommand{\numPerfPRs}{3181} %

\newcommand{\numTasksContainers}{1232}
\newcommand{\numReposStageThree}{75}

\newcommand{\numFCVProblems}{108}

\newcommand{\avgWorkloadsPerProblem}{264.58}
\newcommand{\avgCoveragePercent}{41.24}
\newcommand{\earliestTaskDate}{2017-10-21}
\newcommand{\latestTaskDate}{2025-11-21}
\newcommand{\percentTasksRecentYears}{55.88} %
\newcommand{\avgProblemsPerMonth}{27.00}
\newcommand{\monthlyUpdateDay}{25th}

\begin{abstract}

Large language model (LLM) coding agents increasingly operate at the repository level, motivating benchmarks that evaluate their ability to optimize entire codebases under realistic constraints. Existing code benchmarks largely rely on synthetic tasks, binary correctness signals, or single-objective evaluation, limiting their ability to assess holistic optimization behavior. We introduce \fc, a benchmark for evaluating agentic optimization on large, real-world codebases with fine-grained, multi-objective performance metrics. \fc comprises \fcsize performance bottlenecks mined from scientific Python repositories on GitHub, each paired with expert-authored patches and, on average, \numworkloads community-maintained performance workloads per task, enabling the holistic ability of LLM agents to optimize codebases under realistic correctness and performance constraints. Our evaluations reveal that repository-scale, multi-objective optimization remains a major challenge for frontier LLM agents. Project website at: \url{https://formula-code.github.io}.

\end{abstract}

\section{Introduction}\label{sec:introduction}

Large Language Models (LLMs) for code are rapidly evolving from isolated function-level synthesis to file-level editing, and now, to repository-level optimization \citep{terminalbench, jimenez2024swebenchlanguagemodelsresolve, zhang2025swebenchgoeslive, zhao2024commit0librarygenerationscratch, shetty2025gsochallengingsoftwareoptimization, ma2025swefficiencylanguagemodelsoptimize}. 
These models are now transitioning from assistants into autonomous coding agents, increasingly tasked with navigating complex, interconnected software ecosystems to diagnose bottlenecks and improve performance. 
However, we currently lack frameworks to study these emerging capabilities for the full optimization lifecycle; for example, how agents balance multiple workloads, maintain function integrity, and structure improvements at different levels of the codebase hierarchy. 

While there exist coding benchmarks based on real GitHub repositories~\cite{jimenez2024swebenchlanguagemodelsresolve, zhang2025swebenchgoeslive, zhao2024commit0librarygenerationscratch}, they generally do not fully capture the multi-workload real-world tasks that engineers and researchers face in practice. 
These benchmarks often rely on binary pass/fail feedback, which is insufficient for measuring optimization, or synthetic (e.g., LLM generated) tasks, which lack the complexity and characteristics of real-world coding. For example, real-world optimization is rarely isolated, diagnosing and improving performance often requires reasoning about architectural decisions, component interactions, and design trade-offs on the system-level rather than tuning an isolated function \citep{balsamo2004model, woodside2007future, jin2012understanding}.
Consequently, this requires a new evaluation standard capable of measuring the emerging ability of agents across this entire optimization workflow under realistic software engineering constraints.

We identify several directions for improving agentic coding benchmarking: (1) Fine-grained metrics: evaluation must move beyond binary correctness to capture continuous performance changes and trade-offs; (2) Real-world measurements: metrics should be derived from established execution environments (e.g., standard profiling suites) rather than synthetic proxies; (3) Reliable baselines: agent performance must be assessed against human optimization to provide a meaningful standard; and (4) Repository scale: agents must operate within large, evolving codebases.

\begin{figure*}[t]
    \centering
    \includegraphics[width=\textwidth]{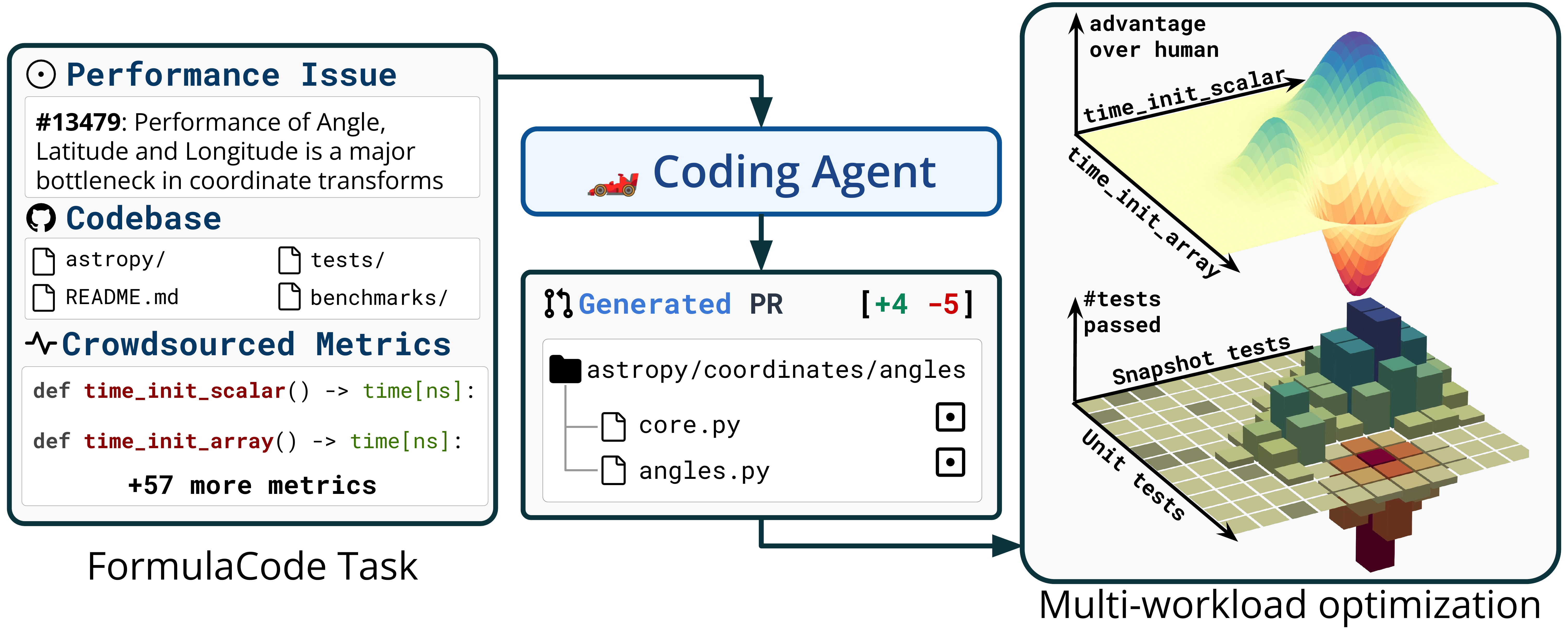}
    \vspace{-2em}
    \caption{
    \fc is a continuously updating benchmark for evaluating the holistic ability of agents to optimize large codebases.
    Each task in \fc comprises a problem description of a performance regression from GitHub, an environment containing a baseline repository snapshot, and multiple expert-written crowdsourced performance workloads, along with the tools to execute them. An agent's performance improving edits are assessed based on their ability to outperform expert-written edits in optimizing \emph{multiple workloads} while meeting multiple forms of correctness guarantees.
    }
    \label{fig:overview}
    \end{figure*}

We introduce \fc
\footnote{\fc draws inspiration from Formula 1, where constructors must optimize entire systems—not just individual components—to achieve peak performance on the track. Similarly, \fc challenges code agents to perform holistic, codebase-level optimizations, reflecting the complexity and interdependence found in real-world software.}, a novel benchmark designed for advancing agentic optimization on large, evolving software ecosystems. 
\fc is constructed from \fcsize real-world performance bottlenecks mined from 70 scientific, open-source Python repositories, like Pandas, Scikit-learn, and SciPy.
Unlike previous datasets, each task in \fc is paired with an average of \numworkloads community-maintained performance workloads alongside expert-authored patches. This unique construction enables the use of the airspeed-velocity (asv) framework to assess the full lifecycle of optimization (triage, diagnosis, and resolution) in a way that isolated coding tasks cannot.

We conduct a large-scale evaluation of frontier and open weights models (GPT-5, Claude 4.0 Sonnet, Gemini 2.5 Pro, Qwen 3 Coder) within multiple agentic frameworks (Terminus 2, OpenHands).  \textbf{Our main findings are:}
\vspace{-0.05in}
\begin{itemize}[
  itemsep=0.15em,
]
    \item Agents generally can improve run-time performance, but perform worse than experts (\S \ref{sec:results-global}).
    \item Agents are better at local or function-level optimization, rather than repository-level optimization (\S \ref{sec:results-scale}).
    \item Agents excel at using specific optimization strategies (e.g., parallelizing or batching) and struggle with others (e.g., vectorized operations) (\S \ref{sec:results-type}). 
    \item Agent performance relative to experts can vary dramatically by popularity of the repository, performing worst on the 4th quintile and best on the 2nd quintile (\S \ref{sec:results-popularity}).
    \item Despite being more expensive per call, agents using frontier LLMs are overall more cost effective than those using open weights models (e.g., due to open weights models having much longer reasoning chains) (\S \ref{sec:results-cost}).
    \item Compared to experts, agents negotiate multi-workload performance trade-offs less effectively (\S\ref{sec:results-tradeoff}).
    \item We observe minimal effects from data leakage (i.e., using LLMs potentially trained on expert solutions) (\S\ref{sec:results-leakage}).
\end{itemize}
We open-source \fc as a community resource\footnote{Project website at \url{https://formula-code.github.io/}.}, to not only measure what code agents can generate, but to understand how they can reliably optimize and maintain complex real-world systems.

\newcommand{\normadv}[1]{\widetilde{\adv}_{#1}}         %

\newcommand{\Glevel}[1]{\mathcal{G}^{(#1)}}   %

\section{\fc Benchmark Design}\label{sec:problem-statement}

Each \fc task evaluates the ability of an $\agent$~to optimize a real-world codebase under strict correctness constraints. A task begins with a baseline repository, denoted $\codebaseline$, which represents the unmodified implementation. The $\agent$~operates on $\codebaseline$~and produces a modified version of the repository, denoted $\codeagent$, by making arbitrary repository-level edits.  

Each task is paired with two forms of evaluation signals:
\vspace{-0.1in}
\begin{itemize}
    \item \textbf{Correctness.} Correctness is measured via a suite of tests on the functional behavior. A proposed code modification is considered valid only if $\codeagent$~passes all tests that $\codebaseline$~passes.
    \vspace{-0.05in}
    \item \textbf{Performance Workloads.} Each task includes a large collection of expert-written performance workloads that exercise known performance-critical execution paths in the codebase. Each workload measures a single performance dimension, such as runtime or memory usage, and may exhibit natural variability due to execution noise.
\end{itemize}
Figure \ref{fig:overview} depicts our benchmark setup. The top half shows a task from the Astropy repository, highlighting a performance issue with three functions: Angle, Latitude, and Longitude.  There are 59 workloads defined by community-sourced expert-written metrics.  The goal of the coding $\agent$~is to modify the repository to optimize these workloads while still maintaining correctness. 

Performance evaluation proceeds by executing the full set of workloads on both $\codebaseline$~and $\codeagent$~and comparing their measured outcomes. Improving performance on one $\workload$~may degrade performance on others \citep{balsamo2004model, woodside2007future, jin2012understanding}. As a result, optimization in \fc is inherently multi-objective: agents must reason about trade-offs across subsystems and deliver improvements that are broad and consistent rather than localized to a single execution path.

\subsection{Metrics}

\textbf{Speedup.}
For each $\workload_i$, we compare the performance ratio of $\codeagent$~versus $\codebaseline$: 
$$
\speedup_{\agent, i} = \frac{\workload_i(\codebaseline)}{\workload_i(\codeagent)}.
$$
Having $\speedup>1$ indicates an improvement. These ratios are dimensionless and allow performance changes to be compared across heterogeneous workloads.  If $\codeagent$ does not pass correctness tests for $\workload_i$, then $\speedup_i=1$ (i.e., the modifications were reverted).

For $n$ workloads, the overall speedup is the geometric mean:
\begin{eqnarray}
\speedup_\agent = \left(\prod_{\workload_i} \speedup_{\agent, i}\right)^\frac{1}{n}.\label{eqn:speedup}
\end{eqnarray}

\textbf{Advantage.}  For each task, we also have expert-written code modifications, $\codeexpert$.  For example, the performance issue in Figure \ref{fig:overview} was eventually resolved by an expert.  We use the performance of $\codeexpert$ as a reference point to characterize the difficulty of each task.  We can then define the advantage of each $\workload_i$~as:
$$
\adv_i = \speedup_{\agent, i} - \speedup_{\expert, i}.
$$
and subsequently, $\adv_\agent$ is the geometric mean over $n$ workloads. Notice that if an agent had simply memorized the expert solution (e.g., due to training data contamination), then the advantage is zero.  Indeed, the goal of super-human optimization is to achieve a large positive advantage.
Appendix Figure \ref{fig:advantage-example} provides a geometric intuition for this metric.

\textbf{Stratified Advantage.}  We now turn to measuring advantage aggregated at different levels of granularity. We use $\ell\in \{0,1,\ldots\}$ to denote the code hierarchy level. 
\vspace{-0.1in}
\begin{itemize}[
  itemsep=0.25em,
]
\item At the coarsest level (\(\ell=0\)), we group workloads by entire modules such as \texttt{algorithms.*}. 
\vspace{-0.05in}
\item At finer levels, we group workloads under individual classes or functions
(e.g., \texttt{algorithms.Sorting.*}, \texttt{algorithms.Sorting.time\_sort\_int.*}).
\end{itemize}
\vspace{-0.05in}
Each level $\ell$ thus partitions the workloads into groups: $\Glevel{\ell} = \{g^{(\ell)}_1, \dots, g^{(\ell)}_{K_\ell}\}$,
where each $\workload$~belongs in some  $g_k^{(\ell)}$.  We can then define per-group advantage as:
$$
\adv_{\agent,g} = \speedup_{\agent}(g) - \speedup_{\expert}(g),
$$
where $\speedup_*(g)$ is defined using \Eqref{eqn:speedup} computed only over workloads in $g$.  The \textit{stratified advantage at level $\ell$} is then the average across all groups at that level:
$$
\adv_\agent^{(\ell)} = \frac{1}{|\Glevel{\ell}|}\sum_{g\in\Glevel{{\ell}}}\adv_{\agent,g}.
$$
The family \(\{ \adv_\agent^{(\ell)} | {\ell} \in \mathbb{Z}_{\geq 0} \}\) thus forms a multi-scale profile of an agent's performance.
Because aggregation is performed over multiplicative speedup ratios within each group, \(\adv_\agent^{(\ell)} \) remains in the same metric family as the global advantage, but is sensitive to how performance gains are organized across the codebase hierarchy (Figure \ref{fig:hierarchical-overview}).

\textbf{Normalized Advantage.}
Finally, we introduce a normalized version of advantage that explicitly accounts for noise and heterogeneity across workloads.
Given the variance of the per-workload speedup ratios for an $\agent$, $\sigma^2(\agent)$,  we define the \emph{normalized advantage} of an $\agent$~as:
\begin{align*}
    \normadv{\agent}
    &= \frac{\adv_\agent}{\sqrt{\sigma^2(\agent) + \sigma^2(\expert)}}.
\end{align*}
Conceptually, $\normadv{\agent}$ captures a signal-to-noise ratio of the agent advantage, and rewards consistency across workloads.

\begin{figure*}[t]
    \centering
    \includegraphics[width=\textwidth]{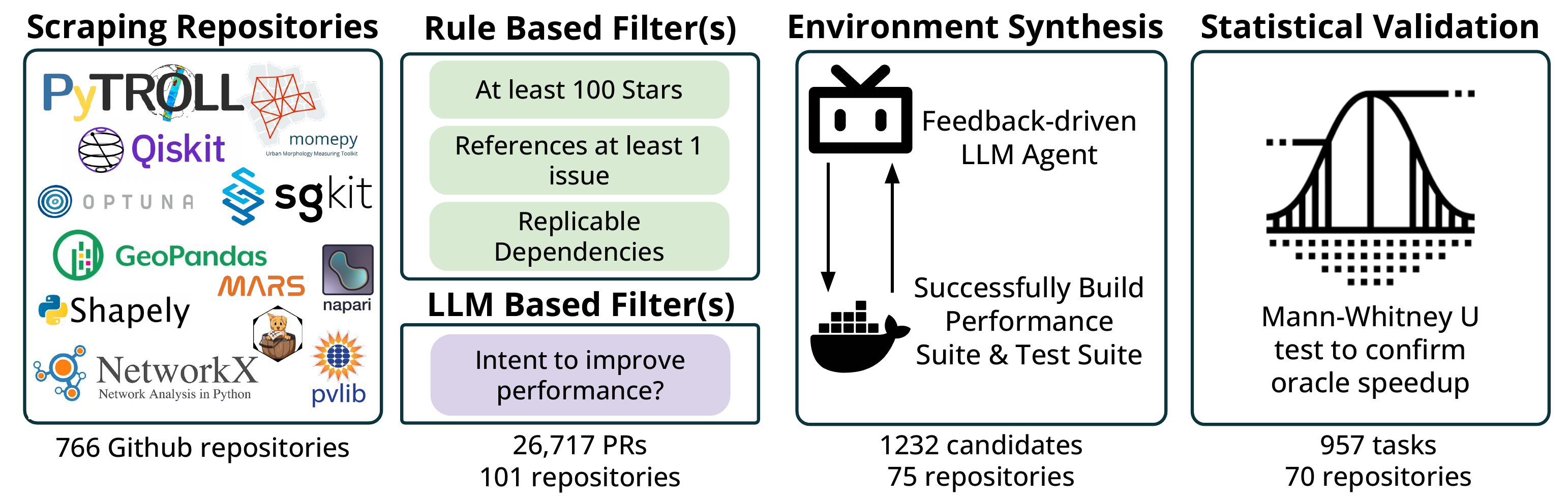}
    \caption{
    Overview of \fc construction pipeline. \fc follows a four stage pipeline to identify real-world performance optimization tasks. (1) Scrape compliant repositories (\S \ref{sec:a.dataset.scraping}). (2) Apply rule-based and LLM-based filters to identify candidate performance improvement pull requests (\S \ref{sec:a.dataset.filtering}). (3) Construct reproducible Docker environments for each candidate (\S \ref{sec:a.dataset.stage3}). (4) Validate each candidate for correctness and statistically significant performance improvement (\S \ref{sec:a.dataset.stat_testing}). The pipeline is fully automated and updates \fc with new tasks every month.
    }
    \label{fig:dataset-overview}
    \vspace{-1em}
\end{figure*}

\textbf{Cost Weighted Metrics.} In practice, we also care about the inference budget of the optimization agent. We estimate the total inference cost as  $\modelcost_\agent = c_{\text{in}} N_{\agent}^{\text{in}} + c_{\text{out}} N_{\agent}^{\text{out}}$ where $N_{\agent}^{\text{in}}$ and $N_{\agent}^{\text{out}}$ denote the total number of input and output tokens, and $c_{\text{in}}$ and $c_{\text{out}}$ are the per-token prices. This allows us to define the cost-weighted advantage:
\[
    \modelcost(\adv_{\agent})
    = \frac{\adv_{\agent}}{\modelcost_{\agent}},
\]
which captures the human-relative improvement obtained per unit of inference budget. We will use these metrics in \S\ref{sec:experiments} to evaluate code optimization agents' performance on real-world codebases.

\section{\fc: Dataset Construction}\label{sec:dataset}

\fc consists of \fcsize multi-workload, real-world code optimization problems drawn from \numFinalRepos{} repositories. In this section, we briefly describe an automated four-stage pipeline that extracts these problems from \numTotalPRs{} pull requests across \numReposDiscovered{} GitHub repositories. The key idea in each stage is to maximize recall as ambiguity is resolved in the last stage when we apply the expert-written patch and validate the performance improvement with statistical testing. Full details are presented in Appendix~\S\ref{sec:a.dataset}, Figure \ref{fig:dataset-overview} provides an overview of the pipeline, and Table~\ref{table:dataset-stats} summarizes key statistics about the final dataset.

\subsection{Construction Pipeline}\label{sec:dataset.construction}

\textbf{Stage 1: Repository Scraping.} Our benchmarking apparatus relies on mature tools developed within the Python performance benchmarking community, for which a package's core developers write customized performance workloads in a pre-specified format. We can therefore identify both crowdsourced workloads and repositories with an established benchmarking procedure by searching for the presence of these tools, which yields \numReposDiscovered{} repositories (Appendix~\S\ref{sec:a.dataset.scraping}).

\textbf{Stage 2: Attribute Filtering.} From these repositories we scrape \numTotalPRs{} merged pull requests; rule-based filters then retain \numPRsScraped{} pull requests from \numReposAttributeFiltering{} repositories that each reference at least one issue. We then construct a knowledge graph of the issues and comments referenced by each pull request (restricted to content created on or before the pull request) and an LLM agent analyzes it to gauge whether the pull request's primary intent is performance-oriented. This yields \numPerfPRs{} candidate performance-improving tasks from \numReposPerfFiltering{} repositories (Appendix~\S\ref{sec:a.dataset.filtering},~\S\ref{sec:a.dataset.llmfilters}).

\textbf{Stage 3: Environment Synthesis.} Before validating a performance improvement, we must build and install a development copy of the package. This proves to be a non-trivial task, since scientific packages often require bespoke build processes, their build processes evolve as the project matures, and their documentation is fragmented across many files. We automate this with a reflexive LLM agent~\citep{shinn2023reflexionlanguageagentsverbal} that iteratively refines a shell script to build an editable environment, and we aggressively cache and reuse previously synthesized scripts to lower the amortized cost. This yields \numTasksContainers{} tasks with reproducible Docker containers from \numReposStageThree{} repositories (Appendix~\S\ref{sec:a.dataset.stage3}).

\textbf{Stage 4: Statistical \& Correctness Validation.} For each task we apply the expert-produced patch and retain only tasks whose workloads exhibit a statistically significant speedup. We additionally provide two kinds of correctness tests: unit tests, which we manually validate for the \numFCVProblems{} problems in \fcv, and automated snapshot tests that compare each workload's return values against a reference, greatly increasing the correctness-verification surface of each task. This yields the final \fcsize{} tasks (Appendix~\S\ref{sec:a.dataset.stat_testing}).

\subsection{Dataset Properties}\label{sec:dataset.statistics}

\textbf{Multi-workload optimization.} Code optimizations rarely have isolated effects: an optimization in one part of the code can slow down another. \fc therefore frames performance optimization as a \emph{multi-workload} problem: each task exposes on average \avgWorkloadsPerProblem{} performance workloads (Table~\ref{table:dataset-stats}) presented to the agent alongside the problem description, and the agent is scored on the aggregate improvement across all of them, so it must reason about competing objectives rather than a single hot path.

\textbf{Live, contamination-resistant benchmark.} Data contamination is known to skew the performance of frontier models on code tasks mined from GitHub~\citep{zhang2025swebenchgoeslive}. To mitigate this, \fc functions as a live dataset, updated with new problems each month. Tasks in \fc span \earliestTaskDate{} to \latestTaskDate{}, and their creation timestamps enable the temporal out-of-distribution analysis in \S\ref{sec:results-leakage}.

\textbf{Hierarchical workloads.} Following the benchmark directory structure assigned by core developers, workloads are organized at three levels of granularity---module, class, and function (Figure~\ref{fig:hierarchical-overview})---which lets us aggregate workloads by semantic grouping for the scale-resolved, stratified advantage analysis in \S\ref{sec:results-scale}.

\textbf{Task diversity.} Restricting collection to a hand-curated set of repositories creates a cumulative ``Matthew effect''~\citep{koch2021reduced} that disconnects a benchmark from the broader task distribution. \fc instead samples performance benchmarks from a large set of repositories, based on whether they satisfy the four pipeline stages above. We report the composition of the dataset in Appendix~\S\ref{sec:a.dataset.composition}.

\enablevspace

\begin{table}[t]
    \centering
    \small
    \begin{tabular}{llrr}
    \toprule
        & & Mean & Max \\
    \midrule
        Issue Text & Length (Tokens) & 2718.03 & 15781 \\
        \midrule
        \multirow{3}{*}{Gold Patch} & \# Lines edited & 38.088 & 526 \\
                                    & \# Files edited & 3.93 & 34 \\
                                    & \# Func. edited & 6.06 & 54 \\
        \midrule
        \multirow{2}{*}{Workloads}
        & \# Eval. Fns & 264.58 & 1364 \\
        & \% Coverage & 41.24\% & 97.86\% \\
    \bottomrule
    \end{tabular}
    \caption{
    (Micro-averaged) statistics characterizing different attributes of a \fc task instance. The average \fc gold patch requires $5.2$ more lines of code spread over $1.29\times$ more files and $1.01\times$ more functions than the average SWE-Bench \citep{jimenez2024swebenchlanguagemodelsresolve} patch.
    }
    \label{table:dataset-stats}
    \vspace{-2.5em}
\end{table}

\disablevspace

\section{Experiments}\label{sec:experiments}

We organize our experimental findings into three categories.
\begin{itemize}[
  leftmargin=1.4em,
  labelsep=0.5em,
  itemsep=0.15em,
  topsep=0.25em,
  parsep=0pt,
  partopsep=0pt
]
\vspace{-0.2in}
    \item First, we present overall performance metrics to investigate whether agents can achieve meaningful runtime speedups and whether they can outperform experts.
    \item  Second, we provide detailed breakdown of agent capabilities, examining performance across optimization strategies, optimization scope, and repository popularity.
    \item Third, we present findings on cost-effectiveness, multi-workload optimization, and data leakage. 
    \vspace{-0.05in}
\end{itemize}

We compare four frontier LLMs -- GPT-5 \citep{singh2025openaigpt5card}, Claude 4.0 Sonnet \citep{anthropic2025_claude4_systemcard}, Gemini 2.5 Pro \citep{comanici2025gemini25pushingfrontier}, and Qwen 3 Coder \citep{yang2025qwen3technicalreport} -- under two LLM Frameworks -- Terminus 2 \citep{terminalbench} and OpenHands \citep{wang2025openhandsopenplatformai}. Terminus~2 is evaluated with all four models, while OpenHands is evaluated with GPT-5, Claude~4.0~Sonnet, and Qwen~3~Coder. Additional discussion of model and framework choices appears in Appendix~\S\ref{sec:a.expt.models}. 
Evaluations are conducted on \fcv due to compute constraints, using the metrics defined in \S\ref{sec:problem-statement}. Full experimental details and additional analyses are provided in Appendix~\S\ref{sec:a.expt}.

\newcommand{\RP}{\textsc{RP}}

\newcommand{\bestAdvModel}{\textsc{Terminus~2 + Gemini~2.5~Pro}}          %
\newcommand{\bestNormAdvModel}{\textsc{OpenHands + GPT-5}}               %
\newcommand{\bestCostModel}{\textsc{Terminus~2 + Gemini~2.5~Pro}}        %
\newcommand{\bestCostNormModel}{\textsc{Terminus~2 + GPT-5}}    %

\newcommand{\bestModuleModel}{\textsc{Terminus~2 + Gemini~2.5~Pro}}      %

\newcommand{\ModelX}{\textsc{Terminus~2 + Gemini~2.5~Pro}}      %
\newcommand{\ModelY}{\textsc{Terminus~2 + Claude~4.0~Sonnet}}     %

\newcommand{\ModelXDelta}{\text{-94.7}}     %
\newcommand{\ModelYDelta}{\text{-95.3}}     %
\newcommand{\LongTailSpeedupDrop}{-23.0}

\newcommand{\PopularStarThreshold}{20}
\newcommand{\PopularFrac}{25.0\%}
\newcommand{\UnpopularFrac}{75.0\%}

\newcommand{\LevelOneName}{Module}
\newcommand{\LevelTwoName}{Class}
\newcommand{\LevelThreeName}{Function}

\newcommand{\OODWindowPre}{\text{TBD-PreWindow}}       %
\newcommand{\OODWindowPost}{\text{TBD-PostWindow}}     %

\newcommand{\DropGPTFive}{\text{TBD-GPT5DropPercent}}      %
\newcommand{\DropClaudeSonnet}{\text{TBD-ClaudeDropPercent}} %
\newcommand{\DropQwenCoder}{\text{TBD-QwenDropPercent}}     %

\newcommand{\DropAvgAll}{\text{TBD-AvgDropPercent}}

\newcommand{\WorstViolationsConfig}{\textsc{Terminus 2 + Claude~4.0~Sonnet}} %

\newcommand{\LongestTrajConfig}{\textsc{OpenHands + Qwen~3~Coder}} %
\newcommand{\ShortestTrajConfig}{\textsc{OpenHands + GPT-5}}%
\newcommand{\LongestTrajMean}{633.1}
\newcommand{\LongestTrajStd}{228.1}
\newcommand{\ShortestTrajMean}{68.6}
\newcommand{\ShortestTrajStd}{28.0}
\newcommand{\LengthPerfCorr}{-0.005}
\newcommand{\BestTrajNormConfig}{\textsc{Terminus~2 + Claude~4.0~Sonnet}}%
\newcommand{\NumTrajectories}{424}

\newcommand{\MostPeripheralToolConfig}{\textsc{Terminus~2 + GPT-5}} %
\newcommand{\TopProfilerConfig}{\textsc{Terminus~2 + GPT-5}}           %
\newcommand{\ToolCallPerfCorr}{-0.005}
\newcommand{\ProfilerSpeedupCorr}{0.039}

\newcommand{\numFCVTasksTemporal}{\text{TBD-NumTasksTemporal}}      %
\newcommand{\numFCVTasksDev}{\text{TBD-NumTasksDev}}                %

\newcommand{\SnapshotTestsName}{snapshot tests}
\newcommand{\PytestSuiteName}{PyTest suite}

\newcommand{\figTemporal}{Figure~\ref{fig:temporal-ood}}
\newcommand{\tabCorrectness}{Table~\ref{tab:correctness-violations}}
\newcommand{\tabTraj}{Table~\ref{tab:trajectory-length}}
\newcommand{\tabTools}{Table~\ref{tab:tool-usage}}

\newcommand{\tabRobustness}{Table~\ref{tab:apparatus-robustness}}

\newcommand{\NumRobustTasks}{\texttt{TBD-NumRobustTasks}}

\newcommand{\WorstRobustLevel}{\textsc{TBD-WorstRobustLevel}} %
\newcommand{\WorstRobustSE}{\texttt{TBD-WorstRobustSE}}       %

\newcommand{\tighttable}{%
  \small
  \setlength{\tabcolsep}{3.5pt}
  \renewcommand{\arraystretch}{1.15}
}

\subsection{Global Leaderboard}
\label{sec:results-global}

For each agent--model configuration, we compute the human-relative advantage $\adv{}$ and normalized advantage $\normadv{}$ defined in \S\ref{sec:problem-statement}. We then aggregate configurations into a global leaderboard using the Ranked Pairs (\RP) method \citep{tideman1987independence}, yielding a transitive ordering. Table~\ref{tab:advantage-leaderboard} summarizes the resulting rankings.

\begin{table*}[t]
\begin{minipage}[c]{0.72\linewidth}
    \centering
    \adjustbox{max width=0.95\textwidth}{
    \begin{tabular}{l l r r r r}
        \toprule
        Agent & Model & \RP\ Rank ($\adv$) $\downarrow$ & $\adv{}$ $\uparrow$ & $\normadv{}$ $\uparrow$ & $\speedup{}$ $\uparrow$
        \\
        \midrule
        Terminus 2 & GPT-5 & 7 & -0.0504 & -0.1387 & 1.0585 \\
        & Claude 4.0 Sonnet & 4 & -0.0410 & -0.1065 & 1.0987 \\
        & Gemini 2.5 Pro & 6 & -0.0433 & -0.1138 & 1.0963 \\
        & Qwen 3 Coder & 5 & -0.0454 & -0.1257 & 1.0677 \\
        OpenHands & GPT-5 & 3 & -0.0209 & -0.0702 & 1.0825 \\
        & Claude 4.0 Sonnet & 1 & -0.0112 & -0.0483 & 1.0539 \\
        & Qwen 3 Coder & 2 & -0.0301 & -0.1529 & 1.0346 \\
        Human Expert & - & - & \textbf{0.0000} & \textbf{0.0000} & \textbf{1.1040} \\
        \bottomrule
    \end{tabular}
    }
    \end{minipage}
    \hfill
      \begin{minipage}[c]{0.28\linewidth}
    \caption{
        Global leaderboard of agent-model configurations on \fcv. We report the Ranked Pairs (RP) position induced by human-relative advantage (\adv), the normalized advantage ($\normadv{}$), and the speedup (\speedup) as defined in \S\ref{sec:problem-statement}.
    }\label{tab:advantage-leaderboard}
    \end{minipage}
\end{table*}

\textit{Observation: Agents achieve non-trivial speedups over the baseline.} All evaluated configurations attain $\speedup > 1$ on \fcv relative to the baseline codebase (associated with the issue), indicating that agents can successfully identify and implement runtime-relevant changes.

\textit{Observation: Agents underperform experts on performance optimization tasks.} For all agents, the overall advantage, \adv,~is negative, indicating a fundamental performance gap. We also notice a disagreement between the \adv~and \speedup~metrics for many configurations, where large performance gains on certain `easier' tasks have a disproportionate influence on the global \speedup~score. The influence of such tasks is diminished in the \adv~score, which compares each agent improvement to the corresponding expert improvement; since tasks that are ``easier'' typically also admit larger expert speedups, this relative metric yields a more consistent difficulty reference.

\subsection{Large-Scale vs.\ Small-Scale Refactors}
\label{sec:results-scale}

\begin{figure}
    \centering
    \includegraphics[width=\linewidth]{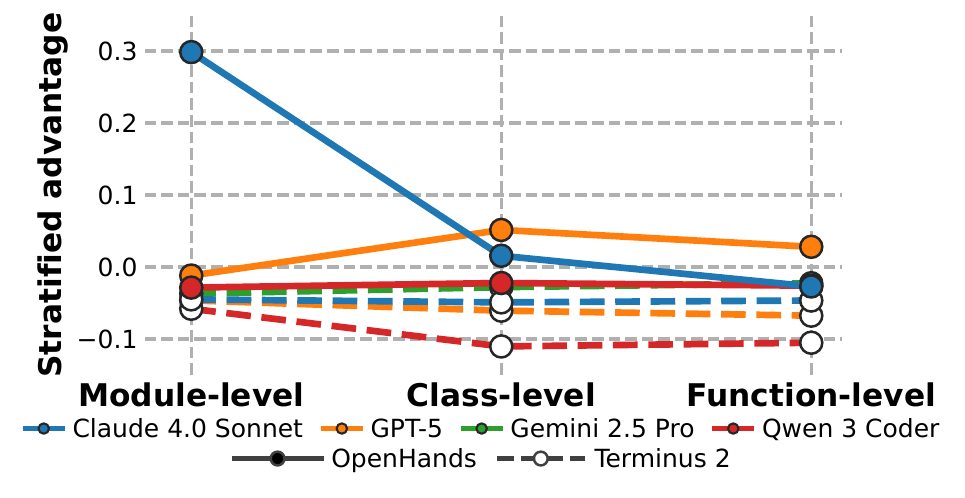}
    \vspace{-0.25in}
    \caption{
        Showing stratified advantage across hierarchy levels for each agent--model configuration.
        Each line traces the stratified advantage ($\adv_{\agent}^{(\ell)}$) over $\ell \in \{1,2,3\}$, revealing whether a configuration prefers coarse module-level changes or fine-grained function-level edits.
    }
    \label{fig:agg-advantage-ladders}
\end{figure}

To disentangle performance by optimization scale, we use the hierarchical structure of \fcv workloads (Figure~\ref{fig:hierarchical-overview}) and  stratified advantage  $\adv_{\agent}^{(\ell)}$ from \S\ref{sec:problem-statement}. We construct per-configuration profiles across three strata: \LevelOneName~level aggregation ($\ell=1$), \LevelTwoName~level aggregation ($\ell=2$), and \LevelThreeName~level aggregation ($\ell=3$). For each configuration and level $\ell$, we compute group-level speedups and advantages, shown in Figure~\ref{fig:agg-advantage-ladders}.

\textit{Observation: Agents demonstrate characteristic performance profiles.} In Figure~\ref{fig:agg-advantage-ladders}\, models exhibit diverse performance profiles. OpenHands + Claude 4.0 Sonnet performs best at the module-level optimization but underperforms at the function-level, indicating that this configuration can overlook small-scale optimizations in favor of large-scale ones. Conversely, OpenHands + GPT-5 performs best at the function-level but loses effectiveness at the module-level.

\textit{Observation: Agents are comparatively stronger on local optimizations.} With few exceptions (notably Claude 4.0 Sonnet + OpenHands), configurations achieve higher stratified advantage at function-level aggregation.

\subsection{Type of Optimization Problem}
\label{sec:results-type}

\begin{table*}[t]
    \centering
    \caption{
        Per-tag advantage for each agent--model configuration.
        Columns correspond to optimization tags (see \ref{tab:a.dataset.construction.question_composition}), and cells report the human-relative advantage restricted to workloads whose patches are annotated with the respective tag.
        OpenHands + GPT-5 shows strong advantage on algorithmic rewrites and data-structure changes, while other models perform comparatively better on micro-optimizations or caching.
    }
    \vspace{-0.1in}
    \label{tab:tag-analysis}
    \adjustbox{max width=\textwidth}{%
    \begin{tabular}{l l
                    r r r r r r r r r r r r r r}
        \toprule
        Agent & Model &
        Algo & Data & Lower & Approx & Parallel & Reduce & Cache & Batch & Scale & DB & Micro & I/O & Higher & Uncat \\
        \midrule
Terminus 2 & GPT-5 & -0.064 & -0.112 & -0.233 & -- & 0.010 & -0.006 & -0.054 & 0.028 & -- & -- & 0.001 & -- & -0.002 & -- \\
 & Claude 4.0 Sonnet & -0.019 & 0.011 & -0.720 & -- & 0.013 & -0.028 & -0.048 & 0.041 & -- & -- & -0.038 & -- & -0.009 & -- \\
 & Gemini 2.5 Pro & -0.029 & 0.011 & -0.676 & -- & 0.013 & -0.028 & -0.048 & 0.041 & -- & -- & -0.038 & -- & -0.007 & -- \\
 & Qwen 3 Coder & -0.023 & 0.007 & -0.455 & -- & 0.007 & -0.079 & -0.027 & 0.042 & -- & -- & -0.066 & -- & 0.005 & -- \\
OpenHands & GPT-5 & 0.015 & -0.052 & -0.211 & -- & 0.015 & -0.051 & -0.018 & 0.040 & -- & -- & -0.018 & -- & -0.008 & -- \\
 & Claude 4.0 Sonnet & -0.028 & 0.023 & -0.180 & -- & 0.007 & -0.049 & -0.017 & 0.047 & -- & -- & 0.086 & -- & -0.005 & -- \\
 & Qwen 3 Coder & -0.020 & -0.004 & -0.203 & -- & 0.012 & -0.016 & -0.019 & 0.051 & -- & -- & -0.063 & -- & 0.013 & -- \\
        \bottomrule
    \end{tabular}
    }
\end{table*}

We investigate whether models can outperform human experts on particular classes of optimizations. For each problem in \fcv, we label the optimization attempted by the expert patch using an LLM (see \S\ref{sec:a.dataset.composition} for details). Next, we aggregate the advantage of each agent-model pair within each optimization class. Table~\ref{tab:tag-analysis} summarizes the results.

\textit{Observation: Some optimization classes remain systematically difficult for agents.} We observe certain optimization categories where agents outperform experts. Specifically, all agents were able to find faster solutions in tasks where the expert attempted a parallelization or batching based solution.  Conversely, all agents struggle when the human solutions require delegating to lower-level system implementations (C extensions, vectorized operations).

\subsection{Long-Tail Generalization Across Repository Popularity}
\label{sec:results-popularity}

\begin{table}[t]
    \centering
    \small
    \setlength{\tabcolsep}{4pt}
    \renewcommand{\arraystretch}{1.15}
    \caption{
        Performance across repository popularity quintiles (by GitHub stars).
        We report $\adv_\agent$ for workloads drawn from repositories in each quintile, from least popular (Q1) to most popular (Q5). Red signifies worse performance.
    }
    \vspace{-0.1in}
    \label{tab:long-tail}
    \adjustbox{max width=0.48\textwidth}{%
    \begin{tabular}{l l r r r r r}
        \toprule
        Agent & Model &
        Q1 &
        Q2 &
        Q3 &
        Q4 &
        Q5 \\
        \midrule
        Terminus 2 & GPT-5
          & \trendnum{-0.0194}
          & \trendnum{0.0423}
          & \trendnum{-0.0045}
          & \trendnum{-0.2754}
          & \trendnum{-0.0123} \\
        & Claude 4.0 Sonnet
          & \trendnum{-0.0450}
          & \trendnum{-0.0062}
          & \trendnum{0.0025}
          & \trendnum{-0.3529}
          & \trendnum{-0.0220} \\
        & Gemini 2.5 Pro
          & \trendnum{0.0077}
          & \trendnum{-0.0062}
          & \trendnum{0.0024}
          & \trendnum{-0.3311}
          & \trendnum{-0.0445} \\
        & Qwen 3 Coder
          & \trendnum{-0.0691}
          & \trendnum{0.0052}
          & \trendnum{-0.0179}
          & \trendnum{-0.1669}
          & \trendnum{-0.0332} \\
        OpenHands & GPT-5
          & \trendnum{-0.0387}
          & \trendnum{0.0315}
          & \trendnum{0.0072}
          & \trendnum{-0.0769}
          & \trendnum{-0.0068} \\
        & Claude 4.0 Sonnet
          & \trendnum{-0.1041}
          & \trendnum{0.0291}
          & \trendnum{-0.0200}
          & \trendnum{-0.0378}
          & \trendnum{0.0263} \\
        & Qwen 3 Coder
          & \trendnum{-0.0159}
          & \trendnum{0.0137}
          & \trendnum{0.0227}
          & \trendnum{-0.0878}
          & \trendnum{-0.0270} \\
        \bottomrule
    \end{tabular}
    }
\end{table}

We next study how performance varies by repository popularity (measured using GitHub stars).
We compute advantage statistics for each popularity quintile. 

\textit{Observation: Agents underperform experts on tail repositories.} Agent performance is lower in the first popularity quintile (Q1; bottom 20\%), which comprises repositories with 133-202 GitHub stars. Expert patches, however, yield comparatively large gains in this regime: $\speedup_{\expert}(\mathrm{Q1}) = 1.1104$, the second-largest speedup across quintiles. One hypothesis is that smaller repositories contain more heterogeneous, high-impact micro-optimizations that may have already been discovered in larger, more mature repositories, leading to more variable (but sometimes high-impact) optimization opportunities.  A second plausible hypothesis is distribution shift: smaller repositories may be less represented in training corpora, reducing agent effectiveness.

\textit{Observation: Agents are most competitive on mid-popularity repositories.} In the 20th to 60th percentile range, mean advantages are closest to expert performance, and some configurations perform comparably with experts. We hypothesize that this is due to two reasons.  First, moderately popular repositories more closely match the agent's training distribution than tail repositories. Second, these repositories have more unexploited optimization avenues relative to highly popular projects.

\textit{Observation: Performance dips in high-popularity repositories.} Agent performance is lowest in the fourth quintile (Q4; 6,371-10,343 stars). In this regime, expert patches also yield the smallest gains: $\speedup_{\expert}(\mathrm{Q4}) = 1.0822$, the lowest expert speedup across all quintiles. This pattern indicates reduced remaining optimization headroom in these repositories, where many simpler improvements may have already been realized. Additionally, slight distribution shift may persist and limit agent effectiveness.

\subsection{Practical Considerations}

\subsubsection{Cost Efficiency.}
\label{sec:results-cost}

\begin{figure}
    \centering
    \includegraphics[width=\linewidth]{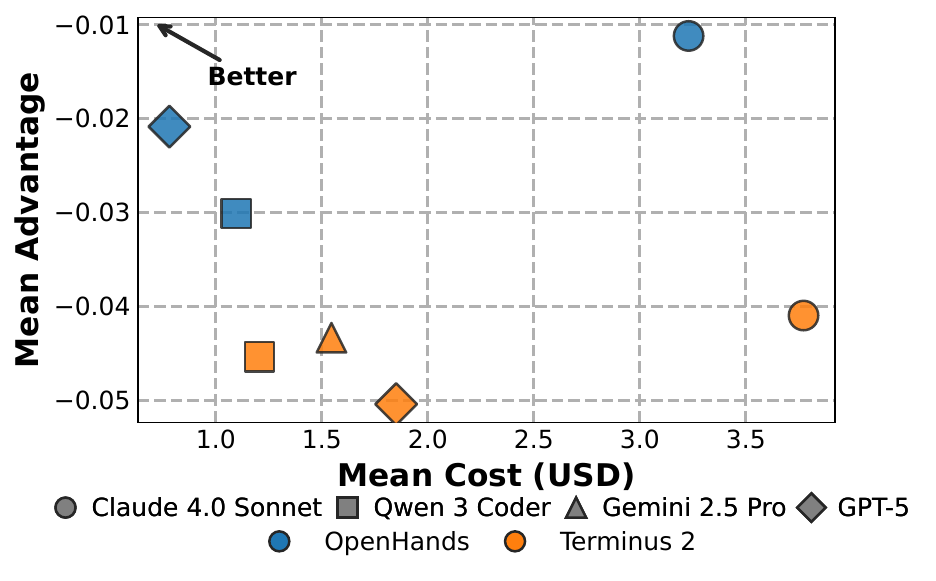}
    \caption{
        Cost-Performance tradeoff of agent-model configurations. As most agents struggle on code optimizations tasks, the pareto set is primarily dominated by the most expensive model (Claude 4.0 Sonnet).
    }
    \label{fig:cost-performance-tradeoff}
\end{figure}

Frontier models differ substantially in end-to-end inference cost due to provider pricing and the number of tokens consumed by a given agent configuration. In this experiment, we consider the cost--performance tradeoff within our agent configurations using the cost-weighted objectives defined in \S\ref{sec:problem-statement}. Table~\ref{tab:cost-advantage-leaderboard} reports a leaderboard based on cost-weighted normalized advantage, and Figure~\ref{fig:cost-performance-tradeoff} summarizes the resulting trade-off.

\textit{Observation: Higher-priced models rank best under the cost-weighted objective.} When weighted by cost, top-ranked configurations tend to use the higher-priced (and more capable) models. A contributing factor is that lower-capability models often consume more tokens within the agent loop, which can offset lower per-token prices. This might also indicate that smaller models lack the capabilities to reason effectively about performance optimizations.

\subsubsection{Multi-Workload Tradeoff Performance.}
\label{sec:results-tradeoff}
\begin{figure}
    \centering
    \includegraphics[width=\linewidth]{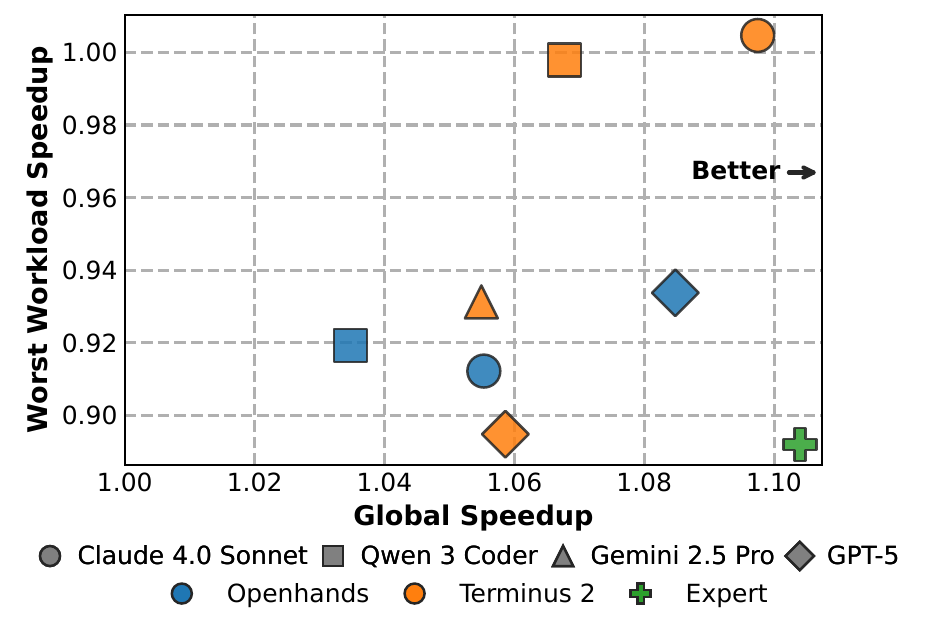}
    \caption{
        Multi-workload tradeoff performance of agent-model configurations. We quantify a model's speedup performance as a function of its worst regression.
        The expert patch achieves the highest speedup while negotiating considerably high workload regressions.
    }
    \label{fig:optimization-tradeoff}
\vspace{-2em}
\end{figure}

Performance optimization necessitates a holistic understanding of competing workloads. In this experiment, we compare the global speedup achieved by a model with the largest regression it causes. For each agent--model configuration, we compute (i) global speedup aggregated across tasks and workloads, and (ii) the average \emph{worst-workload} speedup, defined as follows: for each task, we take the minimum speedup across the task workloads, and then average this minimum across tasks. Figure~\ref{fig:optimization-tradeoff} plots these two quantities.

\textit{Observation: Multi-workload optimization remains challenging for agents.} Despite causing large regressions, human code edits achieve the best global speedup, indicating a superior ability to negotiate multi-workload performance tradeoffs than our configurations.

\subsubsection{Temporal Generalization.}
\label{sec:results-leakage}

\textbf{Motivation.} \fc is a live benchmark: tasks are continuously added and include creation timestamps. This enables us to probe the temporal out-of-distribution behavior of agents on performance optimization tasks. Related work on code correctness finds large gains when tasks are present in training corpora \citep{jain2024livecodebench}.

\begin{table}[t]
    \centering
    \small
    \setlength{\tabcolsep}{4pt}
    \renewcommand{\arraystretch}{1.15}
    \caption{Temporal analysis of model performance across knowledge cutoff boundaries.
    Each column represents a temporal bin defined by distance (in months) from the model's training data cutoff; values indicate mean global speedup ($\speedup_\agent$) within each bin. We find no consistent drop in performance.}
    \label{tab:temporal-distance}
    \adjustbox{max width=0.48\textwidth}{%
    \begin{tabular}{l r r r r r r}
    \toprule
    & \multicolumn{3}{c}{Before Cutoff} & \multicolumn{3}{c}{After Cutoff} \\
    \cmidrule(lr){2-4} \cmidrule(lr){5-7}
    Model & 6+ mo & 3-6 mo & 0-3 mo & 0-3 mo & 3-6 mo & 6+ mo \\
    \midrule
    Claude 4.0 Sonnet & 1.0892 & 1.0564 & 0.9966 & 1.0915 & 1.0951 & 1.0519 \\
        GPT-5             & 1.1708 & 1.0454 & 0.9871 & 1.0378 & 1.0679 & 1.0500 \\
        Gemini 2.5 Pro    & 1.1071 & 0.9989 & 1.0219 & 1.0523 & 1.1063 & 1.0251 \\
        \bottomrule
    \end{tabular}
    }
\end{table}

We bucket tasks by their month of creation and compute mean global speedup in windows defined by the temporal distance to each model's knowledge cutoff (\S\ref{sec:a.expt.models}). We use 3-month bins and consider bins up to 6 months before/after the cutoff. Table~\ref{tab:temporal-distance} summarizes results.

\textit{Observation: Limited evidence of a cutoff-aligned leakage effect.} Performance shows no consistent shift when moving from pre-cutoff to post-cutoff task creation dates, suggesting the gap is capability-based rather than data-based.

\section{Related Work}\label{sec:related-work}

{

\newcommand{\good}{\cellcolor{green!10}}
\newcommand{\warn}{\cellcolor{yellow!10}}
\newcommand{\bad}{\cellcolor{red!10}}

\newcommand{\maybe}{\textbf{\ensuremath{\cdot}}}

\newcolumntype{Y}{>{\raggedright\arraybackslash}X}

\newcommand{\icon}[1]{\raisebox{-0.2\height}{\includegraphics[height=1.05em]{#1}}}
\newcommand{\githubicon}{\icon{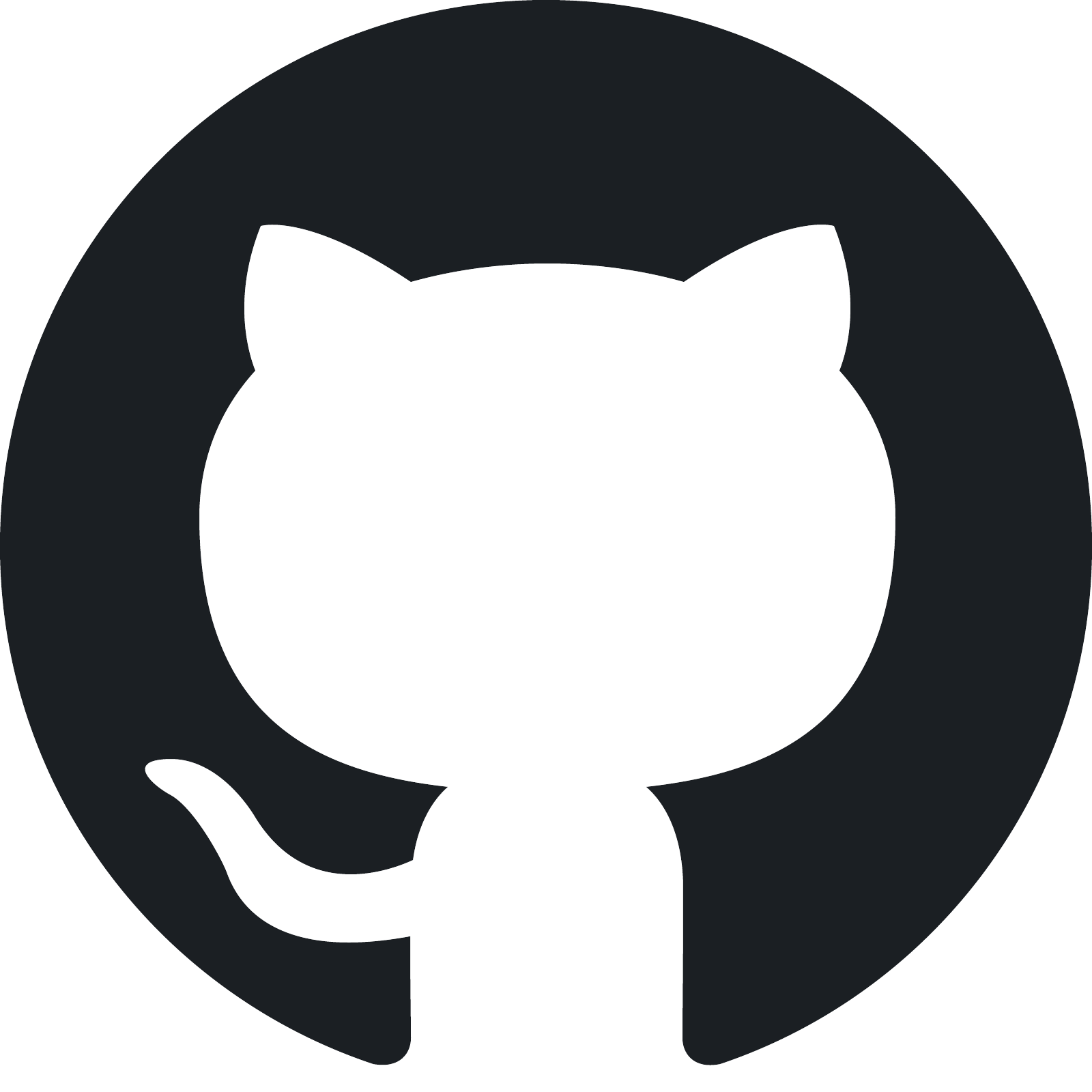}}
\newcommand{\atcodericon}{\icon{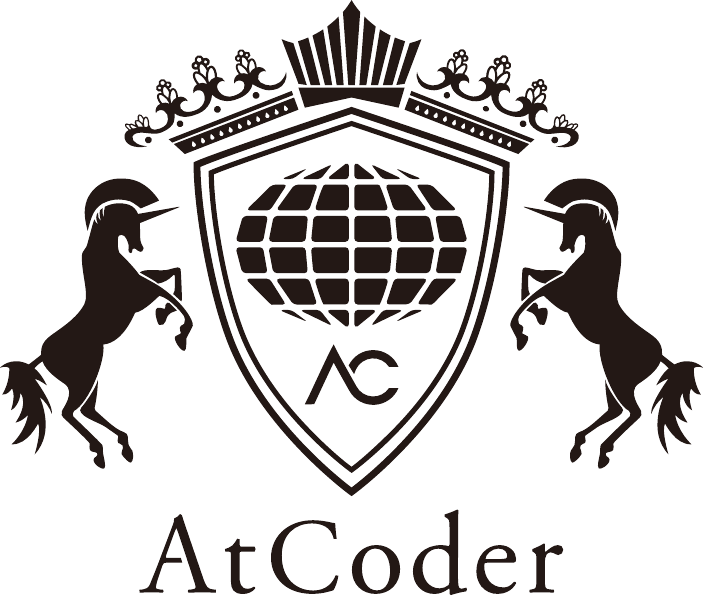}}
\newcommand{\leetcodeicon}{\icon{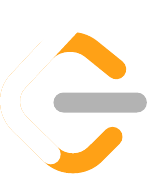}}
\newcommand{\codeforcesicon}{\icon{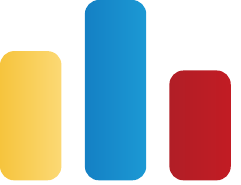}}
\newcommand{\autoicon}{\icon{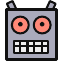}}

\begin{table*}
\caption{Comparing \fc with related codebase benchmarks. \fc is the only benchmark that satisfies the desired properties for evaluating LLM agents on real-world code optimization tasks.
  {++} denotes continually updating benchmarks.
  Data is sampled from real distributions like GitHub (\protect\githubicon), Leetcode (\protect\leetcodeicon), AtCoder (\protect\atcodericon), and Codeforces (\protect\codeforcesicon); and LLM-generated or synthetic distributions (\protect\autoicon). An extended analysis is presented in \S\ref{sec:related-work}.}
  \vspace{-0.1in}
  \centering
  \footnotesize
  \setlength{\tabcolsep}{4pt}
  \renewcommand{\arraystretch}{1.15}

  \adjustbox{max width=\textwidth}{%
  \begin{tabularx}{\textwidth}{
    l  %
    Y  %
    l  %
    c  %
    c  %
    c  %
    c  %
    c  %
    c} %
    \toprule
    \textbf{Benchmark} &
    \makecell{\textbf{Evaluation}\\\textbf{framework}} &
    \textbf{\# Tasks} &
    \makecell{\textbf{\# Workloads}\\\textbf{/ Task}} &
    \makecell{\textbf{Live}\\\textbf{updates}} &
    \textbf{Data source} &
    \makecell{\textbf{Search}\\\textbf{space}} &
    \makecell{\textbf{Synthesis}\\\textbf{scope}} &
    \makecell{\textbf{Leakage}\\\textbf{resistant?}} \\[2pt]
    \midrule

    GSO-Bench             & Performance & 102    & \bad Single &
    \bad \xmark &
    \warn \githubicon~;~\autoicon &
    \good Large &
    \good Repo &
    \bad \xmark \\[2pt]

    SWE-Bench        & Unit Tests & 2292   & \bad - &
    \bad \xmark &
    \good \githubicon &
    \bad Small &
    \good Repo &
    \bad \xmark \\[2pt]

    LiveCodeBench    & Unit Tests & 300~\textsuperscript{++}  & \bad - &
    \good \cmark &
    \good \leetcodeicon~;~\atcodericon~;~\codeforcesicon &
    \bad Small &
    \bad File &
    \good \cmark \\[2pt]

    SWEfficiency     & Performance \& Unit Tests & 400    & \bad Single &
    \bad \xmark &
    \good \githubicon &
    \good Large &
    \good Repo &
    \bad \xmark \\[2pt]

    SWE-Perf         & Performance \& Unit Tests & 140    & \bad Single &
    \bad \xmark &
    \good \githubicon &
    \good Large &
    \good Repo &
    \bad \xmark \\[2pt]

    CruxEval         & Unit Tests & 800~\textsuperscript{++}  & \bad - &
    \bad \xmark &
    \bad \autoicon &
    \bad Small &
    \bad File &
    \good \cmark \\

    FormulaCode      & Performance \& Unit Tests & \fcsize\textsuperscript{++}  & \good \avgWorkloadsPerProblem{} &
    \good \cmark &
    \good \githubicon &
    \good Large &
    \good Repo &
    \good \cmark \\[2pt]

    \bottomrule
  \end{tabularx}
  }
  \label{tab:related-work}
\end{table*}
}

\textbf{Algorithms for Code optimization.} 
There is a long history of research on iterative code optimization using execution feedback. 
Classical approaches to this problem were based on stochastic search and constraint solving~\citep{schkufza2013stochastic, souper}.
Among deep-learning based approaches, 
AlphaTensor and AlphaDev produce super-optimized matrix multiplication and sorting routines, respectively \citep{fawzi2022discovering, mankowitz2023faster}. These systems combine large, publicly sourced pretraining datasets with carefully chosen inductive biases to make optimization faster.  
The more general 
\textit{agentic Optimization} workflows 
operate by 
iteratively running LLM-generated code, evaluating the output, and feeding the output back to the model. Terminus 2 and OpenHands represent two such configurations out of many that benefit from iterative feedback \citep{yao2024language,yang2024sweagentagentcomputerinterfacesenable, terminalbench, wang2025openhandsopenplatformai, merrillShaw2025Terminus}.
\fc is the first benchmark purpose built to assess the multi-workload optimization ability of such agentic AI algorithms in real-world codebases and provides the fine-grained evaluation functions needed for iterative optimization.

\textit{Evolutionary Optimization} algorithms equipped with LLMs~\citep{romera2024mathematical,grayeli2024symbolic} iteratively improve a candidate pool of programs using execution feedback. 
Systems like AlphaEvolve~\citep{novikov2025alphaevolve} and OpenEvolve~\citep{openevolve} demonstrate that such agents can efficiently discover and refine novel, high-performance code-based heuristics across diverse scientific domains. These methods are scalable but require high quality evaluation functions to penalize degenerate solutions. While \fc provides the necessary evaluation functions, we could not benchmark evolutionary methods due to their substantial compute needs.

\textbf{Code Generation Benchmarks.} Coding benchmarks can be differentiated by their synthesis scope. For a list of differences, consult Table~\ref{tab:related-work}.

\textit{Function and file level.} 
HumanEval~\citep{chen2021evaluating} and MBPP~\citep{austin2021program} present hand-written programming problems in Python with corresponding unit tests. Many contributions extend these benchmarks to have more testing~\citep{evalplus}, broader scope~\citep{yin2022natural,yang2023intercode}, and more task diversity~\citep{muennighoff2023octopack, lai2022ds1000,zan2022cert}. CruxEval~\citep{gu2024cruxeval} benchmarks the code execution and reasoning ability of LLMs more deeply. LiveCodeBench~\citep{jain2024livecodebench} attempts to mitigate data-leakage by annotating problems with release dates. All these benchmarking efforts utilize unit testing suites to gauge program correctness. \fc 
\textit{supplements} the evaluation signal provided by the above datasets by using community-maintained evaluation functions that continually update with each commit. 

\textit{Repository level.} Function and file level benchmarks evaluate coding ability on self-contained coding tasks. However, real software issues typically span multiple modules and files. Repository level benchmarks~\citep{jimenez2024swebenchlanguagemodelsresolve, tang2024mlbenchevaluatinglargelanguage, jain2024r2e, shetty2025gsochallengingsoftwareoptimization} aim to preserve the inherent challenges in real-world software engineering beyond text completion, such as finding relevant files, capturing relationships between modules, tracing information flow, etc. SWE-Bench~\citep{jimenez2024swebenchlanguagemodelsresolve} collects GitHub issues from popular repositories and evaluates coding agents' ability to resolve the issues. Follow-up efforts benchmark agents on repository-conditioned code synthesis~\citep{tang2024mlbenchevaluatinglargelanguage}, scale-up benchmarking by admitting smaller codebases with LLM-generated unit tests~\citep{jain2024r2e}, and introduce continually updating pipelines for the task~\citep{zhang2025swebenchgoeslive}. Such extensions provide valuable insights into LLM agent behavior yet ground their evaluations in correctness tests, that present a discrete optimization surface for the agents. \fc \textit{complements} these benchmarks by assessing agents on community-maintained evaluation functions that present a smoother optimization landscape and higher fidelity than unit tests.

\textbf{Optimization Benchmarks.} There are prior benchmarks for efficient code synthesis on function and file-level tasks. COFFE~\citep{peng2025coffe} samples tasks from HumanEval, MBPP, CodeContests, and APPS~\citep{chen2021evaluating, austin2021program,hendrycks2021measuring} and auto-generates stress tests while ECCO~\citep{waghjale2024ecco} curates a function and file-level efficient synthesis dataset from IBM CodeNet \citep{puri2021codenet} with data-mined test cases. 

Recent repository-level benchmarks like GSO-Bench \citep{shetty2025gsochallengingsoftwareoptimization} and SWEfficiency \citep{ma2025swefficiencylanguagemodelsoptimize} also study LLM agents' ability to optimize code. However, these benchmarks only optimize for a single target function at a time.
SWE-Perf~\citep{he2025sweperflanguagemodelsoptimize} is a closely related concurrent benchmark comprising 140 instances derived from performance-improving PRs across 9 Python repositories, measuring performance improvement via unit test runtime. \fc differs in scale (\fcsize tasks from \numFinalRepos repositories), evaluation signal ($\sim$\avgWorkloadsPerProblem dedicated performance workloads per task vs.\ unit test timing as a proxy), and contamination resistance (live monthly updates). In contrast to all prior optimization benchmarks, \fc focuses on: (1) using community-maintained benchmarks specifically designed to profile performance inefficiencies instead of using hand-curated stress tests or unit test timing, (2) benchmarking on repository-level codebases, which better capture the natural challenges with real-world code optimization, and (3) presenting multiple workloads that can compete with one another to assess the holistic optimization ability of agents.

\section{Conclusion}\label{sec:conclusion}

We present \fc, a comprehensive coding benchmark for repository-level agentic optimization.  In this benchmark, coding agents must not only write code that passes standard correctness tests, but also improve runtime, and our benchmark design enable us to study the impact of repository popularity, temporal cutoffs, and multi-scale optimization to guide the design of future agents capable of surpassing human experts. 
To ensure longevity and prevent saturation, we operate as a live benchmark, continually ingesting new tasks to test agents against an evolving human baseline.
Our evaluations show that \fc~is a challenging benchmark for frontier LLMs and agentic frameworks, leaving open significant room for future agent development.

\paragraph{Limitations.} \fc has several limitations, which we discuss in more detail in Appendix~\S\ref{sec:discussion-limitations}. First, \fc currently covers scientific Python repositories that ship Airspeed Velocity (ASV) benchmarks so many applications like web frameworks, distributed systems, embedded software, and other languages remain out of scope. Extending \fc to these domains is an interesting problem for future work. Second, the fidelity of our performance signal is bounded by the quality of the community-maintained workloads: a patch's measured gain reflects only the code paths that the workloads exercise, so speedups on uncovered paths go unmeasured. The large number of workloads per task (\numworkloads~on average) keeps this coverage broad, though not exhaustive. Third, due to compute constraints our full agent evaluation runs on the \fcv subset (\numFCVProblems~tasks) rather than the complete \fcsize-task benchmark; \fcv preserves the repository and difficulty distribution of \fc, but results may not transfer perfectly to the full benchmark. Fourth, all \fc tasks are run on a specific AWS configuration, so absolute measurements are hardware-specific---advantage, computed as a difference of speedup ratios on identical hardware, is more robust to this than raw timing, but cross-platform behavior remains untested.
\clearpage

\section*{Acknowledgements}

This work was supported in part by a Laude Institute Slingshot Award, NSF awards III-\#2505097, PPoSS-\#2316161, NSF \#2505096, NSF \#2505098, and gifts from Point72 and OpenAI.

We also thank Alex Shaw, Braden Hancock, Miles Cranmer, Neehar Kondapaneni, Rogério Guimarães, Anant Asthana, Arjun Sharma, Alex Farhang, and Markus Marks for helpful discussions.

\section*{Impact Statement}

We have presented \fc: a benchmark for measuring the capabilities of LLM-guided agents to optimize performance on large codebases. \fc is designed to serve two audiences:  researchers (those developing new LLMs / Agents) and practitioners (those using Agents for daily workflows). For researchers, we hope that \fc accelerates the development of coding agents by providing contamination-free training and evaluation signals. 
For practitioners, we hope \fc offers \textit{comparative} metrics that gauge the utility of LLMs and agents in specialized repositories under diverse cost-performance constraints. In this section, we discuss the broader societal impacts and ethical considerations of our work.

\textit{Potential for Misuse.} Benchmark results are only as reliable as the interpretations drawn from them.
To ground evaluations in realistic developer workflows, we use community-maintained workloads that already exist in each repository and attempt to preserve the same information and performance instrumentation available to a human contributor. This design also supports practical impact: strong model-generated changes can, in principle, be merged upstream to reduce maintenance burden, particularly for smaller repositories after thorough manual analysis. At the same time, reliance on repository workloads introduces an attack surface: an adversary could submit pull requests that alter or add workloads to make tasks artificially easier. While such additional workloads can increase regression coverage (thereby providing some downstream utility), practitioners should treat workload provenance and review practices as part of the evaluation's trust boundary.

\textit{Privacy Concerns.} \fc is an `open-book' benchmark and necessarily includes interactions from open-source software developers. We include such context to provide models access to the same information a human would use when solving these tasks. Although we anonymize usernames and remove personally identifiable information to the best of our ability, some contributors may remain indirectly identifiable via secondary cues (e.g., writing style, repeated project-specific references).

\textit{Bias and Fairness.} Benchmarks can incentivize and influence which capabilities are prioritized by the community. We strive to make \fc's metrics explicit and stable, and we apply statistical analyses to reduce unintended measurement artifacts. Yet, \fc inherits limitations from the underlying repository benchmarks. In particular, \fc is susceptible to a form of the Quantitative Fallacy: aspects of agent competence that are difficult to measure may be underweighted or omitted, inflating the true utility of such algorithms. This is a limitation of all execution-based benchmarks. We therefore recommend using \fc as a \textit{complementary} signal rather than as a substitute for careful manual assessment of Agent / LLM behavior.

\bibliography{neurosymbolic,symreg}
\bibliographystyle{icml2026/icml2026}

\clearpage
\enablevspace
\appendix
\onecolumn

\section{Dataset Construction Details}\label{sec:a.dataset}

In this section, we provide details on the dataset construction process. Our core aim is to provide an automated pipeline for constructing a dataset of pull requests that are relevant for performance benchmarking. The dataset was constructed on a single machine with Ubuntu 22.04 LTS running on a machine with 503 GiB RAM, a dual-socket Intel Xeon Platinum 8352Y CPU @ 2.20 GHz (128 hardware threads), equipped with 4xNVIDIA A40 GPUs (46 GiB VRAM each). Making the dataset from scratch takes $\sim 32$ hours, consuming $\sim 100$ GB of disk space for the metadata and $\sim 2$ TB of disk space for the docker image cache. 

We use two LLMs during the dataset construction process. For less complex tasks such as textual classification and extraction, we use \texttt{openai/gpt-oss-120b} model served locally \citep{kwon2023efficient, openai2025gptoss120bgptoss20bmodel}. For complex tasks such as environment build script synthesis, we first attempt to use the local LLM and fallback to the \texttt{anthropic/claude-3-5-sonnet-20241022}  \citep{c3modelfamily} model (with a one-time total cost of $\$446$ for the entire dataset). The additional cost may change if a different locally available LLM is utilized.

\subsection{Repository Scraping}\label{sec:a.dataset.scraping}

We identify compliant repositories by searching for the presence of mature tools developed within the Python performance benchmarking community. To search for these repositories at scale, we develop a CommonSQL script to search for the presence of performance-oriented tools and workloads in the GitHub Public Dataset on Google BigQuery \citep{bigquery_github_repos_2025}, which snapshots about \numGithubRepos{} open-source repositories and \numCodeFiles{} code files. We add additional filters to ensure only mature software packages are considered. Specifically, we ensure that each valid repository has (1) Markers identifying the presence of at least one performance workload (e.g., \texttt{asv.conf.json}); (2) does not fork an existing repository. (3) Presence of PR merges and active maintenance in the last three years. (4) Support for Python 3.8+. This leaves us with \numReposDiscovered repositories.

The CommonSQL script executes in about $48$ seconds and cost $\$9.4$. As an alternative, we can also use the GitHub Search API to query for the repositories. This yields the same number of repositories, but can be much slower due to API rate limits.

\subsection{Rule-based Filtering}\label{sec:a.dataset.filtering}

Once we have a list of compliant repositories, it is technically possible to execute and measure the performance of all pull-requests in the repository. However, as most pull-requests do not primarily intend to improve performance, this leads to unnecessary waste of compute resources. The rule-based filtering stage ensures that we collect performance metrics for only those pull requests where we can ensure that the pull request is suitable for benchmarking. Most filters in this stage aim to identify unambiguous signals that disqualify a pull request from being used for benchmarking. The prominent filters are listed below:

\begin{itemize}
    \item \textbf{Repository Compliance:} We select repositories that have at least \minStars{} GitHub stars. Below \minStars{} stars, we found that repositories often lacked the necessary community engagement to produce good quality pull requests.
    \item \textbf{Pull Request Status:} We strictly filter for pull requests that have been successfully merged (\texttt{state='closed'} with a valid \texttt{merged\_at} timestamp) within the target date range. We also ensure that we can retrieve and successfully apply the patch to the repository.
    \item \textbf{Benchmarking Infrastructure:} The specific commit tree must contain an Airspeed Velocity (ASV) configuration file (\texttt{asv.conf.json}), ensuring the repository supported benchmarking at that point in history.
    \item \textbf{Core Content:} We explicitly exclude commits that only touch non-functional paths, such as \texttt{tests/}, \texttt{docs/}, \texttt{examples/}, \texttt{.github/}, \texttt{dist-info/}, build artifacts, or packaging metadata (e.g., \texttt{pyproject.toml}, \texttt{requirements.txt}).
    \item \textbf{Heuristic Message Filtering:} We apply a regex-based pre-filter to the commit message. Commits matching ``negative'' patterns (e.g., ``revert'', ``release'', ``bump version'', ``fix typo'', ``formatting'') are discarded unless they also contain ``positive'' performance keywords (e.g., ``speed'', ``optimize'', ``latency'', ``throughput'', ``memory'', ``vectorize''). Ambiguous messages are retained for LLM classification.
    \item \textbf{Complexity Constraints:} To ensure feasibility for both the LLM context and the build system, we exclude commits that change more than 500 files or 80,000 lines of code, or where the patch size exceeds an acceptable context window for a capable local LLM (64,000 tokens). These constraints can be adjusted based on the future capabilities of LLMs.
    \item \textbf{Build Environment:} We clone each repository at the specific commit tree and attempt to build it using \texttt{uv}. \texttt{uv} is a fast python package manager that can be used to install dependencies from a project's dependency files (e.g., \texttt{pyproject.toml}, \texttt{requirements.txt}, or \texttt{setup.py}). If the build fails, we discard the pull request. If the build succeeds, we pin the dependencies to ensure that the build environment can be reproduced. This is a compute-intensive process and, after parallelizing the build process, requires $\sim 13$ hours for all pull requests on our machine.
\end{itemize}

After applying these filters, we are able to select \numPRsScraped{} pull requests from \numReposAttributeFiltering{} repositories that are suitable for benchmarking.

\subsection{Performance Intent Filtering}\label{sec:a.dataset.llmfilters}

The previous stage ensures that we only select pull requests that are suitable for benchmarking. However, it is still possible that the pull request does not primarily intend to improve performance.
To ensure that we only select pull requests that are suitable for benchmarking, we utilize a pre-trained local LLM to classify the pull request as performance improving.  

The primary objective of this classifier is to filter out pull requests that pass the regex-based heuristic but are not \textit{bona fide} performance optimizations. Common examples of such false positives include commits that contribute new features instead of improving performance, refactor code structure without runtime impact, or maintainability improvements. The classifier analyzes the pull request description, file change summary, and the code patch to make this determination. 

The classifier is written in DSPy \citep{khattab2023dspycompilingdeclarativelanguage} and the prompt is shown in Figure \ref{fig:problem-extractor}. We explicitly prioritize recall over precision. The prompt is configured to lean towards a ``YES'' classification in ambiguous cases. This design choice is deliberate, as false positives will be symbolically verified in the subsequent benchmark execution stage, and discarded if they yield no measurable speedup. 

\begin{figure}
    \centering
        \centering
        \includegraphics[width=\linewidth]{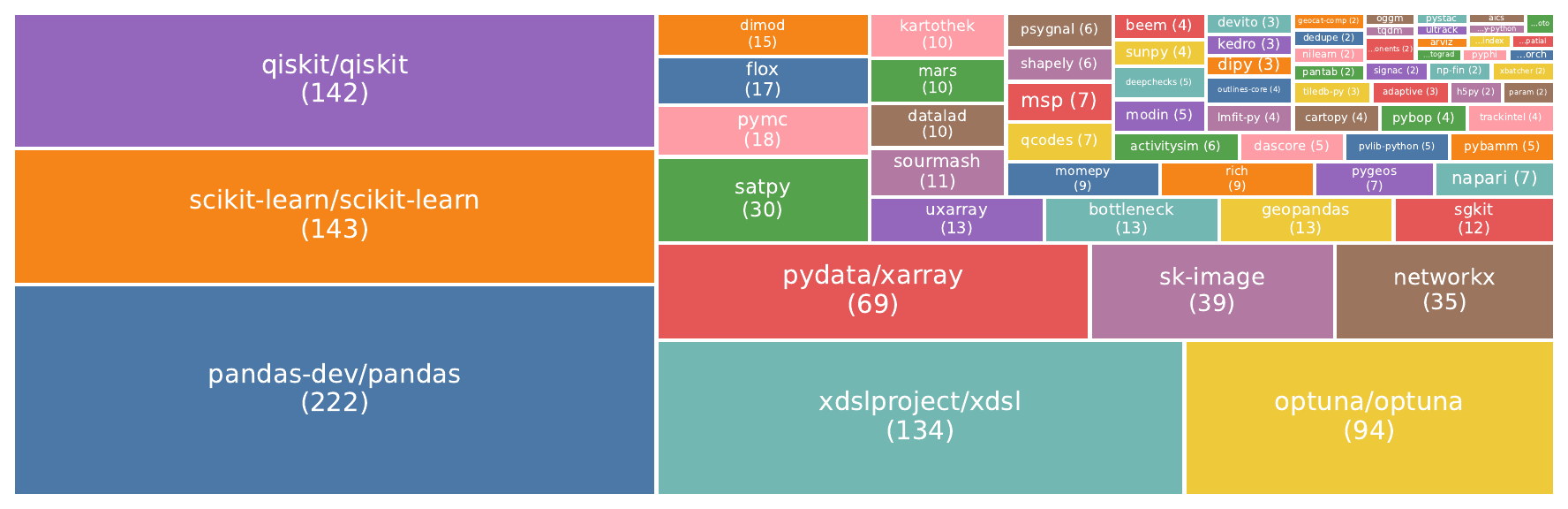}
    \caption{Distribution of tasks across repositories in \fc till November, 2025. \fc comprises of \fcsize tasks sampled from \numFinalRepos{} diverse open source GitHub repositories. Most repositories are software tools used extensively within scientific communities. \fc shows a strong long-tail pattern of bespoke repositories that are rarely covered in contemporary code-generation datasets. Table \ref{tab:a.formulacode-repos} presents a detailed overview.
    }
    \label{fig:a.dataset-statistics.repos}
\end{figure}

\subsection{Problem Statement Construction}\label{sec:a.dataset.problem_construction}

To transform a raw pull request into a benchmark task, we must construct a clear, self-contained problem statement that defines the performance goal. We employ a multi-stage pipeline to aggregate context and extract a structured narrative.

\paragraph{Context Aggregation.} For each candidate pull request, we scrape all available metadata (title, body, labels, and comments, date of creation, and date of merge) that can be used to construct the problem statement. We also fetch the file change summary and the raw patch content to ground the problem statement in the actual code changes. We parse the pull request body and comments to identify linked issues (e.g., \texttt{\#123}, \texttt{owner/repo\#123}). These references are resolved to their full issue descriptions and discussions, which are also parsed and aggregated into the problem statement. We only include information that was available before or at the time the pull request was created to ensure that the problem statement is self-contained.

\paragraph{Context Filtering.} Before attempting extraction, we enforce a strict validity check: a pull request must have at least one linked issue or a descriptive body. The rationale for this constraint is twofold. First, the linked issue typically provides the problem context (the bug report, performance regression analysis, or feature request) that motivated the change. Second, a descriptive pull request provides details of the problem solved, the methodology used, and the solution, which can be helpful for computing metadata for the benchmark task as well as clarifying the overall task goal.

\paragraph{Context Extraction.} We consolidate all linked issues into a single document using a static template (shown in Figure~\ref{fig:performance-classifier-example}). In principle, the issue text alone should sufficiently describe the initial observed performance regression or bottleneck. However, in practice, we find that while an issue provides the initial observed performance regression or bottleneck, they frequently bundle multiple optimization directions that are implemented across several pull requests. As a result, a problem statement derived only from the issue can under-specify the problem statement's starting state, leading to an ambiguous task (an agent may optimize a different aspect than the original change). 

To ensure that each problem statement provides a clear and self-contained description of the problem, we use another specialized LLM-based classifier to extract relevant problem context from the pull request description. We instruct the agent to specifically extract near-verbatim sentences corresponding to the performance goal and constraints relevant to this PR. Each extracted sentence is symbolically verified to maintain a high degree of textual fidelity (High Longest Common Subsequence ratio) to preserve technical terms, error messages, and code snippets. Any pull request that fails to yield a valid problem context is discarded as it lacks a defined starting state for the benchmark. This LLM-based extraction agent is implemented using DSPy, and the prompt is shown in Figure \ref{fig:build_agent_prompt_structure}. 

\paragraph{Examples.} Figure \ref{fig:performance-classifier-example} shows problem statements for some \fc tasks. Each problem statement has an initial set of static instructions, information about the problem extracted from the linked issues, and the initial direction of optimization extracted from the pull request description.

The problem statement construction (\S\ref{sec:a.dataset.problem_construction}) and the performance intent filtering (\S\ref{sec:a.dataset.llmfilters}) stages are applied together to yield \numPerfPRs{} problems.

\subsection{Synthesizing Reproducible Environments}\label{sec:a.dataset.stage3}

\paragraph{Motivation.} A critical challenge in benchmarking historical commits is that the build environment (dependencies, compilers, and system libraries) is often implicit and evolves over time. Simply installing the package via \texttt{pip} is insufficient for performance benchmarking for two main reasons: First, performance-critical Python packages often rely on compiled extensions (C/C++, Cython, Fortran) that must be built from source to accurately reflect the performance characteristics of the code at that specific commit. Installing pre-built binaries (wheels) would benchmark the packaged version rather than the code in the pull request. Second, developers often introduce bespoke dependencies or modify build configurations in a pull request, rendering previous environments obsolete. To address this, we implement an agentic pipeline to synthesize a reproducible Docker environment for each task. 

\paragraph{Setup.} For each task, we first construct a Docker container with the base dependencies installed (Refer to the `Build Environment' subsection under \S\ref{sec:a.dataset.filtering}) containing the source code of the repository at the initial state of the pull request. Our goal is to synthesize a build script that contains shell commands to install an editable version of the package from source. We also want to ensure that certain tools (ASV, PyTest, and our snapshot testing tool) can be successfully run in the container.

\paragraph{Agent.} We employ an iterative, reflexive agent to synthesize a valid build script. The agent is described in Figure \ref{fig:build_agent_prompt_structure} and has four principal components:

\textit{Validation \& Feedback Loop:} The synthesized script is executed in an isolated Docker container. We validate the build using two verification subroutines. (1) A \texttt{profile} check ensures that the package is importable, runnable, and we can run the ASV benchmarks under a generous timeout. (2) A \texttt{pytest} check ensures that we can run the pytest test suite without errors. If the build or validation fails, the \texttt{stderr} and \texttt{stdout} logs are fed back to the agent as observations, allowing it to iteratively refine the script (e.g., installing missing system libraries, fixing syntax errors).

\textit{Chronological Retrieval:} We leverage the insight that build requirements rarely change drastically between adjacent commits. For a given task, we sample $10$ successful build scripts from the same repository, sourced from a database of successfully built tasks, sorted by commit date. We first attempt to build the container using the script from the nearest chronological neighbor. If the build or verification fails, we move to the next neighbor until we either find a successful build or run out of neighbors. The failure logs are preserved and used as observations for the agent.

\textit{Agentic Synthesis:} If the retrieved scripts fail (or no history exists), we instantiate an LLM-based agent to generate a new build script. The agent acts as an interactive planner with access to the failure logs and a set of tools that allows it to inspect the repository state (e.g., list directories, read files, parse \texttt{setup.py} or \texttt{pyproject.toml}). Given $10$ interactive turns, the model can either choose to use one of the tools or prematurely end the turns by synthesizing a build script. The largest model we tried (Claude Sonnet 3.5 and GPT OSS 120b) rarely chooses to use tools as the error messages provide sufficient context while the smallest model (Meta Llama 3.3 8B; \citep{llama3modelcard}) often utilizes many tool interactions before synthesizing the build script.

\textit{LLM Choice and Prompt Design.} We find that a locally hosted \texttt{openai/gpt-oss-120b} provides the best balance of performance and cost. We also implement a fallback to \texttt{anthropic/claude-3-5-sonnet-20241022} if the build script synthesis fails after multiple tries. Overall, the chronological caching and local LLM cascade allows us to successfully synthesize a build script for $\numTasksContainers$ out of $\numPerfPRs$ PRs (across $\numReposStageThree$ repositories) at a cost of $\$446$.

This process yields $\numTasksContainers$ reproducible containers for $\numPerfPRs$ PRs. We elected to stop the synthesis prematurely due to limited resources. However, with more resources, we expect the number of reproducible containers to substantially increase.

\subsection{Statistical Testing and Robustness}\label{sec:a.dataset.stat_testing}

Finally, we must ensure that every retained task reflects a statistically significant and reproducible performance change. Because timing measurements are inherently noisy (e.g., due to OS scheduling, background load, and CPU power management), we adopt the statistical significance validation procedure used by ASV to verify that the observed differences between two code states are significant under repeated measurement on commodity hardware.

\paragraph{Measurement protocol.}
All experiments are run on an AWS EC2 instance specified in \S\ref{sec:a.expt.tbenchmods} to ensures hardware isolation. For each candidate pull request, we execute the expert-selected workloads $\texttt{Workloads}=\{w_1,\dots,w_n\}$ on both the baseline codebase $\texttt{Code}_0$ and the expert-optimized codebase $\texttt{Code}^*_{\texttt{expert}}$ on the same instance.
For each workload $w_i$, ASV repeatedly evaluates the benchmark under a warm-up and multi-sample timing protocol (with interleaved rounds when enabled), yielding independent sample sets of observed runtimes for the baseline and expert-edited codebases:
\[
X_i = \{x_{i1},\dots,x_{im}\}\ \ \text{from}\ \ w_i(\texttt{Code}_0),
\qquad
Y_i = \{y_{i1},\dots,y_{ik}\}\ \ \text{from}\ \ w_i(\texttt{Code}^*_{\texttt{expert}}),
\]
where $X_i$ and $Y_i$ denote the set of measurements for workload $w_i$ from the baseline and expert-edited code respectively. We preserve ASV's default sampling parameters (unless a repository overrides it via workload-specific attributes), so that the resulting statistical decision procedure matches common practice in the Python benchmarking ecosystem. The Python benchmarking community emphasizes collecting sufficiently independent measurements over a finite duration rather than enforcing a fixed repetition count. In practice, ASV defaults to collecting between 2 and 40 measurements per workload through adaptive stopping, where additional samples are drawn until the estimate converges or the maximum is reached. These defaults can be adjusted on a per-workload basis to match the characteristics of their specific benchmarks.

\paragraph{Mann--Whitney U test.}
To test whether $\texttt{Code}_0$ and $\texttt{Code}^*_{\texttt{expert}}$ exhibit different performance distributions for a workload, we use the Mann--Whitney U test \citep{mann1947test}, a non-parametric two-sample test based on rank ordering. Formally, for samples $X_i$ and $Y_i$, the $U$ statistic can be written as
\[
U(X_i,Y_i)
=
\sum_{a=1}^{m}\sum_{b=1}^{k}
\mathbb{I}[x_{ia} > y_{ib}]
\;+\;
\tfrac{1}{2}\mathbb{I}[x_{ia} = y_{ib}],
\]
and the associated two-sided $p$-value quantifies evidence against the null hypothesis.

\paragraph{Null hypothesis.}
For each workload $w_i$, we test
\[
H_0:\ X_i \ \text{and}\ Y_i \ \text{are drawn from the same underlying distribution}
\]
(i.e., the patch does not induce a statistically detectable change in the benchmark outcome), against the two-sided alternative that the distributions differ. We only consider workloads that reject $H_0$.

\paragraph{Implementation.}
In practice, ASV applies a conservative two-stage decision rule. When sufficient raw samples are available, it applies the Mann--Whitney $U$ test and declares a difference only if the resulting $p$-value is below a stringent threshold (default $p < 0.002$).
If the sample sizes are too small for the $U$ test to ever reach this threshold (given the discrete nature of the test), ASV falls back to a pessimistic check based on uncertainty estimates: it computes a $99\%$ confidence interval for each sample distribution and only declares a difference when these intervals do not overlap. This fallback biases towards not claiming a difference unless the separation is unambiguous.

\paragraph{Dataset Inclusion Criterion.}
We discard candidate tasks for which no workload exhibits a statistically significant change between $\texttt{Code}_0$ and $\texttt{Code}_\expert$ under this rule. This ensures that every retained task in \fc corresponds to a clear, reproducible, and statistically supported performance difference. Tasks with no positive significant workloads are also discarded.

This yields the final \fcsize problems used in \fc.

\paragraph{Multiple comparisons correction.}
Because each task involves testing significance across $\sim$\avgWorkloadsPerProblem~workloads, a natural concern is whether the per-workload $p$-values require correction for multiple comparisons.
We re-ran the filtering procedure with a Holm--Bonferroni step-down correction~\citep{holm1979simple, armstrong2014bonferroni} applied within each task.
Under this more conservative procedure, 35 of \fcsize~tasks were removed from \fc, and none of the \numFCVProblems~\fcv tasks were affected. The minimal impact is expected as expert-authored performance PRs merged by upstream maintainers almost always produce strong, detectable signals across multiple workloads, so the correction has little practical effect.  Our codebase incorporates this correction uses this as an additional filter.

\subsection{\fcv Construction}\label{sec:a.dataset.fcv}

The \numFCVProblems~problems in the \fcv subset are drawn from the full \fc benchmark to serve as a computationally tractable evaluation set.
Tasks are selected based on the following criteria: (1) the expert patch must produce a statistically significant speedup that is large enough to distinguish meaningful agent improvements from measurement noise, (2) the task's ASV workloads must execute reliably and with low variance across repeated runs, and (3) the task must have a working PyTest suite for correctness validation.

To support the long-tail generalization experiments (\S\ref{sec:results-popularity}), we stratify the sampling to preserve the repository distribution of the full benchmark. Specifically, \fcv draws from a comparable spread of repository popularity quintiles, ensuring that both popular libraries and niche scientific tools are represented.

Tables~\ref{tab:a.dataset.construction.question_composition} and \ref{tab:a.dataset.construction.difficulty} compare the optimization type and difficulty distributions between \fc and \fcv. The distributions are broadly similar, with minor overrepresentation of algorithmic and data structure optimizations in \fcv due to the selection of tasks with large expert speedups.

\subsection{Dataset Composition Statistics}\label{sec:a.dataset.composition}

To better study the characteristics of \fc, we develop an automated classifier that attempts to infer the kind of optimization based on a curated taxonomy (\S\ref{sec:a.expt.optimizationtypes}). The classifier is similar to the one introduced in \S\ref{sec:a.dataset.llmfilters}. It takes as input a sample pull request along with the expert-written patch and attempts to categorize the expert-written solutions using a manually curated taxonomy (Table \ref{tab:a.expt.optimizationtypes}). Such a methodology allows us to efficiently and scalably study the composition of an continuously growing set of problems. The prompts for this classifier are presented in Figure \ref{fig:perf_type_difficulty_prompt_structure} and an example is presented in Figure \ref{fig:optimization-example}.

The distribution of the types of optimizations is presented in Table \ref{tab:a.dataset.construction.question_composition} and the distribution of the inferred difficulty is presented in Table \ref{tab:a.dataset.construction.difficulty}. Importantly, the distribution of optimization problems and difficulty changes marginally between \fc and \fcv.

\begin{table}[t]
    \centering
    \small
    \caption{
        Patch classification distribution in \fc and \fcv. The problems in \fcv are sampled from the best performing tasks in \fc which is why some categories are overrepresented.
    }
    \label{tab:a.dataset.construction.question_composition}
    \adjustbox{max width=\linewidth}{%
    \begin{tabular}{l r r}
        \toprule
        Inferred Type of Optimization Problem & \% \fc & \% \fcv \\
        \midrule
Accept Less Precise Solution & 0.6584 & - \\
Cache And Reuse & 8.3128 & 4.6296 \\
Database And Storage Tuning & 0.5761 & - \\
Do It Earlier Batch Throttle & 2.4691 & 0.9259 \\
Io And Latency Hiding & 0.0823 & - \\
Micro Optimizations & 20.2469 & 23.1481 \\
Remove Or Reduce Work & 20.0823 & 18.5185 \\
Uncategorized & 1.5638 & - \\
Use Better Algorithm & 20.0823 & 26.8519 \\
Use Better Data Structure And Layout & 9.7119 & 12.9630 \\
Use Higher Level System & 2.9630 & 2.7778 \\
Use Lower Level System & 11.0288 & 9.2593 \\
Use Parallelization & 2.2222 & 0.9259 \\
        \bottomrule
    \end{tabular}
    }
\end{table}

\begin{table}[t]
    \centering
    \small
    \caption{The inferred difficulty of human solutions in \fc and \fcv.
    }
    \label{tab:a.dataset.construction.difficulty}
    \adjustbox{max width=\linewidth}{%
    \begin{tabular}{l r r}
        \toprule
        Inferred Difficulty & \% \fc & \% \fcv \\
        \midrule
        Easy & 54.8971 & 60.1852 \\
        Medium & 44.4444 & 37.0370 \\
        Hard & 0.6584 & 2.7778 \\
        \bottomrule
    \end{tabular}
    }
\end{table}

\paragraph{Difficulty calibration.}
The LLM-inferred difficulty in Table~\ref{tab:a.dataset.construction.difficulty} captures the structural complexity of the expert-written patch (e.g., number of files changed, algorithmic sophistication) rather than how difficult the task is for an agent.
Beyond the LLM-inferred difficulty above, we note two additional difficulty proxies that emerge from our experimental analyses.
First, \textit{repository popularity} (measured by GitHub stars) serves as a proxy for distribution shift: agents are less familiar with the conventions and APIs of niche repositories, and our long-tail analysis (\S\ref{sec:results-popularity}) confirms that performance degrades substantially on lower-popularity quintiles.
Second, the \textit{magnitude of the expert speedup} serves as a proxy for optimization headroom---tasks with small expert speedups leave less room for agent improvement, while tasks with large expert speedups indicate a more impactful optimization opportunity.
The \adv~metric normalizes against the expert speedup, helping control for this variation in available headroom.
These three proxies are complementary: LLM-inferred difficulty captures solution complexity, repository popularity captures data familiarity, and expert speedup captures the magnitude of the optimization opportunity.

\subsection{Limitations}\label{sec:discussion-limitations}

\paragraph{Domain coverage.}
\fc is currently restricted to scientific Python repositories that ship with Airspeed Velocity (ASV) benchmarking infrastructure.
While our corpus of \numFinalRepos~repositories is broader than prior optimization benchmarks, it does not yet cover web frameworks, distributed systems, embedded software, or non-Python languages.
Extending \fc beyond scientific computing is an important direction for future work; the pipeline design is language-agnostic in principle but requires a corresponding performance benchmarking infrastructure for each target ecosystem.

\paragraph{Dependence on ASV workloads.}
The quality and coverage of \fc's evaluation signal is bounded by the workloads maintained by each repository's developers.
If community-maintained workloads do not exercise the performance-critical paths modified by a pull request, the benchmark signal may be incomplete.
We mitigate this partially through our multi-workload design ($\sim$\avgWorkloadsPerProblem~workloads per task), which provides broad coverage, but acknowledge that workload completeness is ultimately determined by upstream maintainers.

\paragraph{Evaluation on a subset.}
Due to compute constraints, our full agent evaluation is conducted on \fcv (\numFCVProblems~tasks) rather than the complete \fc benchmark (\fcsize~tasks).
While \fcv preserves the repository distribution and difficulty profile of the full benchmark (Appendix~\S\ref{sec:a.dataset.composition}), results may not generalize to the full task distribution.
We plan to expand evaluation coverage as additional compute becomes available.

\paragraph{Hardware specificity.}
All performance measurements are collected on standardized AWS EC2 \texttt{c5ad.large} instances (Appendix~\S\ref{sec:a.expt.tbenchmods}).
While this ensures internal consistency and reproducibility, performance characteristics may differ on other architectures (e.g., ARM, GPU-accelerated workloads).
The \adv~metric, computed as a difference of speedup ratios on the same hardware, is more robust to hardware variation than absolute timing, but cross-platform generalization remains an open question.

\begin{figure*}[p]
\centering
\scriptsize
\setlength{\tabcolsep}{6pt}
\renewcommand{\arraystretch}{1.05}

\begin{tabular}{|r>{\raggedright\arraybackslash}p{\pWidth}|}
\hline
\pStart & \\

\pSection{Input Signature}
\pRow{}

\pRow{\textbf{problem\_description : string}}
\pRow{Problem statement and technical context from PR/issue.}
\pRow{}

\pRow{\textbf{git\_patch : string}}
\pRow{Git diff showing actual code changes.}
\pRow{}

\pRow{\textbf{file\_change\_summary : string}}
\pRow{A markdown table summarizing all the files changed in the commit along with lines added/removed.}
\pRow{}

\pSection{Classifier Module}
\pRow{}

\pRow{Decide if this commit's \textbf{PRIMARY} intent is to improve product/runtime performance.}
\pRow{Label \textbf{YES} only when there is \textbf{CLEAR, EXPLICIT} evidence in the description and/or patch that the runtime gets faster (e.g., algorithm change, fewer allocations, caching, vectorization, reduced I/O, async/non-blocking for throughput, latency reduction, memory footprint reduction, fix a speed regression).}
\pRow{}

\pRow{\textbf{Strong positive signals} (weigh these collectively):}
\pRow{• PR title/body contains performance intent (e.g., ``PERF:'', ``speed up'', ``faster'', ``performance'').}
\pRow{• Linked issues/comments include benchmark links or timings demonstrating impact.}
\pRow{• Low-level/hot-path tweaks (e.g., reuse global context, avoid per-call init/teardown, vectorize C/NumPy).}
\pRow{}

\pRow{\textbf{Hard NO (non-performance) examples:}}
\pRow{tests/ASV/harness-only changes; CI/workflows/build/packaging; coverage; pre-commit/format/lints (clippy/ruff/black); docs; version bumps; terminology/renames; pure refactors without performance claims; changes aimed at making perf tests pass but not improving runtime.}
\pRow{}

\pRow{If ambiguous, weigh the concrete code changes and problem description together.}
\pRow{When there are \textbf{specific performance cues} (title keywords, measured timings, fewer allocations, vectorization, caching/reuse) lean \textbf{YES}; otherwise \textbf{NO}.}
\pRow{}

\pSection{Output Signature}
\pRow{}

\pRow{\textbf{reasoning : string}}
\pRow{Deductive reasoning steps leading to the classification.}
\pRow{}

\pRow{\textbf{label : string}}
\pRow{Final label: ``YES'' for performance-related, ``NO'' otherwise.}
\pRow{}

&\\
\hline
\end{tabular}

\caption{Prompt template used by the LLM-based performance intent classifier described in ~\ref{sec:a.dataset.llmfilters}. The prompt defines the input signature (problem description, git patch, and file change summary), the classifier module specifying decision criteria for identifying performance-motivated commits, and the output signature producing a reasoning trace and binary label (“YES”/“NO”).}
\label{fig:perf_prompt_structure}
\end{figure*}

\thispagestyle{empty}
\begin{figure*}[p]
\centering
\scriptsize
\setlength{\tabcolsep}{6pt}
\renewcommand{\arraystretch}{1.05}

\begin{tabular}{|r>{\raggedright\arraybackslash}p{\pWidth}|}
\hline
\pStart & \\

\pRow{{\bfseries Example PR}}
\pRow{}

\pSection{Classifier Input}
\pRow{}

\pRow{\textbf{problem\_description : string}}
\pRow{Labels: performance; Description: Fixes \#14471.}

\pRow{\textbf{Body:} The new \texttt{ParameterExpression.bind\_all} is a fast path for producing a numeric result. This has advantages over \texttt{ParameterExpression.bind}:}
\pRow{• Far fewer Python objects are allocated, since no new \texttt{ParameterExpression} objects need to be constructed and the output is guaranteed to be numeric.}
\pRow{• There is no historical API requirement to scan the incoming mapping for invalid keys or values, yielding a large performance improvement when the same mapping is used to bind many expressions.}
\pRow{• This provides a major complexity improvement when a large values dictionary is reused many times.}
\pRow{There is still room for further gains because the Rust-space \texttt{ParameterExpression} and \texttt{SymbolExpr} interfaces require more heap allocations than strictly necessary, but this already yields substantial speedups.}

\pRow{\textbf{Issues:} Fixes \#14471.}
\pRow{The linked issue reports that \texttt{ParameterExpression.bind} scales with the size of the binding dictionary even when only a single parameter is needed, leading to severe performance penalties for large parameter tables.}

\pRow{\textbf{Comments:}}

\pRow{Currently in draft because there's no tests - I'm just putting it up so Sam and Ian from \#14471 can test it out for their use case. For the explicit example in that issue, a complete comparison on my machine:}

\pRow{\texttt{<details><summary>Out of date timings</summary>}}

\pRow{\texttt{In [1]: from qiskit.circuit import Parameter, ParameterExpression}}
\pRow{\texttt{\ \ \ \ N: int = 100\_000}}
\pRow{\texttt{\ \ \ \ parameter\_values = \{Parameter(f"th\_\{i\}"): 1 for i in range(N)\}}}
\pRow{\texttt{\ \ \ \ parameter\_values[param := Parameter("my\_param")] = 1}}
\pRow{ \ldots \textbf{<TRUNCATED>}}

\pRow{I think it's fine without having the same behavior. For clarity it might be helpful to add a blurb to the \texttt{bind\_all} docstring to say that ``unlike \texttt{bind}, NaN and inf are in the range of expected outputs for this method''.}

\pRow{LGTM, thanks!}
\pRow{}

\pRow{\textbf{git\_patch : string}}

\pRow{\texttt{\detokenize{diff --git a/crates/circuit/src/parameter/parameter_expression.rs b/crates/circuit/src/parameter/parameter_expression.rs}}}
\pRow{\texttt{\detokenize{index 1f0406f62c7e..98da2ee3e9e6 100644}}}
\pRow{\texttt{\detokenize{--- a/crates/circuit/src/parameter/parameter_expression.rs}}}
\pRow{\texttt{\detokenize{+++ b/crates/circuit/src/parameter/parameter_expression.rs}}}
\pRow{\texttt{\detokenize{@@ -1048,6 +1048,40 @@ impl PyParameterExpression {}}}}

\pRow{\texttt{\detokenize{+    #[pyo3(name = "bind_all")]}}}
\pRow{\texttt{\detokenize{+    #[pyo3(signature = (values, *))]}}}
\pRow{\texttt{\detokenize{+    pub fn py_bind_all(&self, values: Bound<PyAny>) -> PyResult<Value> {}}}}
\pRow{\texttt{\detokenize{+        let mut partial_map = HashMap::with_capacity(self.inner.name_map.len());}}}
\pRow{\texttt{\detokenize{+        for symbol in self.inner.name_map.values() {}}}}
\pRow{\texttt{\detokenize{+            let py_parameter = symbol.clone().into_pyobject(values.py())?;}}}
\pRow{\texttt{\detokenize{+            partial_map.insert(symbol, values.get_item(py_parameter)?.extract()?);}}}
\pRow{\texttt{\detokenize{+        }}}
\pRow{\texttt{\detokenize{+        let bound = self.inner.expr.bind(&partial_map);}}}

\pRow{\texttt{... \textbf{<Truncated>}}}
\pRow{}

\pRow{\textbf{file\_change\_summary : string}}
\pRowTable{%
\begin{tabular}{|l|c|c|c|}
\hline
\textbf{File} & \textbf{Lines Added} & \textbf{Lines Removed} & \textbf{Total Changes} \\ \hline
\texttt{crates/circuit/src/parameter/parameter\_expression.rs} & 43 & 0 & 43 \\ \hline
\texttt{crates/circuit/src/parameter/symbol\_expr.rs} & 1 & 1 & 2 \\ \hline
\texttt{releasenotes/notes/parameter-bind-all-22337586429509f9.yaml} & 8 & 0 & 8 \\ \hline
\texttt{test/python/circuit/test\_parameters.py} & 18 & 0 & 18 \\ \hline
\end{tabular}}
\pRow{}
\pSection{Classifier Module}
\pRow{}

\pRow{Decide if this commit's \textbf{PRIMARY} intent is to improve product/runtime performance.}
\pRow{Label \textbf{YES} only when\ldots}
\pRow{\ldots there are \textbf{specific performance cues} (title keywords, measured timings, fewer allocations, vectorization, caching/reuse) lean \textbf{YES}; otherwise \textbf{NO}.}
\pRow{}

\pSection{Classifier Output}
\pRow{}

\pRow{\textbf{reasoning : string}}
\pRow{The pull request introduces a \textbf{new fast-path method} \texttt{ParameterExpression.bind\_all} (and \texttt{Parameter.bind\_all}) to bind parameters to numeric values with \textbf{far fewer allocations and reduced overhead} than the existing \texttt{bind} method. The PR explicitly emphasizes performance, using terms such as \emph{``fast path''} and \emph{``huge performance gain''}, and reports timing improvements from tens of milliseconds to microsecond-scale execution.}

\pRow{The implementation avoids constructing new \texttt{ParameterExpression} objects and skips expensive validation logic, directly producing numeric results. It is labeled with the \texttt{performance} tag and documented as a fast-path method for performance-sensitive use cases, making the primary intent of the change to \textbf{reduce computational overhead and improve runtime performance} of parameter binding.}

\pRow{Therefore, the commit's primary purpose is to make parameter binding substantially faster.}
\pRow{}

\pRow{\textbf{label : string}}
\pRow{\texttt{'YES'}} 
\pRow{}

&\\
\hline
\end{tabular}

\caption{Example of performance intent classification for a real pull request \url{(https://github.com/Qiskit/qiskit/pull/14782)}, illustrating how the prompt in Fig. ~\ref{fig:perf_prompt_structure} is instantiated and how the classifier produces a reasoning trace and binary label.}
\label{fig:performance-classifier-example}
\end{figure*}

\begin{figure*}[p]
\centering
\scriptsize
\setlength{\tabcolsep}{6pt}
\renewcommand{\arraystretch}{1.05}

\begin{tabular}{|r>{\raggedright\arraybackslash}p{\pWidth}|}
\hline
\pStart & \\

\pRow{{\bfseries Performance type \& difficulty classifier}}
\pRow{}

\pSection{Input Signature Prompt}
\pRow{}

\pRow{\textbf{problem\_description : string}}
\pRow{Problem statement and technical context from PR/issue.}
\pRow{}

\pRow{\textbf{git\_patch : string}}
\pRow{Git diff showing actual code changes.}
\pRow{}

\pSection{Classifier Module Prompt}
\pRow{}

\pRow{Decide the \textbf{PRIMARY performance optimization technique} and the \textbf{difficulty level} of the optimization.}
\pRow{}

\pRow{\textbf{Category mapping (when performance-related):}}
\pRow{\textbullet\ \textbf{Algorithm improvements}: complexity reduction; switching to faster algorithms $\rightarrow$ \texttt{use\_better\_algorithm}}
\pRow{\textbullet\ \textbf{Data structures / layout}: sets, maps, indices; memory layout tuning $\rightarrow$ \texttt{use\_better\_data\_structure\_and\_layout}}
\pRow{\textbullet\ \textbf{System-level}: C/Rust/NumPy/Vectorized/Native extensions $\rightarrow$ \texttt{use\_lower\_level\_system}}
\pRow{\textbullet\ \textbf{Approximation / heuristics}: trade accuracy for speed $\rightarrow$ \texttt{accept\_less\_precise\_solution}}
\pRow{\textbullet\ \textbf{Parallelization}: threads, processes, parallel algorithms (not just async I/O) $\rightarrow$ \texttt{use\_parallelization}}
\pRow{\textbullet\ \textbf{Cache \& reuse}: memoization, LRU, materialized results $\rightarrow$ \texttt{cache\_and\_reuse}}
\pRow{\textbullet\ \textbf{Scheduling}: batching, lazy execution, throttling $\rightarrow$ \texttt{do\_it\_earlier\_batch\_throttle}}
\pRow{\textbullet\ \textbf{Database / storage}: indices, query tuning, partitioning $\rightarrow$ \texttt{database\_and\_storage\_tuning}}
\pRow{\textbullet\ \textbf{Micro-optimizations}: hot-path tweaks, guards, inlining $\rightarrow$ \texttt{micro\_optimizations}}
\pRow{\textbullet\ \textbf{I/O / latency hiding}: async or non-blocking I/O, overlap I/O and compute $\rightarrow$ \texttt{io\_and\_latency\_hiding}}
\pRow{\textbullet\ \textbf{Higher-level systems}: using optimized libraries or frameworks $\rightarrow$ \texttt{use\_higher\_level\_system}}
\pRow{\textbullet\ \textbf{Uncategorized}: performance-related but does not fit the above categories $\rightarrow$ \texttt{uncategorized}}
\pRow{}

\pRow{\textbf{Difficulty (when performance-related):}}
\pRow{\textbullet\ \textbf{easy}: localized change ($< 50$ lines), minimal risk}
\pRow{\textbullet\ \textbf{medium}: module-level refactor, data structure changes}
\pRow{\textbullet\ \textbf{hard}: algorithm rewrite or architectural change}
\pRow{}

\pSection{Output Signature Prompt}
\pRow{}

\pRow{\textbf{category : OptimizationType}}
\pRow{The classified optimization category.}
\pRow{}

\pRow{\textbf{difficulty : DifficultyLevel}}
\pRow{The difficulty level of the optimization.}
\pRow{}

\pRow{\textbf{reasoning : string}}
\pRow{Brief explanation of the classification.}
\pRow{}

&\\
\hline
\end{tabular}

\caption{Prompt template used by the LLM-based classifier for assigning each performance task an optimization category and difficulty level (~\ref{sec:a.expt.optimizationtypes}). The prompt defines the input signature, a taxonomy-driven classification module that maps code changes to optimization types, and an output schema that produces the predicted category, difficulty, and a brief reasoning trace.}
\label{fig:perf_type_difficulty_prompt_structure}
\end{figure*}

\begin{figure*}[p]
\centering
\scriptsize
\setlength{\tabcolsep}{6pt}
\renewcommand{\arraystretch}{1.05}

\begin{tabular}{|r>{\raggedright\arraybackslash}p{\pWidth}|}
\hline
\pStart & \\

\pSection{Input Signature}
\pRow{}

\pRow{\textbf{owner\_repo : string}}
\pRow{The repository this commit belongs to (e.g., \texttt{scikit-learn/scikit-learn}).}
\pRow{}

\pRow{\textbf{sha : string}}
\pRow{The commit SHA that is currently checked out.}
\pRow{}

\pRow{\textbf{commit\_date : string}}
\pRow{The commit date in ISO format (e.g., \texttt{2023-10-05T12:34:56Z}).}
\pRow{}

\pRow{\textbf{stderr\_logs : string}}
\pRow{Most recent stderr logs from the last build attempt (up to $\sim$8k tail-end characters).}
\pRow{}

\pRow{\textbf{stdout\_logs : string}}
\pRow{Most recent stdout logs from the last build attempt (up to $\sim$8k tail-end characters).}
\pRow{}

\pRow{\textbf{failure\_more : string}}
\pRow{Describes where the failure occurred (e.g., \texttt{N/A}, \texttt{build failed}, \texttt{asv run failed}).}
\pRow{}

\pRow{\textbf{last\_docker\_build\_script : string}}
\pRow{The previously generated \texttt{docker\_build.sh} script.}
\pRow{}

\pRow{\textbf{repo\_facts\_json : string}}
\pRow{JSON object containing inferred repository facts (paths, package names, versions, etc.).}
\pRow{}

\pRow{\textbf{toolbelt : string}}
\pRow{Human-readable summary of available tools and their usage.}
\pRow{}

\pRow{\textbf{messages\_log : string}}
\pRow{Transcript of prior tool calls, actions, and observations.}
\pRow{}

\pSection{Build Agent Module}
\pRow{}

\pRow{An interactive planner for producing a \texttt{docker\_build.sh} bash script that builds and installs a Python repository inside micromamba environments. The agent may either: (A) Request a tool call with structured JSON arguments, or (B) Output the final executable build script.}
\pRow{If a tool is required, set \textbf{next\_action} to one of: \texttt{probe\_repo} \textbar{} \texttt{list\_tree} \textbar{} \texttt{read\_file} \textbar{} \texttt{try\_import} \textbar{} \texttt{none}.}
\pRow{}

\pRow{\textbf{Tool call formats:}}
\pRow{• \texttt{read\_file}: \{\texttt{"path": "...", "max\_bytes": 65536}\}}
\pRow{• \texttt{list\_tree}: \{\texttt{"depth": 2}\}}
\pRow{• \texttt{try\_import}: \{\texttt{"candidates": ["foo","bar"]}\}}
\pRow{Return \texttt{docker\_build\_script} only when fully satisfied with correctness and completeness.}

\pRow{\textbf{Critical constraints on the generated script:}}
\pRow{• Must be idempotent and safe to run inside Docker.}
\pRow{• Fully non-interactive; no user prompts.}
\pRow{• Must be valid executable Bash with no syntax errors.}
\pRow{• Must use real newline characters (not escaped \texttt{\textbackslash n}).}
\pRow{• Must not output literal \texttt{\textbackslash n}.}

\pRow{\textbf{Post-install readiness requirements:}}
\pRow{• After editable install, the environment must be immediately usable.}
\pRow{• A lightweight profiling sanity check and a lightweight \texttt{pytest} sanity check must start without immediate errors, even for projects that require execution from subdirectories.}
\pRow{• Test/benchmark extras and optional dependencies must be installed as needed for import and test discovery to succeed.}
\pRow{}

\pSection{Output Signature}
\pRow{}

\pRow{\textbf{thought : string}}
\pRow{Brief rationale describing the current decision or plan.}
\pRow{}

\pRow{\textbf{next\_action : string}}
\pRow{One of \texttt{probe\_repo}, \texttt{list\_tree}, \texttt{read\_file}, \texttt{try\_import}, \texttt{none}, or \texttt{finish}.}
\pRow{}

\pRow{\textbf{action\_input : string}}
\pRow{JSON arguments for the selected tool, or empty if no tool is called.}
\pRow{}

\pRow{\textbf{error\_summary : string}}
\pRow{Brief summary of the most recent build failure and its possible causes.}
\pRow{}

\pRow{\textbf{resolution\_steps : string}}
\pRow{Concrete steps required to resolve the failure.}
\pRow{}

\pRow{\textbf{docker\_build\_script : string}}
\pRow{Final executable \texttt{docker\_build.sh} script that successfully builds and installs the project from source.}

&\\
\hline
\end{tabular}

\caption{Prompt structure for the docker build agent (~\ref{sec:a.dataset.stage3}), defining its input state, tool-calling interface, constraints, and executable script output.}
\label{fig:build_agent_prompt_structure}
\end{figure*}

\begin{figure*}[p]
\centering
\scriptsize
\setlength{\tabcolsep}{6pt}
\renewcommand{\arraystretch}{1.05}

\begin{tabular}{|r>{\raggedright\arraybackslash}p{\pWidth}|}
\hline
\pStart & \\

\pRow{{\bfseries Example PR}}
\pRow{}

\pSection{Classifier Input}
\pRow{}

\pRow{\textbf{problem\_description : string}}
\pRow{Labels: performance; Description: Fixes \#14471.}

\pRow{\textbf{Body:} The new \texttt{ParameterExpression.bind\_all} is a fast path for producing a numeric result. This has advantages over \texttt{ParameterExpression.bind}:}
\pRow{• Far fewer Python objects are allocated, since no new \texttt{ParameterExpression} objects need to be constructed and the output is guaranteed to be numeric.}
\pRow{• There is no historical API requirement to scan the incoming mapping for invalid keys or values, yielding a large performance improvement when the same mapping is used to bind many expressions.}
\pRow{• This provides a major complexity improvement when a large values dictionary is reused many times.}
\pRow{There is still room for further gains because the Rust-space \texttt{ParameterExpression} and \texttt{SymbolExpr} interfaces require more heap allocations than strictly necessary, but this already yields substantial speedups.}

\pRow{\textbf{Issues:} Fixes \#14471.}
\pRow{The linked issue reports that \texttt{ParameterExpression.bind} scales with the size of the binding dictionary even when only a single parameter is needed, leading to severe performance penalties for large parameter tables.}

\pRow{\textbf{Comments:}}

\pRow{Currently in draft because there's no tests - I'm just putting it up so Sam and Ian from \#14471 can test it out for their use case. For the explicit example in that issue, a complete comparison on my machine:}

\pRow{\texttt{<details><summary>Out of date timings</summary>}}

\pRow{\texttt{In [1]: from qiskit.circuit import Parameter, ParameterExpression}}
\pRow{\texttt{\ \ \ \ N: int = 100\_000}}
\pRow{\texttt{\ \ \ \ parameter\_values = \{Parameter(f"th\_\{i\}"): 1 for i in range(N)\}}}
\pRow{\texttt{\ \ \ \ parameter\_values[param := Parameter("my\_param")] = 1}}
\pRow{\texttt{\ \ \ \ print("Using the specialised `Parameter` methods:")}}
\pRow{\texttt{\ \ \ \ \%timeit param.bind(parameter\_values, allow\_unknown\_parameters=True)}}

\pRow{\texttt{</details> \ldots \textbf{<TRUNCATED>}}}

\pRow{I think it's fine without having the same behavior. For clarity it might be helpful to add a blurb to the \texttt{bind\_all} docstring to say that ``unlike \texttt{bind}, NaN and inf are in the range of expected outputs for this method''.}

\pRow{LGTM, thanks!}

\pRow{}
\pRow{\textbf{git\_patch : string}}

\pRow{\texttt{\detokenize{diff --git a/crates/circuit/src/parameter/parameter_expression.rs b/crates/circuit/src/parameter/parameter_expression.rs}}}
\pRow{\texttt{\detokenize{index 1f0406f62c7e..98da2ee3e9e6 100644}}}
\pRow{\texttt{\detokenize{--- a/crates/circuit/src/parameter/parameter_expression.rs}}}
\pRow{\texttt{\detokenize{+++ b/crates/circuit/src/parameter/parameter_expression.rs}}}
\pRow{\texttt{\detokenize{@@ -1048,6 +1048,40 @@ impl PyParameterExpression {}}}}

\pRow{\texttt{\detokenize{+    #[pyo3(name = "bind_all")]}}}
\pRow{\texttt{\detokenize{+    #[pyo3(signature = (values, *))]}}}
\pRow{\texttt{\detokenize{+    pub fn py_bind_all(&self, values: Bound<PyAny>) -> PyResult<Value> {}}}}
\pRow{\texttt{\detokenize{+        let mut partial_map = HashMap::with_capacity(self.inner.name_map.len());}}}
\pRow{\texttt{\detokenize{+        for symbol in self.inner.name_map.values() {}}}}
\pRow{\texttt{\detokenize{+            let py_parameter = symbol.clone().into_pyobject(values.py())?;}}}
\pRow{\texttt{\detokenize{+            partial_map.insert(symbol, values.get_item(py_parameter)?.extract()?);}}}
\pRow{\texttt{\detokenize{+        }}}
\pRow{\texttt{\detokenize{+        let bound = self.inner.expr.bind(&partial_map);}}}
\pRow{\texttt{\detokenize{+        bound.eval(true).ok_or_else(|| {}}}}
\pRow{\texttt{\detokenize{+            PyTypeError::new_err(format!("binding did not produce a numeric quantity: {bound:?}"))}}}
\pRow{\texttt{\detokenize{+        })}}
\pRow{\texttt{\detokenize{+    }}}

\pRow{\texttt{... \textbf{<Truncated>}}}
\pRow{}

\pSection{Classifier Module}
\pRow{}

\pRow{Decide the \textbf{PRIMARY performance optimization technique} and the \textbf{difficulty level} of the optimization.}
\pRow{Category mapping (when performance-related): \ldots}
\pRow{\ldots algorithm rewrite or architectural change.}
\pRow{}

\pSection{Classifier Output}
\pRow{}
\pRow{\textbf{category : OptimizationType}}
\pRow{\texttt{REMOVE\_OR\_REDUCE\_WORK}}
\pRow{}

\pRow{\textbf{difficulty : DifficultyLevel}}
\pRow{'medium'}
\pRow{}

\pRow{\textbf{reasoning : string}}
\pRow{The change adds a specialized 'bind\_all' fast‑path that skips the heavy allocation and validation logic of the generic 'bind' method, directly mapping parameters to numeric values and evaluating the expression. This reduces the amount of work performed per binding call, yielding large speedups. The implementation touches Rust code and adds tests but is a localized, moderate‑size change, fitting a medium difficulty level.}
\pRow{}

&\\
\hline
\end{tabular}

\caption{Example application of the optimization type and difficulty classifier (Figure ~\ref{fig:perf_type_difficulty_prompt_structure}), illustrating the predicted category, difficulty level, and reasoning for a real pull request (\url{https://github.com/Qiskit/qiskit/pull/14782)}}
\label{fig:optimization-example}
\end{figure*}

\begin{figure*}[p]
\centering
\scriptsize
\setlength{\tabcolsep}{6pt}
\renewcommand{\arraystretch}{1.05}

\begin{tabular}{|r>{\raggedright\arraybackslash}p{\pWidth}|}
\hline
\pStart & \\

\pRow{{\bfseries Judge performance related PR prompt}}
\pRow{}

\pSection{Input Signature Prompt}
\pRow{}

\pRow{\textbf{problem\_description : string}}
\pRow{Problem statement and technical context from PR/issue.}
\pRow{}

\pRow{\textbf{git\_patch : string}}
\pRow{Git diff showing actual code changes.}
\pRow{}

\pRow{\textbf{file\_change\_summary : string}}
\pRow{A markdown table summarizing all the files changed in the commit along with lines added/removed.}
\pRow{}

\pSection{Judge Signature Prompt}
\pRow{}

\pRow{Decide if this commit's PRIMARY intent is to improve product/runtime performance.}
\pRow{}

\pRow{Label YES only when there is CLEAR, EXPLICIT evidence in the description and/or patch that the runtime gets faster (e.g., algorithm change, fewer allocations, caching, vectorization, reduced I/O, async/non-blocking for throughput, latency reduction, memory footprint reduction, fix a speed regression).}
\pRow{}

\pRow{Strong positive signals (weigh these collectively):}
\pRow{- PR title/body contains performance intent (e.g., "PERF:", "speed up", "faster", "performance").}
\pRow{- Linked issues/comments include benchmark links or timings demonstrating impact.}
\pRow{- Low-level/hot-path tweaks (e.g., reuse global context, avoid per-call init/teardown, vectorize C/NumPy).}
\pRow{}

\pRow{Hard NO (non-performance) examples: tests/ASV/harness-only changes; CI/workflows/build/packaging; coverage; pre-commit/format/lints (clippy/ruff/black); docs; version bumps; terminology/renames; pure refactors without performance claims; changes aimed at making perf tests pass but not improving runtime.}
\pRow{}

\pRow{If ambiguous, weigh the concrete code changes and problem description together. When there are specific performance cues (title keywords, measured timings, fewer allocations, vectorization, caching/reuse) lean YES; otherwise NO.}
\pRow{}

\pSection{Output Signature Prompt}
\pRow{}

\pRow{\textbf{reasoning : string}}
\pRow{Deductive reasoning steps leading to the classification.}
\pRow{}

\pRow{\textbf{label : string}}
\pRow{Final label: "YES" for performance-related, "NO" otherwise.'}
\pRow{}

&\\
\hline
\end{tabular}

\caption{Structured DSPy prompt used to judge whether a pull request is primarily intended to improve runtime or product performance. The prompt specifies the required inputs (problem description, code diff, and file-level change summary), explicit decision criteria and exclusions for performance-related changes, and an output format consisting of a justification and a binary YES/NO label. The design emphasizes conservative, evidence-based classification, prioritizing explicit runtime improvements over incidental or refactoring-only changes.}
\label{fig:judge_signature_prompt}
\end{figure*}

\begin{figure*}[p]
\centering
\scriptsize
\setlength{\tabcolsep}{6pt}
\renewcommand{\arraystretch}{1.05}

\begin{tabular}{|r>{\raggedright\arraybackslash}p{\pWidth}|}
\hline
\pStart & \\

\pRow{{\bfseries Problem Extractor Prompt description}}
\pRow{}

\pSection{Input Signature Prompt}
\pRow{}

\pRow{\textbf{pr\_title : string}}
\pRow{The GitHub PR title}
\pRow{}

\pRow{\textbf{pr\_body : string}}
\pRow{The GitHub PR description}
\pRow{}

\pRow{\textbf{pr\_comments : string}}
\pRow{Comments on the PR thread.}
\pRow{}

\pSection{Problem Extractor Signature}
\pRow{}

\pRow{What problem is this Github PR trying to solve? Extract near-verbatim relevant text following the given JSON output. If no relevant context exists for a field, return an empty string for it.}
\pRow{}

\pSection{Output Signature Prompt}
\pRow{}

\pRow{\textbf{\texttt{initial\_observations: string | list[Any] | None}}}
\pRow{Objective symptoms of the problematic behavior, described in the present tense. Focus strictly on what is happening (metrics, user impact, frequency). Do not include causes, hypotheses, or explanations.}
\pRow{}

\pRow{\textbf{\texttt{triage\_attempts: string | list[Any] | None}}}
\pRow{The investigative steps and reasoning used to narrow down contributing factors—what you checked, what you ruled out, and what evidence you gathered to understand where the issue originates.}
\pRow{}

\pRow{\textbf{\texttt{solution\_overview: string | list[Any] | None}}}
\pRow{A concise description of the change(s) made and how they address the identified bottleneck or constraint.}
\pRow{}

\pRow{\textbf{\texttt{solution\_observations: string | list[Any] | None}}}
\pRow{What you observe after applying the change—new measurements, behavior differences, and any regressions or trade-offs that appeared.}
\pRow{}

&\\
\hline
\end{tabular}

\caption{Structured DSPy prompt used to extract the underlying problem and resolution context from a GitHub pull request. The prompt consumes the PR title, description, and discussion, and produces a structured summary capturing observed symptoms, triage steps, the implemented solution, and post-change observations. The design emphasizes near-verbatim extraction and separation of observations, investigation, and outcomes.}
\label{fig:problem-extractor}
\end{figure*}

\begin{figure}[t]
    \centering
    \includegraphics[width=0.9\linewidth]{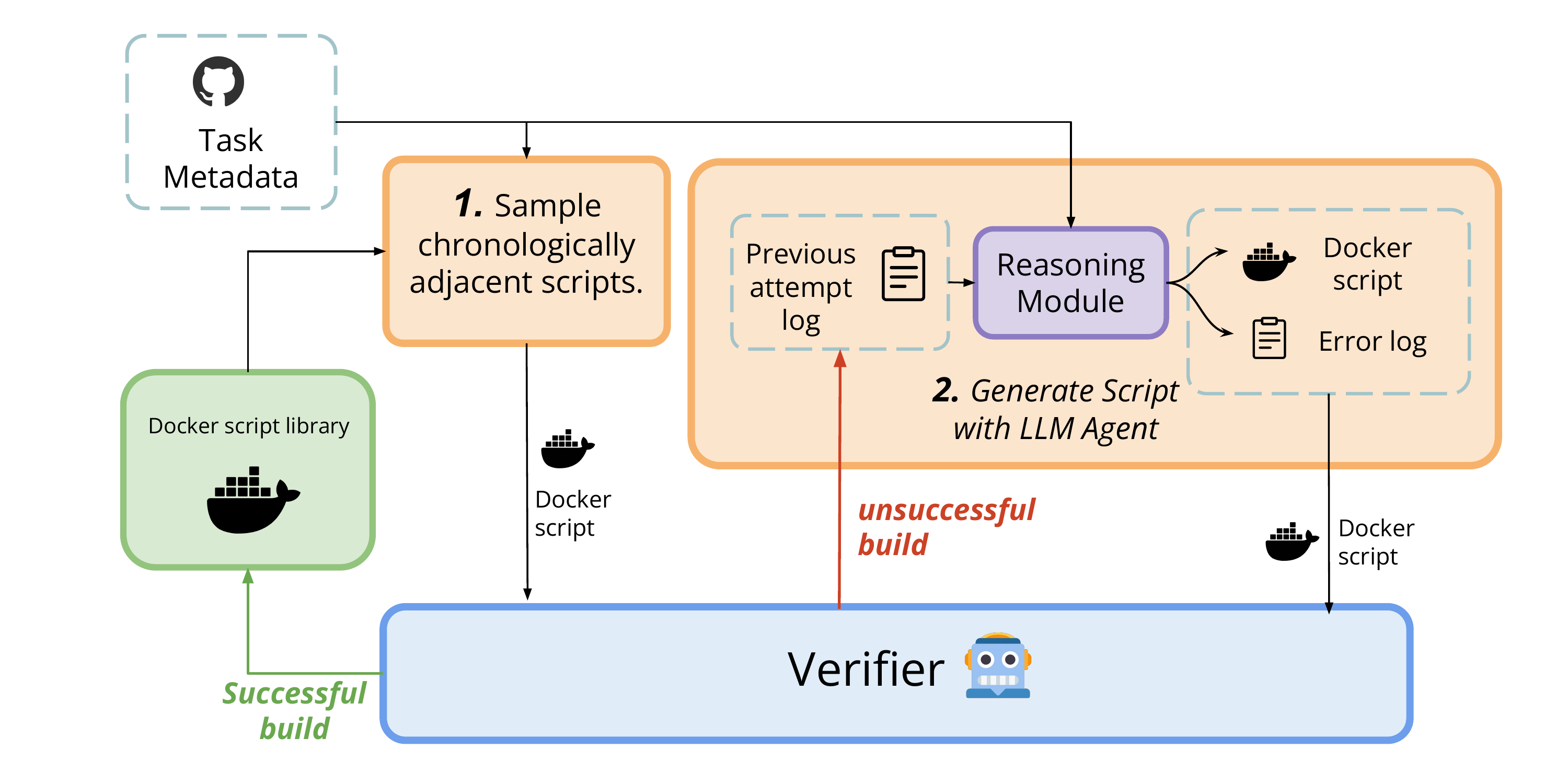}
    \caption{Overview of the pipeline for Docker environment synthesis. The system reuses chronologically adjacent build scripts when possible, otherwise invoking an LLM agent that generates and refines Docker scripts using build logs and repository context until a verifier confirms a successful, reproducible build.}
    \label{fig:dataset_generation_algorithm}
\end{figure}

\newcounter{reponum}
\newcolumntype{N}[1]{>{\textbf{\stepcounter{reponum}\thereponum}.\ \RaggedRight\arraybackslash}p{#1}}
\newcommand{\rowsep}{\\\arrayrulecolor{black!15}\midrule\arrayrulecolor{black}}

\setlength\LTcapwidth{\textwidth}
\begin{small}
\setcounter{reponum}{0}%
\begin{longtable}{N{0.22\textwidth}|>{\RaggedRight\arraybackslash}p{0.07\textwidth}>{\RaggedRight\arraybackslash}p{0.07\textwidth}|>{\RaggedRight\arraybackslash}p{0.04\textwidth}>{\RaggedRight\arraybackslash}p{0.04\textwidth}|>{\RaggedRight\arraybackslash}p{0.11\textwidth}|>{\RaggedRight\arraybackslash}p{0.30\textwidth}}
\caption{Repositories and Tasks after applying rule-based filters (Filter Stage 1) and LLM-based filters (Filter Stage 2) as described in \S\ref{sec:a.dataset.filtering}. We also showcase the number of tasks, the date of creation of the latest task, and additional information about the functionality and popularity of the repository. Most repositories are software tools used extensively within scientific communities. }\label{tab:dataset.construction.stage2}\\
\toprule
\multicolumn{1}{>{\RaggedRight\arraybackslash}p{0.22\textwidth}}{\textbf{Repository Name}} &
\textbf{\#Stars} & \textbf{\#Forks} & \textbf{Filter Stage 1} & \textbf{Filter Stage 2} & \textbf{Latest Task Date} & \textbf{Description} \\
\midrule
\endfirsthead

\toprule
\multicolumn{1}{>{\RaggedRight\arraybackslash}p{0.22\textwidth}}{\textbf{Repository Name}} &
\textbf{\#Stars} & \textbf{\#Forks} & \textbf{Filter Stage 1} & \textbf{Filter Stage 2} & \textbf{Latest Task Date} & \textbf{Description} \\
\midrule
\endhead

\midrule
\multicolumn{7}{r}{\small\itshape Continued on next page}\\
\bottomrule
\endfoot

\bottomrule
\endlastfoot

scikit-learn/\allowbreak{}scikit-learn & 63792 & 26359 & 2434 & 243 & 2025-10-31 & scikit-learn: machine learning in Python \rowsep
pandas-dev/\allowbreak{}pandas & 46922 & 19184 & 3298 & 560 & 2025-11-11 & Flexible and powerful data analysis / manipulation library for Python, providing labeled data structures similar to R data.frame objects, statistical functions, and much more \rowsep
scipy/\allowbreak{}scipy & 14120 & 5516 & 1454 & 209 & 2025-10-29 & SciPy library main repository \rowsep
apache/\allowbreak{}arrow & 16089 & 3884 & 1988 & 267 & 2025-07-22 & Apache Arrow is the universal columnar format and multi-language toolbox for fast data interchange and in-memory analytics \rowsep
networkx/\allowbreak{}networkx & 16277 & 3415 & 288 & 44 & 2025-09-16 & NetworkX is a Python package for the creation, manipulation, and study of the structure, dynamics, and functions of complex networks. \rowsep
Qiskit/\allowbreak{}qiskit & 6598 & 2659 & 717 & 212 & 2025-11-19 & Qiskit is an open-source SDK for working with quantum computers at the level of pulses, circuits, and application modules. \rowsep
scikit-image/\allowbreak{}scikit-image & 6371 & 2320 & 458 & 54 & 2025-11-18 & Image processing in Python \rowsep
pymc-devs/\allowbreak{}pymc & 9322 & 2146 & 685 & 45 & 2025-09-23 & PyMC (formerly PyMC3) is a Python package for Bayesian statistical modeling focusing on advanced Markov chain Monte Carlo (MCMC) and variational inference (VI) algorithms. \rowsep
Textualize/\allowbreak{}rich & 54172 & 1920 & 165 & 11 & 2025-07-25 & Rich is a Python library for rich text and beautiful formatting in the terminal. \rowsep
tqdm/\allowbreak{}tqdm & 30580 & 1402 & 12 & 1 & 2022-03-24 & Fast, extensible progress bar for Python and CLI \rowsep
pydata/\allowbreak{}xarray & 4004 & 1192 & 609 & 101 & 2025-11-21 & N-D labeled arrays and datasets in Python \rowsep
optuna/\allowbreak{}optuna & 12922 & 1177 & 719 & 112 & 2025-11-05 & A hyperparameter optimization framework \rowsep
quantumlib/\allowbreak{}Cirq & 4772 & 1151 & 10 & 3 & 2025-11-18 & Python framework for creating, editing, and invoking Noisy Intermediate-Scale Quantum (NISQ) circuits. \rowsep
pvlib/\allowbreak{}pvlib-python & 1424 & 1126 & 110 & 8 & 2025-10-03 & A set of documented functions for simulating the performance of photovoltaic energy systems. \rowsep
ipython/\allowbreak{}ipyparallel & 2626 & 1006 & 65 & 6 & 2024-10-28 & IPython Parallel: Interactive Parallel Computing in Python \rowsep
geopandas/\allowbreak{}geopandas & 4940 & 981 & 314 & 22 & 2025-05-22 & Python tools for geographic data \rowsep
kedro-org/\allowbreak{}kedro & 10593 & 971 & 41 & 4 & 2025-07-17 & Kedro is a toolbox for production-ready data science. It uses software engineering best practices to help you create data engineering and data science pipelines that are reproducible, maintainable, and modular. \rowsep
HIPS/\allowbreak{}autograd & 7379 & 928 & 13 & 1 & 2017-10-21 & Efficiently computes derivatives of NumPy code. \rowsep
MDAnalysis/\allowbreak{}mdanalysis & 1477 & 733 & 196 & 23 & 2025-10-13 & MDAnalysis is a Python library to analyze molecular dynamics simulations. \rowsep
pybamm-team/\allowbreak{}PyBaMM & 1387 & 692 & 218 & 17 & 2025-04-29 & PyBaMM (Python Battery Mathematical Modelling) is an open-source battery simulation package written in Python. \rowsep
modin-project/\allowbreak{}modin & 10332 & 669 & 50 & 8 & 2025-09-30 & Speed up your Pandas workflows by changing a single line of code \rowsep
nilearn/\allowbreak{}nilearn & 1322 & 631 & 138 & 2 & 2025-10-09 & Machine learning for NeuroImaging in Python \rowsep
sunpy/\allowbreak{}sunpy & 971 & 626 & 663 & 22 & 2025-05-16 & sunpy is a Python software package that provides fundamental tools for accessing, loading and interacting with solar physics data in Python. \rowsep
shapely/\allowbreak{}shapely & 4284 & 600 & 150 & 21 & 2025-05-03 & Manipulation and analysis of geometric objects \rowsep
dedupeio/\allowbreak{}dedupe & 4387 & 568 & 25 & 4 & 2023-12-19 & A python library for accurate and scalable data deduplication and entity-resolution. \rowsep
h5py/\allowbreak{}h5py & 2174 & 547 & 263 & 35 & 2025-08-10 & h5py is a thin, pythonic wrapper around HDF5 \rowsep
PyWavelets/\allowbreak{}pywt & 2294 & 517 & 12 & 1 & 2024-07-16 & PyWavelets - Wavelet Transforms in Python \rowsep
pydicom/\allowbreak{}pydicom & 2070 & 508 & 86 & 7 & 2025-05-12 & Read, modify and write DICOM files with python code \rowsep
arviz-devs/\allowbreak{}arviz & 1737 & 458 & 107 & 5 & 2025-10-21 & Exploratory analysis of Bayesian models \rowsep
napari/\allowbreak{}napari & 2512 & 454 & 849 & 69 & 2025-09-30 & napari: a fast, interactive, multi-dimensional image viewer for python \rowsep
tardis-sn/\allowbreak{}tardis & 225 & 446 & 268 & 13 & 2025-09-16 & TARDIS - Temperature And Radiative Diffusion In Supernovae \rowsep
dipy/\allowbreak{}dipy & 787 & 446 & 194 & 16 & 2025-11-18 & DIPY is the paragon 3D/4D+ medical imaging library in Python. Contains generic methods for spatial normalization, signal processing, machine learning, statistical analysis and visualization of medical images. Additionally, it contains specialized methods for computational anatomy including diffusion, perfusion and structural imaging. \rowsep
python-control/\allowbreak{}python-control & 1908 & 444 & 117 & 6 & 2025-06-21 & The Python Control Systems Library is a Python module that implements basic operations for analysis and design of feedback control systems. \rowsep
SciTools/\allowbreak{}cartopy & 1545 & 389 & 74 & 6 & 2025-04-26 & Cartopy is a Python package designed for geospatial data processing in order to produce maps and other geospatial data analyses. \rowsep
holoviz/\allowbreak{}datashader & 3467 & 377 & 90 & 19 & 2025-10-09 & Quickly and accurately render even the largest data. \rowsep
microsoft/\allowbreak{}Qcodes & 396 & 335 & 187 & 10 & 2025-09-05 & Modular data acquisition framework \rowsep
mars-project/\allowbreak{}mars & 2748 & 326 & 164 & 51 & 2023-02-16 & Mars is a tensor-based unified framework for large-scale data computation which scales numpy, pandas, scikit-learn and Python functions. \rowsep
pytroll/\allowbreak{}satpy & 1146 & 320 & 520 & 45 & 2025-08-02 & Python package for reading, manipulating and writing satellite data \rowsep
SciTools/\allowbreak{}iris & 692 & 297 & 109 & 23 & 2025-10-31 & A powerful, format-agnostic, and community-driven Python package for analysing and visualising Earth science data \rowsep
lmfit/\allowbreak{}lmfit-py & 1164 & 290 & 205 & 8 & 2022-09-05 & Non-Linear Least Squares Minimization, with flexible Parameter settings, based on scipy.optimize, and with many additional classes and methods for curve fitting. \rowsep
deepchecks/\allowbreak{}deepchecks & 3924 & 286 & 99 & 9 & 2023-12-06 & Deepchecks: Tests for Continuous Validation of ML Models \& Data. Deepchecks is a holistic open-source solution for all of your AI \& ML validation needs, enabling to thoroughly test your data and models from research to production. \rowsep
devitocodes/\allowbreak{}devito & 632 & 242 & 99 & 7 & 2025-07-24 & DSL and compiler framework for automated finite-differences and stencil computation \rowsep
danielgtaylor/\allowbreak{}python-betterproto & 1733 & 233 & 42 & 1 & 2023-12-07 & Better Protobuf / gRPC code generator and library for Python \rowsep
scikit-learn-contrib/\allowbreak{}metric-learn & 1425 & 229 & 6 & 1 & 2017-11-27 & Metric Learning in Python \rowsep
pydicom/\allowbreak{}pynetdicom & 551 & 188 & 24 & 1 & 2025-05-24 & A Python implementation of the DICOM networking protocol \rowsep
scverse/\allowbreak{}anndata & 667 & 175 & 142 & 17 & 2025-07-23 & Annotated data matrix for single-cell genomics \rowsep
apache/\allowbreak{}arrow-adbc & 498 & 160 & 571 & 63 & 2025-11-07 & Database connectivity API standard and libraries for Apache Arrow \rowsep
man-group/\allowbreak{}ArcticDB & 2102 & 153 & 11 & 2 & 2025-11-19 & ArcticDB is a high performance data store for time series and tick data \rowsep
stac-utils/\allowbreak{}pystac & 412 & 127 & 48 & 1 & 2023-03-31 & Python library for working with SpatioTemporal Asset Catalog (STAC) \rowsep
xdslproject/\allowbreak{}xdsl & 433 & 125 & 2136 & 236 & 2025-11-04 & A Python compiler design toolkit. \rowsep
ActivitySim/\allowbreak{}activitysim & 217 & 117 & 51 & 10 & 2025-11-12 & An open platform for activity-based travel behavior modeling \rowsep
OGGM/\allowbreak{}oggm & 245 & 115 & 484 & 36 & 2025-04-01 & Open Global Glacier Model (OGGM): a modular framework for glacier modeling \rowsep
datalad/\allowbreak{}datalad & 613 & 115 & 426 & 31 & 2024-09-10 & Keep code, data, containers under control with git and git-annex \rowsep
pydata/\allowbreak{}bottleneck & 1144 & 112 & 61 & 20 & 2025-04-29 & Fast NumPy array functions written in C \rowsep
wmayner/\allowbreak{}pyphi & 406 & 100 & 25 & 1 & 2024-09-24 & A toolbox for integrated information theory. \rowsep
django-components/\allowbreak{}django-components & 1463 & 100 & 53 & 3 & 2025-09-30 & Reusable, composable components for Django templates \rowsep
sourmash-bio/\allowbreak{}sourmash & 524 & 88 & 297 & 27 & 2025-01-09 & Quickly search, compare, and analyze genomic and metagenomic data sets. \rowsep
tskit-dev/\allowbreak{}msprime & 201 & 88 & 209 & 9 & 2025-07-24 & Simulate genealogical trees and genomic sequence data using population genetic models \rowsep
numpy/\allowbreak{}numpy-financial & 384 & 87 & 13 & 4 & 2024-04-04 & Financial functions for NumPy \rowsep
makepath/\allowbreak{}xarray-spatial & 894 & 85 & 38 & 9 & 2023-02-16 & Spatial analysis algorithms for xarray implemented in numba \rowsep
dwavesystems/\allowbreak{}dimod & 135 & 84 & 152 & 20 & 2024-06-13 & dimod is a shared API for samplers. \rowsep
python-hyper/\allowbreak{}h11 & 530 & 83 & 18 & 2 & 2025-01-12 & A pure-Python, bring-your-own-I/O implementation of HTTP/1.1 \rowsep
bjodah/\allowbreak{}chempy & 611 & 81 & 69 & 1 & 2018-03-24 & A package useful for chemistry written in Python \rowsep
holoviz/\allowbreak{}param & 497 & 79 & 85 & 10 & 2025-02-27 & Declarative parameters for robust Python classes and a rich API for reactive programming \rowsep
inducer/\allowbreak{}loopy & 615 & 78 & 172 & 15 & 2023-07-27 & A code generator for array computations on CPUs and GPUs \rowsep
holgern/\allowbreak{}beem & 138 & 75 & 75 & 5 & 2020-12-22 & A Python library for Hive and Steem \rowsep
scverse/\allowbreak{}spatialdata & 329 & 75 & 20 & 2 & 2025-09-29 & An open and interoperable data framework for spatial omics data \rowsep
pysb/\allowbreak{}pysb & 188 & 71 & 107 & 7 & 2021-01-20 & PySB is a framework for building mathematical models of biochemical systems as Python programs \rowsep
xorbitsai/\allowbreak{}xorbits & 1199 & 70 & 186 & 22 & 2024-11-16 & Xorbits is an open-source computing framework that makes it easy to scale data science and machine learning workloads — from data preprocessing to tuning, training, and model serving. \rowsep
pysal/\allowbreak{}momepy & 563 & 67 & 80 & 12 & 2024-07-16 & Urban Morphology Measuring Toolkit \rowsep
python-adaptive/\allowbreak{}adaptive & 1203 & 62 & 28 & 5 & 2025-08-21 & :chart\_with\_upwards\_trend: Adaptive: parallel active learning of mathematical functions \rowsep
probabilistic-numerics/\allowbreak{}probnum & 459 & 61 & 52 & 7 & 2023-05-04 & Probabilistic numerics in Python \rowsep
neurostuff/\allowbreak{}NiMARE & 197 & 60 & 14 & 1 & 2025-06-13 & Coordinate- and image-based meta-analysis in Python \rowsep
NCAR/\allowbreak{}geocat-comp & 140 & 56 & 18 & 2 & 2025-08-18 & GeoCAT-comp provides implementations of computational functions for operating on geosciences data. Many of these functions originated in NCL and were translated into Python. \rowsep
mie-lab/\allowbreak{}trackintel & 243 & 53 & 55 & 5 & 2024-01-07 & trackintel is a library for the analysis of spatio-temporal tracking data with a focus on human mobility. \rowsep
JDASoftwareGroup/\allowbreak{}kartothek & 160 & 53 & 152 & 31 & 2021-03-17 & A dataset library for partitioned datasets stored in Parquet \rowsep
AllenCellModeling/\allowbreak{}aicsimageio & 220 & 51 & 50 & 3 & 2023-04-05 & Image Reading, Metadata Conversion, and Image Writing for Microscopy Images in Python \rowsep
dottxt-ai/\allowbreak{}outlines-core & 254 & 50 & 44 & 5 & 2025-03-31 & Core library for Outlines, providing structured text generation utilities \rowsep
apache/\allowbreak{}arrow-nanoarrow & 207 & 47 & 109 & 8 & 2025-10-27 & nanoarrow: a (C) library for the Apache Arrow C Data interface \rowsep
pangeo-data/\allowbreak{}climpred & 252 & 47 & 9 & 2 & 2021-11-20 & :earth\_americas: Verification of weather and climate forecasts :earth\_africa: \rowsep
pybop-team/\allowbreak{}PyBOP & 152 & 45 & 78 & 8 & 2025-07-15 & A parameterisation and optimisation package for battery models. \rowsep
UXARRAY/\allowbreak{}uxarray & 202 & 44 & 99 & 22 & 2025-09-11 & Python library for working with unstructured grid model data in xarray \rowsep
pygeos/\allowbreak{}pygeos & 388 & 43 & 101 & 17 & 2021-11-30 & Wraps GEOS geometry functions in numpy ufuncs \rowsep
innobi/\allowbreak{}pantab & 120 & 41 & 79 & 7 & 2024-10-31 & Read/Write pandas DataFrames with Tableau Hyper Extracts \rowsep
xarray-contrib/\allowbreak{}xskillscore & 237 & 41 & 23 & 1 & 2021-11-20 & Metrics for verifying forecasts \rowsep
glotzerlab/\allowbreak{}signac & 135 & 37 & 17 & 2 & 2025-04-04 & Manage large and heterogeneous data spaces on the file system. \rowsep
sgkit-dev/\allowbreak{}sgkit & 265 & 37 & 113 & 21 & 2025-09-30 & Scalable genetics toolkit \rowsep
TileDB-Inc/\allowbreak{}TileDB-Py & 198 & 36 & 51 & 5 & 2025-08-01 & Python API for TileDB \rowsep
IntelPython/\allowbreak{}dpctl & 117 & 31 & 37 & 2 & 2025-10-02 & Data Parallel Control (dpctl) - Python device control and USM memory for SYCL \rowsep
tensorwerk/\allowbreak{}hangar-py & 205 & 29 & 19 & 1 & 2019-12-04 & Hangar is version control for tensor data. Commit, branch, merge, revert, and collaborate in the data-defined software era. \rowsep
xarray-contrib/\allowbreak{}xbatcher & 184 & 28 & 20 & 3 & 2023-07-31 & Batch generation from xarray objects. \rowsep
DASDAE/\allowbreak{}dascore & 121 & 26 & 122 & 11 & 2025-09-20 & DASCore: A Python package for the analysis of distributed acoustic sensing data. \rowsep
IntelPython/\allowbreak{}dpnp & 116 & 23 & 680 & 26 & 2025-10-14 & Data Parallel Extension for NumPy \rowsep
not522/\allowbreak{}ac-library-python & 230 & 23 & 5 & 2 & 2021-11-19 & Python implementation of AtCoder Library \rowsep
xarray-contrib/\allowbreak{}flox & 133 & 21 & 150 & 39 & 2025-07-17 & Fast groupby reductions for dask and xarray \rowsep
scipp/\allowbreak{}scipp & 136 & 21 & 268 & 26 & 2025-03-17 & Python library for multi-dimensional data analysis \rowsep
pyapp-kit/\allowbreak{}psygnal & 115 & 21 & 70 & 10 & 2025-09-24 & Python observer pattern (callback/event system). Modeled after Qt Signals \& Slots (but independent of Qt) \rowsep
royerlab/\allowbreak{}ultrack & 149 & 21 & 68 & 5 & 2025-09-23 & Cell tracking and segmentation software \rowsep
xitorch/\allowbreak{}xitorch & 155 & 21 & 9 & 2 & 2024-05-24 & Differentiable scientific computing for PyTorch \rowsep
Quansight-Labs/\allowbreak{}ndindex & 107 & 16 & 12 & 3 & 2025-05-14 & A Python library for manipulating N-dimensional array indices \rowsep
jkjkil4/\allowbreak{}JAnim & 189 & 14 & 3 & 1 & 2025-03-28 & Programmatic animation engine for creating precise and smooth animations with real-time feedback \\
\end{longtable}
\end{small}
\clearpage

\newcounter{fcnum}
\newcolumntype{N}[1]{>{\textbf{\stepcounter{fcnum}\thefcnum}.\ \RaggedRight\arraybackslash}p{#1}}

\setlength\LTcapwidth{\textwidth}
\begin{small}
\setcounter{fcnum}{0}%
\begin{longtable}{N{0.20\textwidth}|>{\RaggedRight\arraybackslash}p{0.07\textwidth}|>{\RaggedRight\arraybackslash}p{0.10\textwidth}|>{\RaggedRight\arraybackslash}p{0.06\textwidth}|>{\RaggedRight\arraybackslash}p{0.09\textwidth}|>{\RaggedRight\arraybackslash}p{0.07\textwidth}|>{\RaggedRight\arraybackslash}p{0.25\textwidth}}
\caption{Repositories and Tasks represented in \fc (as of November 30, 2025). We showcase a repository level breakdown of the number of tasks, the latest task (by PR merge date), the average difficulty (0-5, with 0 being easiest), the average number of tokens in the human patch and in the prompt instructions, and the most common optimization type of the human patch.
 }\label{tab:a.formulacode-repos}\\
\toprule
\multicolumn{1}{>{\RaggedRight\arraybackslash}p{0.22\textwidth}}{\textbf{Repository}} &
\textbf{\#Tasks} & \textbf{Latest Task} & \textbf{Avg. Difficulty} & \textbf{Avg. Patch Size (Tokens)} & \textbf{Avg. PR Size (Tokens)} & \textbf{Most Common Optimization} \\
\midrule
\endfirsthead

\toprule
\multicolumn{1}{>{\RaggedRight\arraybackslash}p{0.22\textwidth}}{\textbf{Repository}} &
\textbf{\#Tasks} & \textbf{Latest Task} & \textbf{Avg. Difficulty} & \textbf{Avg. Patch Size (Tokens)} & \textbf{Avg. PR Size (Tokens)} & \textbf{Most Common Optimization} \\
\midrule
\endhead

\midrule
\multicolumn{7}{r}{\small\itshape Continued on next page}\\
\bottomrule
\endfoot

\bottomrule
\endlastfoot

pandas-dev/\allowbreak{}pandas & 222 & 2025-10-21 & 0.77 & 1842.85 & 489.35 & Micro Optimizations (26.6\%) \rowsep
scikit-learn/\allowbreak{}scikit-learn & 143 & 2025-10-31 & 1.0 & 2735.29 & 491.49 & Micro Optimizations (23.1\%) \rowsep
Qiskit/\allowbreak{}qiskit & 142 & 2025-10-03 & 1.73 & 4438.38 & 505.02 & Use Lower Level System (28.2\%) \rowsep
xdslproject/\allowbreak{}xdsl & 134 & 2025-10-09 & 1.36 & 3567.76 & 463.46 & Remove Or Reduce Work (37.3\%) \rowsep
optuna/\allowbreak{}optuna & 94 & 2025-11-05 & 0.96 & 546.29 & 471.81 & Use Better Algorithm (24.5\%) \rowsep
pydata/\allowbreak{}xarray & 69 & 2025-11-21 & 0.98 & 1929.9 & 474.04 & Micro Optimizations (30.4\%) \rowsep
scikit-image/\allowbreak{}scikit-image & 39 & 2024-11-20 & 0.83 & 2271.46 & 481.36 & Remove Or Reduce Work (28.2\%) \rowsep
networkx/\allowbreak{}networkx & 35 & 2025-09-16 & 1.0 & 1809.74 & 480.46 & Use Better Algorithm (42.9\%) \rowsep
pytroll/\allowbreak{}satpy & 30 & 2024-11-20 & 1.42 & 777.4 & 483.7 & Use Better Data Structure And Layout (30.0\%) \rowsep
pymc-devs/\allowbreak{}pymc & 18 & 2025-06-16 & 1.81 & 2589.89 & 479.89 & Use Better Algorithm (33.3\%) \rowsep
xarray-contrib/\allowbreak{}flox & 17 & 2025-07-17 & 1.47 & 2149.24 & 485.18 & Use Better Algorithm (29.4\%) \rowsep
dwavesystems/\allowbreak{}dimod & 15 & 2024-06-13 & 1.33 & 2322.93 & 476.4 & Use Better Algorithm (26.7\%) \rowsep
geopandas/\allowbreak{}geopandas & 13 & 2025-05-22 & 0.77 & 2231.62 & 497.15 & Use Better Algorithm (46.2\%) \rowsep
UXARRAY/\allowbreak{}uxarray & 13 & 2025-09-11 & 1.73 & 4722.15 & 489.38 & Remove Or Reduce Work (23.1\%) \rowsep
pydata/\allowbreak{}bottleneck & 13 & 2020-11-25 & 1.54 & 1293.23 & 492.0 & Use Lower Level System (38.5\%) \rowsep
sgkit-dev/\allowbreak{}sgkit & 12 & 2025-09-30 & 1.25 & 2231.67 & 469.0 & Do It Earlier Batch Throttle (25.0\%) \rowsep
sourmash-bio/\allowbreak{}sourmash & 11 & 2022-07-20 & 1.36 & 2561.45 & 491.91 & Use Better Algorithm (27.3\%) \rowsep
JDASoftwareGroup/\allowbreak{}kartothek & 10 & 2020-10-01 & 0.5 & 1026.8 & 466.5 & Micro Optimizations (40.0\%) \rowsep
datalad/\allowbreak{}datalad & 10 & 2021-03-19 & 0.25 & 597.5 & 492.8 & Remove Or Reduce Work (40.0\%) \rowsep
mars-project/\allowbreak{}mars & 10 & 2023-02-16 & 1.75 & 3936.5 & 495.1 & Micro Optimizations (30.0\%) \rowsep
pysal/\allowbreak{}momepy & 9 & 2024-07-16 & 1.39 & 3021.56 & 469.33 & Use Better Algorithm (77.8\%) \rowsep
Textualize/\allowbreak{}rich & 9 & 2025-07-25 & 0.56 & 391.11 & 471.67 & Micro Optimizations (55.6\%) \rowsep
tskit-dev/\allowbreak{}msprime & 7 & 2025-07-24 & 1.43 & 3013.43 & 468.86 & Micro Optimizations (28.6\%) \rowsep
pygeos/\allowbreak{}pygeos & 7 & 2021-11-30 & 2.14 & 5001.57 & 483.43 & Use Lower Level System (42.9\%) \rowsep
microsoft/\allowbreak{}Qcodes & 7 & 2025-08-27 & 0.71 & 800.43 & 467.71 & Do It Earlier Batch Throttle (28.6\%) \rowsep
napari/\allowbreak{}napari & 7 & 2025-07-29 & 1.79 & 2595.86 & 485.71 & Cache And Reuse (28.6\%) \rowsep
shapely/\allowbreak{}shapely & 6 & 2025-05-03 & 0.83 & 2131.5 & 480.17 & Use Better Algorithm (33.3\%) \rowsep
pyapp-kit/\allowbreak{}psygnal & 6 & 2025-09-24 & 0.83 & 1647.33 & 482.83 & Remove Or Reduce Work (50.0\%) \rowsep
ActivitySim/\allowbreak{}activitysim & 6 & 2024-08-09 & 1.25 & 833.83 & 465.17 & Remove Or Reduce Work (33.3\%) \rowsep
pvlib/\allowbreak{}pvlib-python & 5 & 2025-10-03 & 1.5 & 7490.2 & 482.6 & Use Better Algorithm (40.0\%) \rowsep
pybamm-team/\allowbreak{}PyBaMM & 5 & 2025-04-29 & 1.5 & 1637.6 & 496.8 & Cache And Reuse (20.0\%) \rowsep
DASDAE/\allowbreak{}dascore & 5 & 2025-09-20 & 1.5 & 5505.6 & 469.2 & Cache And Reuse (40.0\%) \rowsep
deepchecks/\allowbreak{}deepchecks & 5 & 2023-12-06 & 1.5 & 3384.6 & 505.0 & Use Better Algorithm (60.0\%) \rowsep
modin-project/\allowbreak{}modin & 5 & 2025-09-30 & 2.0 & 5533.0 & 481.0 & Micro Optimizations (60.0\%) \rowsep
mie-lab/\allowbreak{}trackintel & 4 & 2024-01-07 & 0.62 & 1404.75 & 471.75 & Use Better Algorithm (50.0\%) \rowsep
lmfit/\allowbreak{}lmfit-py & 4 & 2022-09-05 & 0.0 & 411.75 & 497.0 & Do It Earlier Batch Throttle (25.0\%) \rowsep
dottxt-ai/\allowbreak{}outlines-core & 4 & 2025-03-31 & 0.62 & 5003.75 & 480.75 & Remove Or Reduce Work (25.0\%) \rowsep
pybop-team/\allowbreak{}PyBOP & 4 & 2025-07-15 & 1.88 & 3863.0 & 464.5 & Uncategorized (75.0\%) \rowsep
sunpy/\allowbreak{}sunpy & 4 & 2025-05-12 & 1.25 & 1852.25 & 486.25 & Cache And Reuse (50.0\%) \rowsep
SciTools/\allowbreak{}cartopy & 4 & 2025-04-26 & 1.88 & 1000.0 & 475.75 & Cache And Reuse (50.0\%) \rowsep
holgern/\allowbreak{}beem & 4 & 2018-11-30 & 0.62 & 1302.5 & 462.0 & Use Better Algorithm (50.0\%) \rowsep
dipy/\allowbreak{}dipy & 3 & 2025-03-12 & 0.83 & 803.67 & 523.67 & Micro Optimizations (33.3\%) \rowsep
kedro-org/\allowbreak{}kedro & 3 & 2025-07-17 & 0.83 & 1764.67 & 526.33 & Cache And Reuse (66.7\%) \rowsep
python-adaptive/\allowbreak{}adaptive & 3 & 2025-08-21 & 0.0 & 1400.0 & 462.33 & Cache And Reuse (33.3\%) \rowsep
devitocodes/\allowbreak{}devito & 3 & 2025-07-22 & 2.5 & 2156.67 & 484.33 & Cache And Reuse (66.7\%) \rowsep
TileDB-Inc/\allowbreak{}TileDB-Py & 3 & 2025-07-29 & 0.83 & 1823.33 & 482.0 & Remove Or Reduce Work (33.3\%) \rowsep
numpy/\allowbreak{}numpy-financial & 2 & 2024-04-04 & 1.25 & 423.0 & 457.5 & Use Lower Level System (100.0\%) \rowsep
xarray-contrib/\allowbreak{}xbatcher & 2 & 2023-01-03 & 2.5 & 2981.0 & 502.5 & Do It Earlier Batch Throttle (50.0\%) \rowsep
django-components/\allowbreak{}django-components & 2 & 2025-09-30 & 0.0 & 6528.0 & 463.0 & Cache And Reuse (50.0\%) \rowsep
glotzerlab/\allowbreak{}signac & 2 & 2025-04-04 & 1.25 & 3955.0 & 532.5 & Cache And Reuse (50.0\%) \rowsep
dedupeio/\allowbreak{}dedupe & 2 & 2023-02-17 & 2.5 & 709.0 & 503.0 & Micro Optimizations (50.0\%) \rowsep
NCAR/\allowbreak{}geocat-comp & 2 & 2025-08-18 & 2.5 & 2615.0 & 498.5 & Remove Or Reduce Work (50.0\%) \rowsep
innobi/\allowbreak{}pantab & 2 & 2024-01-22 & 0.0 & 650.5 & 446.5 & Use Better Data Structure And Layout (50.0\%) \rowsep
h5py/\allowbreak{}h5py & 2 & 2025-05-23 & 2.5 & 550.5 & 548.5 & Remove Or Reduce Work (50.0\%) \rowsep
nilearn/\allowbreak{}nilearn & 2 & 2025-10-09 & 0.0 & 4810.0 & 486.5 & Micro Optimizations (50.0\%) \rowsep
holoviz/\allowbreak{}param & 2 & 2025-02-27 & 0.0 & 1287.0 & 473.5 & Do It Earlier Batch Throttle (50.0\%) \rowsep
AllenCellModeling/\allowbreak{}aicsimageio & 1 & 2022-04-13 & 2.5 & 6813.0 & 505.0 & Use Higher Level System (100.0\%) \rowsep
HIPS/\allowbreak{}autograd & 1 & 2017-10-21 & 0.0 & 525.0 & 463.0 & Micro Optimizations (100.0\%) \rowsep
OGGM/\allowbreak{}oggm & 1 & 2022-09-07 & 0.0 & 511.0 & 442.0 & Micro Optimizations (100.0\%) \rowsep
arviz-devs/\allowbreak{}arviz & 1 & 2024-05-10 & 0.0 & 299.0 & 458.0 & Micro Optimizations (100.0\%) \rowsep
danielgtaylor/\allowbreak{}python-betterproto & 1 & 2023-12-07 & 0.0 & 2995.0 & 507.0 & Use Lower Level System (100.0\%) \rowsep
makepath/\allowbreak{}xarray-spatial & 1 & 2022-05-12 & 2.5 & 3774.0 & 436.0 & Use Lower Level System (100.0\%) \rowsep
Quansight-Labs/\allowbreak{}ndindex & 1 & 2024-09-20 & 2.5 & 375.0 & 476.0 & Use Lower Level System (100.0\%) \rowsep
not522/\allowbreak{}ac-library-python & 1 & 2021-11-19 & 0.0 & 388.0 & 441.0 & Micro Optimizations (100.0\%) \rowsep
royerlab/\allowbreak{}ultrack & 1 & 2025-04-22 & 2.5 & 1816.0 & 437.0 & Do It Earlier Batch Throttle (100.0\%) \rowsep
stac-utils/\allowbreak{}pystac & 1 & 2023-03-31 & 0.0 & 1593.0 & 461.0 & Micro Optimizations (100.0\%) \rowsep
tqdm/\allowbreak{}tqdm & 1 & 2022-03-24 & 0.0 & 372.0 & 448.0 & Micro Optimizations (100.0\%) \rowsep
wmayner/\allowbreak{}pyphi & 1 & 2024-09-24 & 2.5 & 1057.0 & 480.0 & Remove Or Reduce Work (100.0\%) \rowsep
xitorch/\allowbreak{}xitorch & 1 & 2024-05-24 & 0.0 & 4352.0 & 479.0 & Micro Optimizations (100.0\%) \\

\end{longtable}
\end{small}
\clearpage

\begin{figure*}[t]
    \centering
    \includegraphics[width=\textwidth]{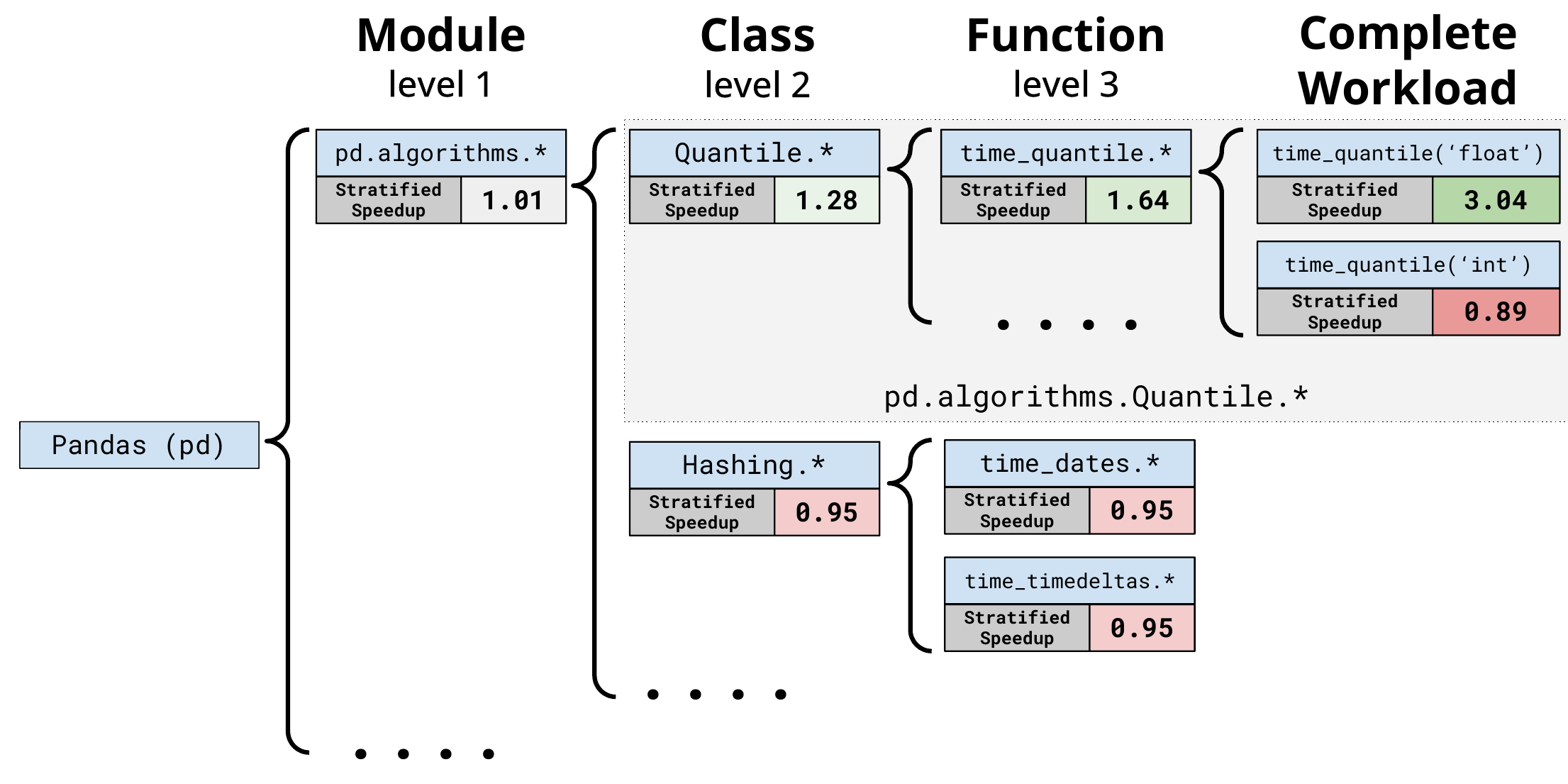}
    \caption{Illustration of Hierarchical Grouping of Pandas Workloads. By construction, each workload in \fc is organized hierarchically based on three levels: $\ell=1$ (\LevelOneName), $\ell=2$ (\LevelTwoName), and $\ell=3$ (\LevelThreeName). Metrics (like $\speedup_{\agent}$ and $\adv_{\agent}$) are computed for each complete workload (leaf nodes).  We can semantically aggregate workloads by
stratification of workloads based on this hierarcy. For instance, in this example, the stratified speedup of \texttt{pd.algorithms.Quantile.*} can be calculated by computing the geometric mean of all leaf nodes that share the same the prefix string (depicted in the gray dotted box; \texttt{pd.algorithms.Quantile.time\_quantile(`float')}, \texttt{pd.algorithms.Quantile.time\_quantile(`int')}, and other complete workloads not shown.). The example also illustrates how highly localized optimizations are diluted by stratification, and underscores that, at higher levels of stratification, \textit{consistent} speedups across a large number of workloads is required to achieve a significant stratified speedup.
}
    \label{fig:hierarchical-overview}
\end{figure*}
\clearpage

\thispagestyle{empty}
\begin{figure*}[p]
\centering
\scriptsize
\setlength{\tabcolsep}{6pt}
\renewcommand{\arraystretch}{1.05}

\begin{tabular}{|r>{\raggedright\arraybackslash}p{\pWidth}|}
\hline
\pStart & \\

\pSection{Objective}
\pRow{\textbf{You are a performance optimization expert. Speed up the repository \emph{while maintaining correctness}.}}
\pRow{}

\pSection{Tooling}
\pRow{The micromamba environment includes \textbf{Pytest} for correctness testing and \textbf{Airspeed Velocity (ASV)} for benchmarking measurements and profiling.}
\pRow{}

\pSection{Process}

\pRow{\textbf{1. Scan \& Baseline}}
\pRow{Read the code and any hints. Map likely bottlenecks. Establish a \textbf{baseline} by running the \textbf{relevant} ASV benchmarks.}
\pRow{\textbf{2. Benchmark (ASV)}}
\pRow{Read through relevant benchmarks. Prefer targeted runs using \texttt{'--bench=<regex>'}; full-suite runs are discouraged.}
\pRow{\hspace{2em}\textbf{Command:}}
\pRow{\hspace{5em}\texttt{ ''' asv run --python=same --bench="<regex>" ''' }}
\pRow{Find benchmarks via \texttt{asv\_benchmarks.txt} or within the ASV benchmarks directory. You may run multiple benchmarks at once using regexes.}

\pRow{\textbf{3. Profile Hotspots}}
\pRow{Profile \textbf{relevant} benchmarks to locate hot paths. Use ASV's built-in profiling support.}
\pRow{\hspace{2em}\textbf{Command:}}
\pRow{\hspace{5em} \texttt{''' asv profile --python=same --config=<path-to-asv.*.json> <benchmark\_name> '''}}

\pRow{\textbf{4. Optimize}}
\pRow{Make targeted changes that address the hot paths while maintaining correctness. Follow the Operating Principles below.}
\pRow{}

\pSection{Operating Principles}
\pRow{• \textbf{One change/command at a time} (code edit, ASV run, profiling).}
\pRow{• \textbf{Baseline first}, then iterate.}
\pRow{• \textbf{Target the hot paths} shown by profiling.}
\pRow{• \textbf{Evidence-driven}: justify changes with benchmark/profile data.}
\pRow{• \textbf{Correctness first}: never trade correctness for speed.}
\pRow{}

\pSection{Repository Description}
\pRow{This repository is called Qiskit/qiskit. Qiskit/qiskit is written primarily in Python and is described as a "Qiskit is an open-source SDK for working with quantum computers at the level of extended quantum circuits, operators, and primitives.".}
\pRow{}

\pSection{Task Description}
\pRow{Your main goal is to optimize the code to run as fast as possible. Use the following information if needed to understand the problem:}
\pRow{}

\pSection{Initial Observations}
\pRow{Binding parameters with `ParameterExpression.bind` is slow, allocating many Python objects and taking tens of milliseconds per call when binding large dictionaries (e.g., 100k parameters).}
\pRow{}

\pSection{Relevant Issues}
\pRow{}

\pRow{\textbf{Issue \#14471: Addressing performance bottlenecks in \texttt{ParameterExpression.bind}}}
\pRow{\textbf{Environment:} \textbf{Qiskit version}: 2.0.0}

\pRow{\textbf{Summary: } Let us consider a parameter expression \texttt{'expr'} and a dictionary \texttt{'parameter\_values: dict[Parameter, float]'} with \texttt{'M'} key, value pairs. Consider the following code to bind the expression:}
\pRow{\hspace{5em}\texttt{''' expression.bind(parameter\_values)'''}}
\pRow{As it turns out, this line takes time that grows with \texttt{len(M)}. As far as I can tell, this is because qiskit applies some checks to all of the parameters in \texttt{parameter\_values}. Even if it turns out that expression only needs one of them, all the parameters are checked and then only one of them is used.}

\pRow{\textbf{Why this needs fixing: } Sometimes, it is useful to maintain a log of parameters outside of a circuit (e.g., in a parameter table) and bind these parameters when needed agains a \texttt{'parameter\_values'} dict. In this case, the \texttt{'QuantumCircuit.assign\_parameters'} method (which does some tricks to speed things up) is not available, and users take a hit in performance when they bind.}

\pRow{\textbf{Some suggestions on how to fix this:} Provide an option for users so that they can choose to check only the 'relevant' parameter values (i.e., those present in \texttt{expression}), so that the runtime of \texttt{bind} becomes independent of \texttt{len(M)}. Review the checks and remove those that are not needed.}

\pRow{\textbf{How can we reproduce the issue?}}
\pRow{\hspace{3em}\texttt{''' from qiskit.circuit import Parameter}}
\pRow{\hspace{5em}\texttt{N: int = ...}}
\pRow{\hspace{5em}\texttt{parameter\_values = \{Parameter(f"th\_\{i\}"): 1 for i in range(N)\}}}
\pRow{\hspace{5em}\texttt{parameter\_values[param := Parameter("my\_param")] = 1}}
\pRow{\texttt{\hspace{5em}\%timeit param.bind(parameter\_values, allow\_unknown\_parameters=True)'''}}
\pRow{On my laptop, with \texttt{N=1} bind takes \texttt{\textasciitilde 2.5 \(\mu\)s}, but with \texttt{N=10**5} it takes \texttt{17.8 ms}.}
\pRow{\textbf{Comments}}
\pRow{ I'd generally be supportive of removing huge tracts of the error-checking code from all the \texttt{ParameterExpression} methods.}
\pRow{Fwiw, there are a couple of tricks we ought to figure out: the \texttt{ParameterExpression.bind} method \emph{either} has to be linear in the number of unbound parameters in the expression, or in the number of elements in the binding dictionary. \ldots}
\pRow{ \hspace{2em} \ldots \textbf{<TRUNCATED>} be cheaper even than adding fast-paths through `ParameterExpression.bind`: we don't need to maintain the QPY replay log and we don't need to allocate a new `ParameterExpression` (which is quite heavy)}

&\\
\hline
\end{tabular}

\caption{Example task in \fc for \texttt{Qiskit/qiskit} (PR: \url{https://github.com/Qiskit/qiskit/pull/14782}). The prompt presents a complete optimization task, including the performance goal, the benchmarking and profiling tools (Pytest and ASV), a structured optimization workflow, and concrete repository context with motivating performance observations. The “Relevant Issues” section contains GitHub issues that are directly related to the performance problem addressed by the PR (describing the underlying bottlenecks the PR aims to fix). These issues provide important background context that mimics a real, human-authored PR setting. Issue discussions are truncated only in this figure for brevity, while the full issue content is provided to the agent during execution.}
\label{fig:qiskit_optimization_task_spec}
\end{figure*}

\clearpage
\thispagestyle{empty}
\begin{figure*}[p]
\centering
\scriptsize
\setlength{\tabcolsep}{6pt}
\renewcommand{\arraystretch}{1.05}

\begin{tabular}{|r>{\raggedright\arraybackslash}p{\pWidth}|}
\hline
\pStart & \\

\pSection{Objective}
\pRow{\textbf{You are a performance optimization expert. Speed up the repository \emph{while maintaining correctness}.}}
\pRow{}

\pSection{Tooling}
\pRow{The micromamba environment includes \textbf{Pytest} for correctness testing and \textbf{Airspeed Velocity (ASV)} for benchmarking measurements and profiling.}
\pRow{}

\pSection{Process}

\pRow{\textbf{1. Scan \& Baseline}}
\pRow{Read the code and any hints. Map likely bottlenecks. Establish a \textbf{baseline} by running the \textbf{relevant} ASV benchmarks.}

\pRow{\textbf{2. Benchmark (ASV)}}
\pRow{Read through relevant benchmarks. Prefer targeted runs using \texttt{'--bench=<regex>'}; full-suite runs are too time-consuming and are discouraged.}
\pRow{\hspace{2em}\textbf{Command:}}
\pRow{\hspace{5em}\texttt{ ''' \# Always pin to current interpreter  asv run --python=same --bench="<regex>" ''' }}
\pRow{Find benchmarks via \texttt{asv\_benchmarks.txt} or in the directory containing the ASV benchmarks. You may run multiple benchmarks at once using regexes.}

\pRow{\textbf{3. Profile Hotspots}}
\pRow{Profile \textbf{relevant} benchmarks to locate hot paths. Use ASV's built-in profiling support.}
\pRow{\hspace{2em}\textbf{Command:}}
\pRow{\hspace{5em}\texttt{''' asv profile --python=same --config=<path-to-asv.*.json> <benchmark\_name> '''}}

\pRow{\textbf{4. Optimize}}
\pRow{Make targeted changes that address the hot paths while maintaining correctness. Always follow the Operating Principles below.}
\pRow{}

\pSection{Operating Principles}
\pRow{• \textbf{One change/command at a time} (code edit, ASV run, profiling).}
\pRow{• \textbf{Baseline first}, then iterate.}
\pRow{• \textbf{Target the hot paths} shown by profiling.}
\pRow{• \textbf{Evidence-driven}: justify changes with benchmark/profile data.}
\pRow{• \textbf{Correctness first}: never trade correctness for speed.}
\pRow{}

\pSection{Repository Description}
\pRow{This repository is called shapely/shapely. shapely/shapely is written primarily in Python and is described as a "Manipulation and analysis of geometric objects".}
\pRow{}

\pSection{Task Description}
\pRow{Your main goal is to optimize the code to run as fast as possible. Use the following information if needed to understand the problem:}
\pRow{}

\pSection{Initial Observations}
\pRow{The \texttt{deprecate\_positional} decorator incurred a noticeable runtime penalty because it invoked the full \texttt{inspect.signature} machinery on every call, leading to slow polygon construction (e.g., \textasciitilde 107\,ms per 1000 iterations in the main branch). Users also experienced repeated deprecation-warning processing overhead.}
\pRow{}

\pSection{Relevant Issues}
\pRow{}

\pRow{\textbf{Issue \#2280: 2.1 Polygon creation is much slower than 2.0.7}}
\pRow{\textbf{Summary:} It seems to be that creating Polygons in 2.1 is much slower (roughly 5--10x) slower than 2.0.7. The following script takes roughly 0.1 seconds with Shapely 2.1 and 0.015 with Shapely 2.0.7 on Python 3.12.}
\pRow{\hspace{3em}\texttt{''' import time}}
\pRow{\hspace{3em}\texttt{import shapely}}
\pRow{\hspace{3em}\texttt{if \_\_name\_\_ == "\_\_main\_\_":}}
\pRow{\hspace{5em}\texttt{start\_time = time.time()}}
\pRow{\hspace{5em}\texttt{for \_ in range(1000):}}
\pRow{\hspace{7em}\texttt{coords = ((0., 0.), (0., 1.), (1., 1.), (1., 0.), (0., 0.))}}
\pRow{\hspace{7em}\texttt{polygon = shapely.Polygon(coords)}}
\pRow{\hspace{5em}\texttt{print(time.time() - start\_time) '''}}
\pRow{\textbf{Comments:} Thanks for the report. This slowdown seems to be due to the overhead of the decorator we added to deprecate positional arguments. That decorator does inspect the signature, which in \ldots}
\pRow{\hspace{2em} \ldots \textbf{<TRUNCATED>} I noticed an even greater performance degradation when running under a debugger.}
\pRow{}

\pRow{\textbf{Issue \#2282: \texttt{deprecate\_positional} is a performance bottleneck (300\%--1000\% slowdown) in Shapely 2.1}}
\pRow{\textbf{Summary:} Performance analysis indicates that only 17 seconds from 66 seconds total is the implementation of \texttt{transform}. The remaining time is taken by the \texttt{deprecate\_positional} decorator.}
\pRow{I have the following code: }
\pRow{\hspace{3em}\texttt{''' @overload}}
\pRow{\hspace{3em}\texttt{def compressible\_geometry(geometry: \_GeomT, /) -> \_GeomT: ...}}
\pRow{\hspace{3em}\texttt{@overload}}
\pRow{\hspace{3em}\texttt{def compressible\_geometry(geometry: NDArray[np.float64], /) -> NDArray[np.float64]: ...}}
\pRow{ \hspace{2em} \ldots \textbf{<TRUNCATED>}}
\pRow{ \textbf{Comments:} -}
\pRow{}

&\\
\hline
\end{tabular}

\caption{Example task in \fc for \texttt{shapely/shapely} (PR: \url{https://github.com/shapely/shapely/pull/2283}).}
\label{fig:shapely_optimization_task_spec}
\end{figure*}

\clearpage
\thispagestyle{empty}
\begin{figure*}[p]
\centering
\scriptsize
\setlength{\tabcolsep}{6pt}
\renewcommand{\arraystretch}{1.05}

\begin{tabular}{|r>{\raggedright\arraybackslash}p{\pWidth}|}
\hline
\pStart & \\

\pSection{Objective}
\pRow{\textbf{You are a performance optimization expert. Speed up the repository \emph{while maintaining correctness}.}}
\pRow{}

\pSection{Tooling}
\pRow{The micromamba environment includes \textbf{Pytest} for correctness testing and \textbf{Airspeed Velocity (ASV)} for benchmarking measurements and profiling.}
\pRow{}

\pSection{Process}

\pRow{\textbf{1. Scan \& Baseline}}
\pRow{Read the code and any hints. Map likely bottlenecks. Establish a \textbf{baseline} by running the \textbf{relevant} ASV benchmarks.}

\pRow{\textbf{2. Benchmark (ASV)}}
\pRow{Read through relevant benchmarks. Prefer targeted runs using \texttt{'--bench=<regex>'}; full-suite runs are too time-consuming and are discouraged.}
\pRow{\hspace{2em}\textbf{Command:}}
\pRow{\hspace{5em}\texttt{ ''' \# Always pin to current interpreter\ \ \ asv run --python=same --bench="<regex>" ''' }}
\pRow{Find benchmarks via \texttt{asv\_benchmarks.txt} or in the directory containing the ASV benchmarks. You may run multiple benchmarks at once using regexes.}

\pRow{\textbf{3. Profile Hotspots}}
\pRow{Profile \textbf{relevant} benchmarks to locate hot paths. Use ASV's built-in profiling support.}
\pRow{\hspace{2em}\textbf{Command:}}
\pRow{\hspace{5em}\texttt{''' asv profile --python=same --config=<path-to-asv.*.json> <benchmark\_name> '''}}

\pRow{\textbf{4. Optimize}}
\pRow{Make targeted changes that address the hot paths while maintaining correctness. Always follow the Operating Principles below.}
\pRow{}

\pSection{Operating Principles}
\pRow{• \textbf{One change/command at a time} (code edit, ASV run, profiling).}
\pRow{• \textbf{Baseline first}, then iterate.}
\pRow{• \textbf{Target the hot paths} shown by profiling.}
\pRow{• \textbf{Evidence-driven}: justify changes with benchmark/profile data.}
\pRow{• \textbf{Correctness first}: never trade correctness for speed.}
\pRow{}

\pSection{Repository Description}
\pRow{This repository is called pandas-dev/pandas. pandas-dev/pandas is written primarily in Python and is described as a "Flexible and powerful data analysis / manipulation library for Python, providing labeled data structures similar to R data.frame objects, statistical functions, and much more".}
\pRow{}

\pSection{Task Description}
\pRow{Your main goal is to optimize the code to run as fast as possible. Use the following information if needed to understand the problem:}
\pRow{}

\pSection{Initial Observations}
\pRow{The \texttt{DataFrame.to\_csv()} call with \texttt{index=False} on a Multi-Index DataFrame was extremely slow (\(\approx\) 869 seconds for 10M rows \(\times\) 20 cols), while resetting the index first and then calling \texttt{to\_csv()} took only \(\approx\) 42 seconds. The performance gap was observed consistently in the benchmark.}
\pRow{}

\pSection{Relevant Issues}
\pRow{}

\pRow{\textbf{Issue \#59312: PERF: Significant Performance Difference in \texttt{DataFrame.to\_csv()} with and without Index Reset}}
\pRow{\textbf{Description:}}
\pRow{Pandas version checks: I have checked that this issue has not already been reported. I have confirmed this issue exists on the latest version of pandas. I have not confirmed this issue exists on the main branch of pandas.}
\pRow{\textbf{Reproducible Example}}
\pRow{Below is a toy DataFrame example with 10M rows and 20 columns. The CSV write speed differ significantly between whether the multi-index is dropped first or not, even if the resulting CSV files are essentially the same. The benchmark for PyArrow is also attached for reference. Notice that the CSV generated from PyArrow has column names and column values additionally double-quoted.}
\pRow{\hspace{3em}\texttt{''' import pandas as pd}}
\pRow{\hspace{3em}\texttt{import pyarrow as pa}}
\pRow{\hspace{3em}\texttt{import pyarrow.csv as csv}}
\pRow{\hspace{3em}\texttt{import time}}
\pRow{\hspace{3em}\texttt{NUM\_ROWS = 10000000}}
\pRow{\hspace{3em}\texttt{NUM\_COLS = 20}}
\pRow{\hspace{3em}\texttt{df = pd.DataFrame(\{f"col\_\{col\_idx\}": range(col\_idx * NUM\_ROWS, (col\_idx + 1) * NUM\_ROWS) for col\_idx in range(NUM\_COLS)\})} \ldots \textbf{<TRUNCATED>}}
\pRow{\textbf{Comments}}
\pRow{Thanks for the report! It seems to me the issue is here:}
\pRow{\hspace{3em}\texttt{''' https://github.com/pandas-dev/pandas/blob/642d2446060afb11f9860c79a7339eb6ec96fea7/pandas/io/formats/csvs.py\#L323 '''}}
\pRow{A significant amount of time on that line is spent getting the index values, only to be ignored because \texttt{self.nlevels} is 0 when \texttt{index=False}. In addition, it seems to me that there may \ldots \textbf{<TRUNCATED>}  }
\pRow{}

&\\
\hline
\end{tabular}

\caption{Example task in \fc for \texttt{pandas-dev/pandas} (PR: \url{https://github.com/pandas-dev/pandas/pull/59608}).}
\label{fig:pandas_optimization_task_spec}
\end{figure*}

\section{Experiment Details}\label{sec:a.expt}

In this section, we provide additional details on the methodology used to evaluate agents on \fc. All experiments ran on a single Ubuntu 22.04 LTS machine with 503 GiB RAM, Intel Xeon Platinum 8352Y CPU @ 2.20 GHz (128 hardware threads), 4 NVIDIA A40 GPUs (46 GiB VRAM each). Making the dataset from scratch takes $\sim 32$ hours, consuming $\sim 100$ GB of disk space for the metadata and $\sim 2$ TB of disk space for the docker image cache. Our evaluation protocol is grounded in Terminal Bench \citep{terminalbench}. Unless explicitly indicated otherwise, all experiments use the default hyperparameters defined by Terminal Bench.

\subsection{Airspeed Velocity Methodology}\label{sec:a.expt.asv-details}
To benchmark a new function with Airspeed Velocity, a developer supplies a \texttt{setup(\ldots)} routine and one or more time profiling functions (e.g.\ \texttt{time\_foo(\ldots)}, \texttt{time\_bar(\ldots)}) and memory profiling functions (e.g.\ \texttt{mem\_foo(\ldots)}, \texttt{mem\_bar(\ldots)}). \texttt{asv} then clones the repository, creates an isolated virtual environment, and records the performance characteristics for \emph{all} commits.  The tool ships with best-practice safeguards (CPU affinity, warm-ups, repeated trials, etc.) to control system variance. Section \ref{sec:problem-statement} includes additional safeguards to further minimize system variance.

Airspeed velocity offers many advantages towards our goal of making a benchmark for code optimization:
\setlist{nolistsep}
\begin{itemize}[noitemsep]
    \item \textbf{Low barrier to entry.}  The minimalist interface means developers routinely add new benchmarks, expanding coverage over time. Asv ships with a robust regression-detection functionality which further motivates developers to ensure that the asv benchmarks maximally cover all performance critical parts of their software.  
    \item \textbf{Maturity and reliability.}  First released on 1~May~2015, \texttt{asv} encapsulates nearly a decade of community experience in timing and memory profiling code on commodity hardware.  Most common pitfalls have documented solutions, and well established platform-specific best practices, ensuring results are both accurate and precise.
    \item \textbf{CI integration.}  \texttt{asv} co-exists naturally with other continuous-integration tools, so each commit carries both performance \emph{and} correctness metadata.
\end{itemize}

\begin{figure}
    \centering
    \centering
    \includegraphics[width=0.7\linewidth]{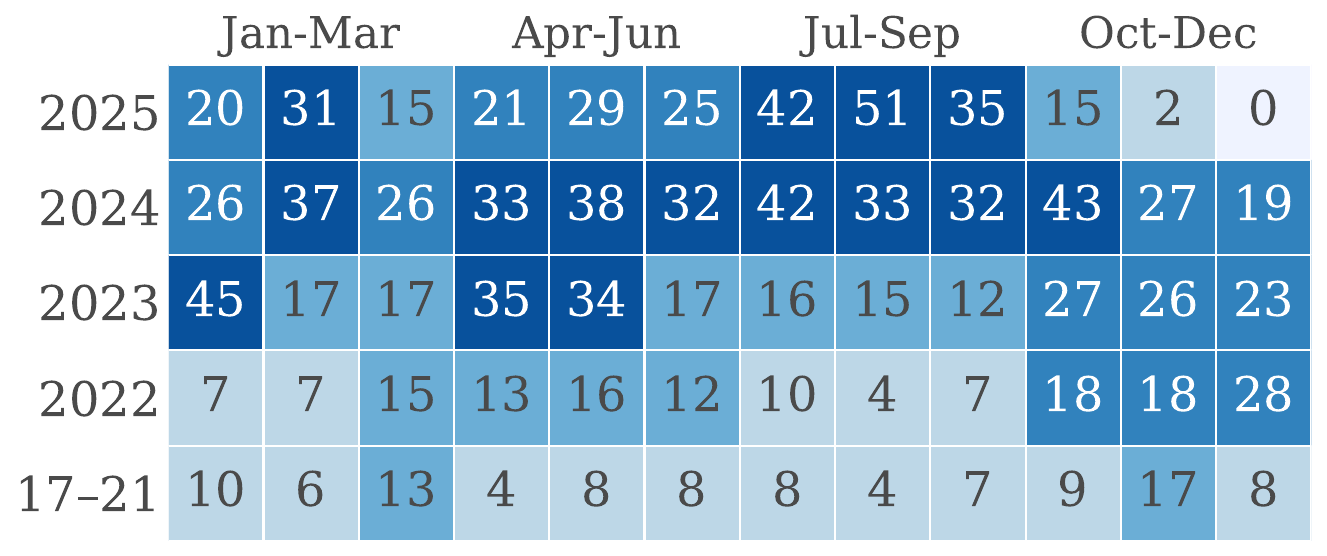}
    \caption{Timeline of \fc tasks organized by the date the expert-patch was merged till November, 2025. Each box represents the number of expert-patch tasks merged during a particular month/year. \fc is updated on the \monthlyUpdateDay{} of each month, and our most recent task is from \latestTaskDate{}. The dataset grew by \avgProblemsPerMonth{} tasks per month on average in 2025, facilitating contamination analyses for performance-optimization agents. Table \ref{tab:a.formulacode-repos} presents a detailed overview.
    }
    \label{fig:a.dataset-statistics.timeline}
\end{figure}

\subsection{Model and Agent Choices}\label{sec:a.expt.models}

\paragraph{Models.} Our experimental design centers on four models -- GPT-5, Claude 4.0 Sonnet, Gemini 2.5 Pro, and Qwen 3 Coder -- that represent the strongest generally available systems for coding and tool-use workloads at the time of paper writing. We selected these models because they are natively integrated with our inference provider and support long context windows, function calling, and multi-turn interactions at a cost profile compatible with large-scale benchmarking. We treat these models as representative of the frontier capability regime against which different agent architectures can be fairly compared.

\begin{enumerate}[leftmargin=*,label=\textbf{\arabic*.}]
    \item \textbf{GPT-5.}~GPT-5 \citep{singh2025openaigpt5card} is OpenAI's flagship general-purpose model in this study, and we use the standard API configuration with built-in “thinking” enabled. It is a multimodal, tool-using model with strong performance on code, math, and long-context reasoning benchmarks, and is widely deployed in agentic coding systems. We use the \texttt{gpt-5-2025-08-07} version specifically with a documented knowledge cutoff of late September 2024.
    
    \item \textbf{Claude 4.0 Sonnet.}~ Claude 4.0 Sonnet \citep{anthropic2025_claude4_systemcard} is Anthropic's top-end general-purpose model at the time of our experiments, designed for complex reasoning, long-form generation, and tool-heavy workloads such as software development. Public reports place Claude 4.0 Sonnet at or near the frontier on a wide range of coding and reasoning benchmarks. We use the \texttt{claude-sonnet-4-20250514} version specifically with a documented knowledge cutoff date of January 2025, with training data extending to March 2025. 
    
    \item \textbf{Gemini 2.5 Pro.}~ Gemini 2.5 Pro \citep{comanici2025gemini25pushingfrontier} is Google DeepMind's latest high-end model at the time of writing, introduced as the first member of the Gemini 2 series and optimized for complex multimodal reasoning. It offers a very large context window (up to 1M tokens in the preview configuration) and supports advanced tool-calling and code execution. It has a documented knowledge cutoff date of January 2025. We include Gemini 2.5 Pro to ensure that our agentic analysis covers three distinct provider ecosystems under comparable frontier-model conditions.

    \item \textbf{Qwen 3 Coder.}~Qwen 3 Coder is a large open Mixture-of-Experts model explicitly optimized for agentic coding tasks rather than general conversation. Qwen 3 Coder (in particular, the \texttt{qwen3-coder-480b-a35b-instruct} model) combines 480 B total parameters with sparse expert activation (35 B active parameters per forward pass) and a context window of roughly 262k tokens, enabling it to reason over entire repositories and multi-file refactors in a single pass. Third-party model cards list a knowledge cutoff of 23 January 2025 \citep{langdb_qwen3_coder_480b_2025}. Empirically, Qwen 3 Coder claims strong results on SWE-Bench and related agentic coding and browser-use benchmarks \citep{yang2025qwen3technicalreport}.
    
\end{enumerate}

\paragraph{Agents.} We evaluate two agent frameworks within \fc: Terminus 2, the default harness for Terminal-Bench, and an agent implemented with OpenHands, a popular open-source framework for AI-driven software development. We intentionally omit more complex agent families such as tree-structured search agents and evolutionary or population-based methods. Tree agents that branch over alternative command sequences must maintain multiple snapshots of the terminal state, which quickly leads to exponential blowup in cloud compute usage. Evolutionary agents that track a Pareto frontier across many workloads are similarly expensive: given that the average \fc task exposes roughly \numworkloads workloads, the number of candidate solutions required to reasonably explore the frontier is beyond our evaluation budget.

\begin{enumerate}[leftmargin=*,label=\textbf{\arabic*.}]
    \item \textbf{Terminus 2.}~ Terminus 2 is a reference agent for Terminal-Bench \citep{terminalbench}. It is intentionally minimal: the agent spawns a single tmux session and exposes the raw shell to the model, which issues commands as plain text and receives the terminal output verbatim, without additional structured tools or high-level abstractions. This architecture can be viewed as a reflexive, single-trajectory agent that repeatedly observes the current terminal state, updates its internal plan implicitly in the model's hidden state, and emits the next command. Despite its simplicity, Terminus 2 is competitive with more elaborate systems, making it a natural baseline for \fc.
    \item \textbf{OpenHands.}~OpenHands is a widely used open-source framework for AI-driven software development \citep{wang2025openhandsopenplatformai}. OpenHands exposes a flexible SDK that allows defining agents as compositions of tools and routines that can clone repositories, edit files, run tests, and manage long-running coding sessions, with support for swapping out the underlying LLM. In our experiments, we utilize a single-trajectory terminal-plus-editor agent implemented in the OpenHands SDK, following a default configuration used in terminal bench \citep{terminalbench}.
\end{enumerate}

\subsection{Kinds of Optimization Problems}\label{sec:a.expt.optimizationtypes}

We categorize expert-written solutions in \fc into thirteen optimization classes gathered from various online sources. We reviewed these sources, normalized overlapping suggestions into standard terminology, and used them to define the categories, which are then applied consistently in our analysis. 
This taxonomy is intentionally non-exhaustive: it serves as a practical baseline for analysis, capturing the principal codebase optimizations that developers typically consider when improving performance, rather than offering an authoritative catalog of all systems optimizations.

\begin{table*}[t]
    \centering
    \small
    \caption{Optimization categories used to categorize human solutions in \fc. The taxonomy is derived from various online sources, listed in the primary references for each category.}
    \label{tab:a.expt.optimizationtypes}
    \adjustbox{max width=\textwidth}{%
    \begin{tabular}{l l l}
        \toprule
        Category Abbreviation & Category Description & Source \\
        \midrule
        Algo & Use a better algorithm & \citep{tratt-four-kinds-optimisation-2023} \\
        Data & Use a better data structure (and layout) & \citep{tratt-four-kinds-optimisation-2023} \\
        Lower & Use a lower-level system & \citep{tratt-four-kinds-optimisation-2023} \\
        Approx & Accept a less-precise solution (approximation/heuristics) & \citep{tratt-four-kinds-optimisation-2023} \\
        Parallel & Use parallelization & \citep{tratt-fifth-kind-optimisation-2025} \\
        Reduce & Remove or reduce work (requirements \& UX) & \citep{hn-fifth-kind-thread-43555311,hn-four-kinds-thread-38262251} \\
        Cache & Cache \& reuse & \citep{hn-fifth-kind-thread-43555311} \\
        Batch & Do it earlier / batch it / throttle it & \citep{hn-fifth-kind-thread-43555311} \\
        Scale & Scale the platform & \citep{hn-fifth-kind-thread-43555311} \\
        DB & Database \& storage tuning & \citep{hn-fifth-kind-thread-43555311} \\
        Micro & Micro-optimizations (hot path tweaks) & \citep{hn-fifth-kind-thread-43555311} \\
        I/O & I/O and latency hiding (async, overlap I/O/compute) & \citep{hn-fifth-kind-thread-43555311,hn-four-kinds-thread-38262251} \\
        Higher & Use a higher-level system that optimizes for you & \citep{hn-fifth-kind-thread-43555311} \\
        Uncat & Uncategorized & -- \\
        \bottomrule
    \end{tabular}
    }
\end{table*}

\subsection{Qualitative Examples}\label{sec:a.expt.examples}

Qualitative examples are presented in Figure \ref{fig:qiskit_optimization_task_spec}, Figure \ref{fig:shapely_optimization_task_spec}, and Figure \ref{fig:pandas_optimization_task_spec}.

\subsection{Terminal Bench Modifications}\label{sec:a.expt.tbenchmods}

Terminal-Bench \citep{terminalbench} is a widely used harness for benchmarking terminal-based software development tasks. It is actively maintained, well understood by the agent development and benchmarking community, and already designed around end-to-end agent execution in a containerized shell environment. However, Terminal-Bench primarily targets correctness-oriented evaluations. In \fc, the evaluation target shifts: tasks are optimization-centric and require measuring performance improvements reliably, comparing multiple agent/model configurations under matched conditions, and auditing performance-oriented behavior and cost. We therefore extend Terminal-Bench along four capability axes.

\textit{Standardized execution for low-variance measurement.} To complement the variance-control safeguards in Section~\ref{sec:problem-statement}, we add support for executing runs in standardized isolated environments (e.g., fixed cloud machines). This reduces machine-to-machine drift and makes speedup measurements more comparable across runs, which is essential when the benchmark signal is a relative performance change rather than a binary pass/fail outcome. Operationally, we extend Terminal Bench to support running tasks on compute optimized Amazon Web Services (AWS) EC2 instances. Such instances are guaranteed to have a finite amount of isolated hardware resources situated in professionally-managed data centers, ensuring third-party reproducibility of \fc's experiments \citep{hardwareisolationAWS}. We use the \texttt{c5ad.large} instance with 2 vCPUs, 4GiB RAM, and a dedicated 75 GiB SSD for storage. This instance is chosen specifically because it is extremely cost efficient (on-demand price of \$0.086 per hour at the time of writing). Importantly, remote execution is a reproducibility convenience rather than a methodological prerequisite. The ASV-based protocol (warm-ups, repeated trials, and the variance controls in Section~\ref{sec:problem-statement}) is designed to yield reliable estimates on well-managed local commodity machines. We use EC2 primarily to eliminate avoidable confounds -- resource contention, background load, and hardware heterogeneity -- to provide a clean gold-standard reference for subsequent experiments.

\textit{Sequential agent evaluation.} We add controls to evaluate multiple agent/model configurations sequentially within the same standardized environment. For each \fc task, we provision a single instance and evaluate agent/model configurations in separate fresh containers: we measure the baseline implementation ($\texttt{Code}_0$), then the expert-written optimized solution ($\texttt{Code}_{\texttt{expert}}$), and then each agent-produced candidate in turn, resetting the container state between configurations. This design ensures that comparisons are statistically matched by construction (same hardware and near-identical runtime conditions) while preventing cross-run interference from accumulated state.

\begin{figure}
    \centering
    \includegraphics[width=0.8\linewidth]{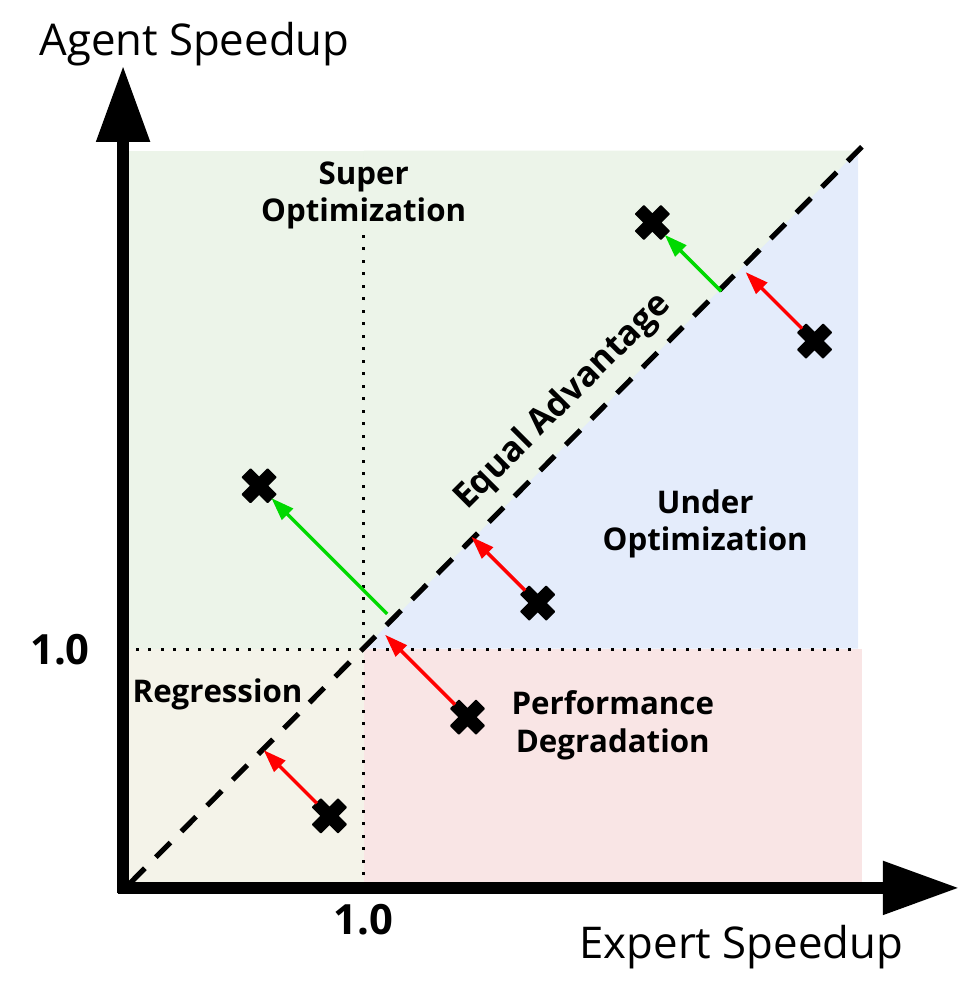}
    \caption{
    Visual intuition for Agent Advantage ($\adv_{\agent}$; \S\ref{sec:problem-statement}). Each cross (\xmark) represents an individual workload using the expert-derived speedup ($\speedup_{\texttt{expert}}$) and the agent-derived speedup ($\speedup_{\agent}$). The identity function line represents \textit{equal advantage} (i.e., $\speedup_{\texttt{expert}} = \speedup_{\agent}$). Then, the agent advantage is the mean weighted deviation from the equal advantage line. The plot also showcases four optimization regions clockwise from top: (1) \textit{Super Optimization}: workloads where an agent's code performs better than the expert's code and the baseline. (2) \textit{Under Optimization}: workloads where the agent's code and the expert's code both deliver a positive speedup, but the expert outperforms the agent. (3) \textit{Performance Degredation}: workloads where the expert discovers a speedup while the agent slows down the code. (4) \textit{Regression}: workloads where neither the expert nor the agent slow down the code; usually an intentional tradeoff to optimize other workloads. Figure \ref{fig:advantage-example} showcases an example of workload distribution for various agents on \fc.
    }
    \label{fig:advantage-example}
\end{figure}

\textit{Optimization-centric metrics.} Terminal-Bench natively aggregates discrete outcomes (e.g., test pass/fail). We extend the measurement and analysis layers to parse and summarize continuous optimization signals (e.g., speedup, advantage, and variance) and to support custom aggregation procedures (e.g., stratification by difficulty, as described in Figure~\ref{fig:advantage-example}).

\textit{Additional Accounting metrics.} Finally, we add explicit support for token-usage and API-cost accounting, as well as other observability metrics (improved logging, robust timeout handling, and comprehensive interactive traces). These additions enable the cost-aware and failure-mode analysis reported in Section~\ref{sec:experiments}.

Overall, these modifications enable the use of Terminal-Bench as a stable evaluation harness for \fc.

\paragraph{Evaluation Hyperparameters.}
Tables~\ref{tab:hyperparams} and~\ref{tab:eval-model-specs} report the full evaluation configuration: Table~\ref{tab:hyperparams} covers decoding, the agent loop, tools, and the execution environment, while Table~\ref{tab:eval-model-specs} lists per-model API specifications. Following Terminal-Bench~\citep{terminalbench} conventions, each framework runs at its pinned-version defaults (entries marked $^{a}$, verified from the frameworks' source at the pinned versions), with two FormulaCode-specific changes (marked $^{b}$): context compaction at a 200k-token limit and a six-hour wall-clock cap (never reached; average runtime 38 minutes per task).

\begin{small}
\setlength{\LTpre}{6pt}\setlength{\LTpost}{6pt}
\begin{longtable}{@{}>{\raggedright\arraybackslash}p{0.23\textwidth} >{\raggedright\arraybackslash}p{0.34\textwidth} >{\raggedright\arraybackslash}p{0.34\textwidth}@{}}
\caption{Evaluation configuration for all agent--model runs. Entries marked $^{a}$ are framework defaults verified from source at the pinned versions (Terminus~2 at commit \texttt{d71d8fc}; OpenHands \texttt{v0.51.0}, run through the Terminal-Bench adapter); entries marked $^{b}$ are FormulaCode-specific changes. Per-model API limits, reasoning modes, and prices are in Table~\ref{tab:eval-model-specs}.}
\label{tab:hyperparams}\\
\toprule
\textbf{Parameter} & \textbf{Terminus 2} & \textbf{OpenHands} \\
\midrule
\endfirsthead
\multicolumn{3}{@{}l}{\footnotesize\itshape Table~\ref{tab:hyperparams}, continued from previous page}\\
\toprule
\textbf{Parameter} & \textbf{Terminus 2} & \textbf{OpenHands} \\
\midrule
\endhead
\midrule
\multicolumn{3}{r@{}}{\footnotesize\itshape continued on next page}\\
\endfoot
\bottomrule
\endlastfoot
\multicolumn{3}{@{}l}{\textbf{Decoding and sampling}}\\
Temperature & $0.7^{a}$ & $0.0^{a}$ \\
Top-$p$ & provider default (unset)$^{a}$ & $1.0^{a}$ \\
Top-$k$ / penalties & provider default & provider default \\
Reasoning effort & per model (Table~\ref{tab:eval-model-specs}) & per model (Table~\ref{tab:eval-model-specs}) \\
Max output tokens & model cap (Table~\ref{tab:eval-model-specs}) & model cap (Table~\ref{tab:eval-model-specs}) \\
\midrule
\multicolumn{3}{@{}l}{\textbf{Agent loop and orchestration}}\\
Max iterations / steps & unlimited (\texttt{max\_episodes}\,$=10^{6}$)$^{a}$ & $500^{a}$ \\
Context management & compaction at 200k-token limit$^{b}$ & LLM-summarizing condenser ($>$100 events) $+$ history truncation$^{a}$ \\
Per-output truncation & 10\,000 bytes, head$+$tail kept$^{a}$ & 30\,000 chars / observation$^{a}$ \\
Stopping criterion & self-termination (double-confirmed) or timeout & self-termination (\texttt{finish}) or timeout \\
Wall-clock cap / task & \multicolumn{2}{@{}>{\raggedright\arraybackslash}p{0.68\textwidth}}{6\,h hard cap, never reached; average 38 min per task$^{b}$} \\
Retry policy & retry on malformed response & \texttt{num\_retries}\,$=5$ (LiteLLM)$^{a}$ \\
Tool / token budget & none; fully sandboxed & none; no budget cap$^{a}$ \\
\midrule
\multicolumn{3}{@{}l}{\textbf{Tools and interface}}\\
Available tools & raw \texttt{tmux} keystrokes (bash) & bash, \texttt{str\_replace} editor, IPython, \texttt{think}, \texttt{finish}$^{a}$ \\
Browsing & none & disabled$^{a}$ \\
Response parsing & JSON plain-text parser, XML fallback$^{a}$ & native tool calling, auto-detected$^{a}$ \\
LLM backend & LiteLLM (provider-agnostic)$^{a}$ & LiteLLM (provider-agnostic)$^{a}$ \\
Agent / version & \texttt{terminus-2} at \texttt{d71d8fc} (2025-08-23) & \texttt{CodeActAgent}, \texttt{openhands-ai}~v0.51.0 (2025-07-31) \\
\midrule
\multicolumn{3}{@{}l}{\textbf{Execution environment (shared)}}\\
Hardware & \multicolumn{2}{@{}>{\raggedright\arraybackslash}p{0.68\textwidth}}{AWS EC2 \texttt{c5ad.large}: 2 vCPU, 4\,GiB RAM, 75\,GiB SSD; on-demand \$0.086/h (\S\ref{sec:a.expt.tbenchmods})} \\
Isolation & \multicolumn{2}{@{}>{\raggedright\arraybackslash}p{0.68\textwidth}}{fresh Docker container per configuration; container state reset between runs} \\
Measurement & \multicolumn{2}{@{}>{\raggedright\arraybackslash}p{0.68\textwidth}}{ASV warm-up $+$ 2--40 adaptive samples per workload; significance by Mann--Whitney~U ($p<0.002$) with Holm--Bonferroni correction (\S\ref{sec:a.dataset.stat_testing})} \\
\end{longtable}
\noindent{\footnotesize $^{a}$\,Framework default at the pinned version, verified from source; please confirm against the exact run configuration. $^{b}$\,FormulaCode-specific change relative to the Terminal-Bench defaults.}
\end{small}

\begin{table}[t]
\centering
\small
\caption{Per-model API specifications for the four evaluated models (provider list values at the mid-to-late-2025 experiment era). Prices are reference list prices; the cost analysis (\S\ref{sec:results-cost}) is based on incurred inference costs.}
\label{tab:eval-model-specs}
\adjustbox{max width=\textwidth}{%
\begin{tabular}{l l r r l l r}
\toprule
Model & API snapshot & Context & Max out & Cutoff & Reasoning & Price in/out (\$/M) \\
\midrule
GPT-5 & \texttt{gpt-5-2025-08-07} & 400K & 128K & Sep 2024 & effort levels (high) & 1.25 / 10.00 \\
Claude 4.0 Sonnet & \texttt{claude-sonnet-4-20250514} & 200K & 64K & Jan 2025$^{*}$ & thinking budget & 3.00 / 15.00 \\
Gemini 2.5 Pro & \texttt{gemini-2.5-pro} & 1M & 65.5K & Jan 2025 & thinking (always on) & 1.25--2.50 / 10--15$^{\dagger}$ \\
Qwen3-Coder-480B & \texttt{qwen3-coder-480b-a35b} & 262K$^{\ddagger}$ & 65.5K & 23 Jan 2025$^{\S}$ & none (non-thinking) & host-dependent$^{\S}$ \\
\bottomrule
\end{tabular}
}

{\footnotesize $^{*}$\,Reliable cutoff Jan 2025; training data to Mar 2025. $^{\dagger}$\,Tiered by prompt size ($\leq$200K\,/\,$>$200K input tokens); output price includes thinking tokens. $^{\ddagger}$\,262K native, extendable to $\sim$1M via YaRN. $^{\S}$\,Open-weights model; cutoff per a third-party model card; price varies by inference host (e.g., \$2.00\,/\,\$2.00 per 1M tokens on Together AI).}
\end{table}

\begin{table*}[t]
    \centering
    \caption{
        Cost-aware leaderboard of agent--model configurations.
        We report cost per task, mean advantage $\adv_{\agent}$, cost-weighted advantage $\adv_{\agent}^{\text{cost}}$, and cost-weighted normalized advantage $\normadv{\agent}^{\text{cost}}$.
    }
    \label{tab:cost-advantage-leaderboard}
    \adjustbox{max width=\textwidth}{%
    \begin{tabular}{l l r r r r}
        \toprule
        Agent & Model & Cost/Task $\downarrow$ & $\adv_{\agent}$ $\uparrow$ & $\adv_{\agent}^{\text{cost}}$ $\uparrow$ & $\normadv{\agent}^{\text{cost}}$ $\uparrow$ \\
        \midrule
Terminus 2 & GPT-5 & 1.8508 & -0.0504 & -0.0272 & -0.0750 \\
 & Claude 4.0 Sonnet & 3.7722 & -0.0410 & -0.0109 & -0.0282 \\
 & Gemini 2.5 Pro & 1.5455 & -0.0433 & -0.0280 & -0.0737 \\
 & Qwen 3 Coder & 1.2060 & -0.0454 & -0.0376 & -0.1043 \\
OpenHands & GPT-5 & 0.7814 & -0.0209 & -0.0267 & -0.0899 \\
 & Claude 4.0 Sonnet & 3.2300 & -0.0112 & -0.0035 & -0.0150 \\
 & Qwen 3 Coder & 1.0974 & -0.0301 & -0.0274 & -0.1393 \\
        \bottomrule
    \end{tabular}
    }
\end{table*}

\subsection{Metrics Discussion}\label{sec:discussion-metrics}

\paragraph{Why advantage and speedup can disagree.}
A recurring observation in our results (\S\ref{sec:results-global}) is that the global leaderboard rankings under \speedup~and \adv~can differ.
This disagreement arises because \speedup~treats all tasks equally in absolute terms, while \adv~compares each agent's improvement to the corresponding expert improvement.
In practice, large absolute speedups tend to concentrate on tasks where the expert also achieved large speedups (i.e., tasks with substantial optimization headroom).
An agent that achieves high \speedup~by excelling on these ``easier'' tasks may receive a lower \adv~than an agent whose improvements are distributed toward tasks where experts found only modest gains.
We observe this pattern concretely: under OpenHands, GPT-5 achieves higher \speedup~than Claude 4.0 Sonnet but lower \adv, because GPT-5's gains concentrate on high-headroom tasks while Claude 4.0 Sonnet's gains are more uniformly distributed.
We recommend \adv~(and its Ranked Pairs aggregation) as the primary comparison metric because it controls for task difficulty and is invariant to the magnitude of the expert solution.

\paragraph{Log-space advantage.}
Since \speedup~is a multiplicative quantity aggregated via geometric means, a natural concern is whether comparisons should be conducted in log-space rather than in linear-space.
We define the log-space advantage as $\text{LogAdv}_\agent = \log(\speedup_\agent) - \log(\speedup_\expert)$.
We recomputed the Ranked Pairs ordering under $\text{LogAdv}$ for the OpenHands configurations:

\begin{table}[h]
\centering
\small
\caption{Comparison of advantage metrics under linear and log-space for OpenHands configurations. Rankings are preserved under both formulations.}
\label{tab:logadv}
\begin{tabular}{l c c c}
\toprule
\textbf{Model} & \textbf{Adv} & \textbf{LogAdv} & \textbf{RP Rank} \\
\midrule
Claude 4.0 Sonnet & $-0.0112$ & $-0.0130$ & 1 \\
Qwen 3 Coder & $-0.0301$ & $-0.0280$ & 2 \\
GPT-5 & $-0.0209$ & $-0.0167$ & 3 \\
\bottomrule
\end{tabular}
\end{table}

The relative ordering of models is unchanged, and the magnitudes are similar to the original analysis.
This is because the geometric-mean speedups approach $1$ as the number of workloads per task grows, and in this regime $\text{Adv}$ is approximately equal to $\text{LogAdv}$---this approximation is exactly the first-order Taylor expansion.
We retain the linear-space formulation as the primary metric due to its interpretability, but note that our conclusions are robust to log-space transformation.

\paragraph{Advantage metric design principles.}
The \adv~metric has several desirable properties for evaluating optimization agents.
First, it calibrates for task difficulty by comparing against the best-known expert solution: a speedup of 1.05 on an already highly-optimized codebase is valued more than the same speedup on an under-optimized one.
Second, if an agent memorizes and reproduces the expert patch verbatim (e.g., due to training data contamination), it receives $\adv = 0$ rather than spurious credit.
Third, because \adv~is defined per-workload before aggregation, it captures behavioral similarity between agent and expert strategies even when the surface-level code differs substantially.

\subsection{Additional Analysis}\label{sec:a.expt.additional-analysis}

This section lists additional analysis on \fcv that was not included in the main paper for space reasons. We analyze (1) the rate of correctness constraint violations across agent/model configurations, (2) the relationship between trajectory length and performance, (3) patterns of tool usage across configurations, (4) patch-level memorization, and (5) qualitative examples of agent patches.

\paragraph{Correctness Constraint Violations.}

Each \fcv task is associated with two types of correctness constraints: (1) Snapshot tests, that verify that the optimized codebase preserves each workload's local variables, and (2) the original PyTest suite from the upstream repository which captures broader functional correctness. At initialization, the agent--model configuration receives explicit instructions to maximize performance \emph{while preserving correctness}. If the patch fails either constraint, we `roll back' any performance improvements and revert to the original codebase, ensuring that all reported speedups are strictly correctness-preserving. We therefore ask: how often are candidate performance improving edits rejected solely due to correctness violations?

For each agent--model pair, we count the number of tasks in which the final patch fails at least one test, and then further break this down into PyTest failures and snapshot test failures.
\tabCorrectness\ summarizes these statistics over \numFCVProblems\ attempted solutions per configuration.

\begin{table}[t]
    \centering
    \small
    \setlength{\tabcolsep}{4pt}
    \renewcommand{\arraystretch}{1.15}
    \caption{
        Correctness constraint violations by agent--model configuration.
        For each configuration, we report the total number of rejected solutions (out of 108), along with how many are attributable to PyTest failures versus snapshot test failures.
    }
    \label{tab:correctness-violations}
    \adjustbox{max width=0.45\textwidth}{%
    \begin{tabular}{l l r r r}
        \toprule
        Agent & Model &
        Total $\downarrow$ &
        PyTest $\downarrow$ &
        Snapshot $\downarrow$ \\
        \midrule
Terminus 2 & GPT-5 & 54 & 51 & 32 \\
 & Claude 4.0 Sonnet & 55 & 52 & 36 \\
 & Gemini 2.5 Pro & 55 & 53 & 30 \\
 & Qwen 3 Coder & 56 & 54 & 29 \\
OpenHands & GPT-5 & 47 & 42 & 30 \\
 & Claude 4.0 Sonnet & 50 & 43 & 34 \\
 & Qwen 3 Coder & 50 & 44 & 32 \\
 \bottomrule
    \end{tabular}
    }
\end{table}

\textit{Observation: Correctness violations are common and represent a major source of rollbacks.} We find that models spend most of their budget exploring patches that ultimately fail correctness checks. On average, $\sim$48.5\% of trajectories are rejected due to correctness violations, with the majority of these failures stemming from PyTest suite violations rather than snapshot test failures. We believe this to be a consequence of the multi-objective nature of the optimization problem. A single-objective setting allows verifying new functionalities with a single tool call. However, in a multi-objective setting, the agent--model configuration must strategically allocate interactions towards running either the benchmarking tool, the snapshot verification tool, or the pytest suite depending on the new functionality it introduces. The tool call distribution in \autoref{tab:tool-usage} supports this hypothesis, as most agents demonstrate an inclination towards running performance validation commands rather than correctness validation commands.

\paragraph{Trajectory Length and Performance.} Discovering effective performance optimizations requires a deep understanding of the codebase. Agents must interact with the codebase through a terminal interface to obtain such an understanding. In this experiment, we study the relation between the number of interactions and the global performance achieved by the cumulative trajectory of interactions. 

For each task, we record the number of complete command-line agent interactions (interactions where the agent runs a command and receives a response from the environment) and calculate the mean and median trajectory lengths averaged over all tasks. We then calculate the length-weighted advantage as $ \texttt{len}(\adv_{\agent}) = \frac{\adv_{\agent}}{\texttt{len}_{\agent}}$.  \autoref{tab:trajectory-length} showcases these results.

\textit{Observation: Trajectory lengths can be highly skewed.} Some configurations demonstrate highly skewed trajectories. Specifically, Terminus 2 + GPT-5 and Terminus 2 + Gemini 2.5 Pro have mean lengths substantially larger than the median length, suggesting that these configurations occasionally require very long interactive runs. By contrast, OpenHands + Claude 4.0 Sonnet has more stable trajectory lengths across tasks as the deviation between the mean and median is much smaller.

\textit{Observation: Agent choice has a substantial effect on overall behavior.} The same model behaves very differently depending on the chosen agent. For example, GPT-5 produces much longer trajectories in Terminus 2 than in OpenHands, while Claude 4.0 Sonnet and Qwen 3 Coder show the opposite pattern. This suggests that surrounding agent design heavily shapes search behavior.

\begin{table*}[t]
    \centering
    \caption{
        Trajectory length and length-weighted advantage.
        For each agent--model configuration, we report the mean and median trajectory length (in interaction steps), as well as a length-weighted advantage ($\texttt{len}(\adv_{\agent})$).
    }
    \label{tab:trajectory-length}
    \adjustbox{max width=\textwidth}{%
    \begin{tabular}{l l
                    r r
                    r}
        \toprule
        Agent & Model &
        Mean Length $\downarrow$ & Median Length $\downarrow$ &
        $\texttt{len}(\adv_{\agent})$ $\uparrow$ \\
        \midrule
    Terminus 2 & GPT-5 & 295.53 & 198.50 & -0.000226 \\
     & Claude 4.0 Sonnet & 73.13 & 63.50 & -0.000349 \\
     & Gemini 2.5 Pro & 106.99 & 63.50 & -0.000755 \\
     & Qwen 3 Coder & 99.91 & 90.50 & -0.000557 \\
    OpenHands & GPT-5 & 68.60 & 61.00 & -0.000299 \\
     & Claude 4.0 Sonnet & 222.80 & 219.50 & -0.000106 \\
     & Qwen 3 Coder & 633.10 & 595.00 & -0.000044 \\        \bottomrule
    \end{tabular}
    }
\end{table*}

\paragraph{Tool-Usage Patterns.} In all \fc tasks, agents are given unrestricted access to the \texttt{bash} command line with additional performance profiling and correctness testing tools. In this experiment, we analyze how different configurations employ tools during optimization. For each task, we store the command-line interactions of the agent--model configurations and use an LLM to categorize the input commands based on the primary purpose of the command. The implementation is identical to the performance categorization classifier \S\ref{sec:a.dataset.composition}. We then aggregate the tool type classifications by total tool uses and tool uses per category. \tabTools\ summarizes these statistics across all configurations.

\begin{table}[t]
    \centering
    \small
    \caption{
        Tool categories used in trajectory classification. The classifier's implementation mirrors that of the optimization category classifier (\S\ref{sec:a.expt.optimizationtypes}). 
    }
    \label{tab:tool-categories}
    \adjustbox{max width=\linewidth}{%
    \begin{tabular}{l  l}
        \toprule
        Category & Description \\
        \midrule
        \texttt{editing} & Text editing or transformation commands (e.g., \texttt{sed}, \texttt{awk}, \texttt{ed}). \\
        \texttt{search} & Search/discovery commands for finding files or text (e.g., \texttt{grep}, \texttt{rg}, \texttt{find}, \texttt{fd}). \\
        \texttt{view} & Read-only inspection commands for showing file/output snippets (e.g., \texttt{cat}, \texttt{less}, \texttt{head}, \texttt{tail}). \\
        \texttt{fs\_ops} & Filesystem mutation/metadata operations (e.g., \texttt{cp}, \texttt{mv}, \texttt{rm}, \texttt{mkdir}, \texttt{chmod}). \\
        \texttt{shell\_session} & Shell navigation/session management commands (e.g., \texttt{cd}, \texttt{ls}, \texttt{pwd}, \texttt{clear}, \texttt{exit}). \\
        \texttt{git} & Version-control commands and git-derived shell variable setup (e.g., \texttt{git}, \texttt{diff}, \texttt{reset}). \\
        \texttt{python\_exec} & Python execution plus Python environment/package commands (e.g., \texttt{python}, \texttt{pip}, \texttt{micromamba}). \\
        \texttt{test} & Test-running commands, including snapshot checks (e.g., \texttt{pytest}, \texttt{snapshot-tool}). \\
        \texttt{bench} & Benchmark/profiling commands, primarily ASV workflows (e.g., \texttt{asv run}, \texttt{asv profile}). \\
        \texttt{patching} & Patch/diff application commands or diff-marker lines (e.g., \texttt{patch}, \texttt{applypatch}, \texttt{---}/\texttt{+++}). \\
        \texttt{other} & Commands/fragments that do not match the above classes, including control-flow snippets or terminal noise. \\
        \bottomrule
    \end{tabular}
    }
\end{table}

\begin{table*}
    \centering
    \caption{
        Tool-usage statistics by agent--model configuration.
        Columns report the total number of tool calls and the percentage distribution of calls across tool categories (Judged by \texttt{openai/gpt-oss-120b} using categories in\autoref{tab:tool-categories}). The most effective configurations spend the majority of their tool calls on file-operations (\texttt{editing}, \texttt{search}, and \texttt{view}) and running performance benchmarks (\texttt{bench}), with the remaining calls distributed across a variety of tool categories.
    }
    \label{tab:tool-usage}
    \adjustbox{max width=\textwidth}{%
    \begin{tabular}{l l
                    r r r r r r r r r r r r}
        \toprule
        Agent & Model &
        Total &
        \texttt{editing} & \texttt{search} & \texttt{view} & \texttt{fs\_ops} & \texttt{shell} & \texttt{git} & \texttt{python} & \texttt{test} & \texttt{bench} & \texttt{patching} & \texttt{other} \\
        \midrule
Terminus 2 & GPT-5 & 13370 & 19.40 & 15.70 & 10.29 & 2.76 & 5.00 & 12.14 & 11.20 & 2.89 & 17.51 & 2.07 & 1.02 \\
 & Claude 4.0 Sonnet & 4214 & 6.12 & 8.00 & 11.25 & 7.78 & 8.19 & 0.00 & 16.61 & 2.18 & 6.36 & 0.00 & 33.51 \\
 & Gemini 2.5 Pro & 5641 & 11.35 & 6.29 & 5.96 & 7.43 & 11.45 & 2.80 & 5.48 & 0.62 & 16.04 & 0.27 & 32.32 \\
 & Qwen 3 Coder & 3565 & 17.59 & 16.89 & 8.61 & 12.17 & 12.45 & 0.45 & 8.72 & 1.49 & 9.99 & 0.36 & 11.28 \\
OpenHands & GPT-5 & 4683 & 14.35 & 19.65 & 26.44 & 0.62 & 1.62 & 4.36 & 7.94 & 3.93 & 18.26 & 0.04 & 2.80 \\
 & Claude 4.0 Sonnet & 6323 & 12.92 & 20.94 & 26.51 & 0.76 & 1.57 & 3.56 & 9.90 & 3.67 & 16.86 & 0.03 & 3.29 \\
 & Qwen 3 Coder & 8638 & 10.66 & 20.39 & 25.24 & 0.79 & 1.62 & 4.33 & 9.92 & 3.96 & 19.23 & 0.02 & 3.84 \\
        \bottomrule
    \end{tabular}
    }
\end{table*}

\textit{Observation: Agents invoke benchmarking tools more than testing tools.} All agent--model configurations show a strong preference for running benchmarking and profiling commands over correctness validation commands, with an average of $14.90$\% of tool calls dedicated to benchmarking/profiling and only $2.68$\% of calls dedicated to testing. This proclivity towards performance validation over correctness validation might have a substantial impact on our previous observation that correctness violations are prevalent for all agent--model configurations.

\textit{Observation: Reading dominant tool category.} The most frequently used tool category across all configurations is file-system operations (editing, searching, and viewing files), which accounts for an average of $31.74$\% of all tool calls. This is consistent with the intuition that developing a holistic understanding of the codebase is a prerequisite for synthesizing effective optimizations.

\paragraph{Patch-Level Memorization Analysis.}\label{sec:a.expt.memorization}

A natural concern when evaluating LLM agents on tasks derived from GitHub pull requests is whether agents achieve their performance gains by memorizing expert-authored patches encountered during pre-training.
To assess this, we compute the normalized edit distance between each agent-generated patch and the corresponding expert patch using Python's \texttt{difflib} library.
A score of 0 indicates identical patches, while 1 indicates no character-level similarity.
We stratify results by whether the agent's advantage is close to the expert's ($|\adv| \leq 0.05$) or far ($|\adv| > 0.05$), reasoning that near-expert advantage might correlate with higher patch similarity if memorization were occurring.

\begin{table}[t]
\centering
\small
\caption{Mean normalized edit distance between agent and expert patches on \fcv, stratified by advantage proximity. Values close to 1 indicate low patch similarity. Across all configurations, agent and expert patches share only $\sim$4\% of their content.}
\label{tab:memorization-edit-distance}
\begin{tabular}{l c c}
\toprule
\textbf{Model} & $|\adv| \leq 0.05$ & $|\adv| > 0.05$ \\
\midrule
Claude 4.0 Sonnet & 0.94 & 0.97 \\
GPT-5 & 0.96 & 0.96 \\
Gemini 2.5 Pro & 0.98 & 0.95 \\
Qwen 3 Coder & 0.96 & 0.97 \\
\bottomrule
\end{tabular}
\end{table}

\textit{Observation: Verbatim patch memorization is rare.} Across all configurations, the mean normalized edit distance exceeds 0.94, indicating that agent-generated patches share at most $\sim$6\% of their character-level content with expert patches.
Notably, there is no meaningful difference in edit distance between tasks where agents achieve near-expert advantage and those where they do not, suggesting that similar performance outcomes arise from genuinely different code changes rather than memorization.

This analysis provides a useful signal for ruling out exact or near-exact patch-level memorization.
However, it does not rule out all forms of contamination: an agent may recover the same high-level optimization strategy (e.g., ``switch from a list to a set'') while expressing it with substantially different code.
Detecting such semantic-level overlap requires execution-based comparison, which the \adv~metric already provides by measuring behavioral similarity between agent and expert solutions.

\subsection{Implications of Results}\label{sec:a.expt.implications}

\paragraph{Agent optimization strategies are qualitatively diverse.}
Our stratified advantage analysis (\S\ref{sec:results-scale}) reveals that agents do not simply differ in overall capability---they exhibit structurally distinct optimization strategies.
OpenHands + Claude 4.0 Sonnet achieves its strongest advantage at the module level ($\ell=1$), suggesting a preference for broad architectural refactors, while OpenHands + GPT-5 excels at function-level optimizations ($\ell=3$) but loses effectiveness at coarser granularities.
These characteristic profiles are invisible to scalar leaderboard metrics and suggest that ensemble or routing strategies---where different models are deployed for different optimization scopes---could outperform any single configuration.

\paragraph{Long-tail repositories expose a distribution shift bottleneck.}
Agent performance trails experts on repositories in the lowest popularity quintile (Q1; 133--202 GitHub stars), even though expert patches in this regime yield the second-largest speedups (\S\ref{sec:results-popularity}).
This asymmetry suggests that the primary bottleneck is not task difficulty but rather distribution shift: agents are less familiar with the conventions, APIs, and idioms of niche repositories.
This finding has direct implications for practical deployment---agents are less helpful on precisely the repositories where human developers may need them most, in under-maintained and under-documented codebases.

\paragraph{Multi-workload optimization reveals real-world tradeoffs.}
Expert solutions achieve the best global speedup despite causing an average regression on some workloads (\S\ref{sec:results-tradeoff}).
This tradeoff pattern---improving many execution paths at the cost of modest regressions on others---mirrors real-world optimization practice, where universal improvement across all inputs is rarely achievable~\citep{balsamo2004model, jin2012understanding}.
Single-workload benchmarks cannot detect such tradeoffs.
\fc's multi-workload design (averaging $\sim$\avgWorkloadsPerProblem~workloads per task) provides the resolution needed to study how agents negotiate competing objectives, an aspect critical to understanding their real-world utility.

\paragraph{Tool usage patterns correlate with optimization success.}
Our tool-usage analysis (\S\ref{sec:a.expt.additional-analysis}) shows that all agent--model configurations preferentially invoke benchmarking and profiling tools over correctness testing tools. Combined with our finding that $\sim$48.5\% of candidate solutions are rejected due to correctness violations, this suggests a systematic allocation failure: agents invest disproportionately in performance validation at the expense of verifying functional correctness.
Future agent designs that better balance performance profiling with correctness checking may substantially reduce wasted computation.

\definecolor{fcHumanBg}{RGB}{230,245,255}
\definecolor{fcAgentBg}{RGB}{255,240,230}
\definecolor{fcHumanFg}{RGB}{70,130,180}
\definecolor{fcAgentFg}{RGB}{205,133,63}
\definecolor{fcCodeBg}{RGB}{248,248,248}
\definecolor{fcDiffAdd}{RGB}{0,120,0}
\definecolor{fcDiffDel}{RGB}{180,30,30}
\definecolor{fcDiffComment}{RGB}{120,120,120}

\lstdefinelanguage{fcDiff}{
    morecomment=[f][\color{fcDiffAdd}]{+},
    morecomment=[f][\color{fcDiffDel}]{-},
    morecomment=[f][\color{fcDiffComment}]{\#},
    morecomment=[f][\color{blue}]{diff\ \-\-git},
}

\lstdefinestyle{fcDiffStyle}{
    language=fcDiff,
    basicstyle=\ttfamily\fontsize{6pt}{7.2pt}\selectfont,
    breaklines=true,
    breakatwhitespace=false,
    columns=flexible,
    keepspaces=true,
    showstringspaces=false,
    tabsize=4,
    frame=none,
    backgroundcolor=\color{fcCodeBg},
    aboveskip=0pt,
    belowskip=0pt,
}

\newtcblisting{fcHumanPatch}[1]{%
  colback=fcCodeBg,colframe=fcHumanFg,
  title={\small\bfseries Human Expert Patch},
  fonttitle=\sffamily,boxrule=0.5pt,
  left=1pt,right=1pt,top=1pt,bottom=1pt,
  equal height group=code#1,
  listing only,
  listing options={style=fcDiffStyle}%
}
\newtcblisting{fcAgentPatch}[1]{%
  colback=fcCodeBg,colframe=fcAgentFg,
  title={\small\bfseries AI Agent Patch},
  fonttitle=\sffamily,boxrule=0.5pt,
  left=1pt,right=1pt,top=1pt,bottom=1pt,
  equal height group=code#1,
  listing only,
  listing options={style=fcDiffStyle}%
}
\newcommand{\fcSummaryRow}[3]{%
\noindent
\begin{minipage}[t]{0.485\linewidth}%
\begin{tcolorbox}[colback=fcHumanBg,colframe=fcHumanFg,title={\small\bfseries Summary},fonttitle=\sffamily,boxrule=0.5pt,left=3pt,right=3pt,top=2pt,bottom=2pt,equal height group=summ#1]%
{\small #2}%
\end{tcolorbox}%
\end{minipage}%
\hfill
\begin{minipage}[t]{0.485\linewidth}%
\begin{tcolorbox}[colback=fcAgentBg,colframe=fcAgentFg,title={\small\bfseries Summary},fonttitle=\sffamily,boxrule=0.5pt,left=3pt,right=3pt,top=2pt,bottom=2pt,equal height group=summ#1]%
{\small #3}%
\end{tcolorbox}%
\end{minipage}%
}
\newcommand{\fcTaskCaption}[2]{%
\captionof{figure}{#1}%
\label{#2}%
\clearpage}

\subsection{Qualitative Examples: Human Expert vs.\ AI Agent Patches}
\label{sec:qualitative-examples}

This section presents side-by-side comparisons of human expert and AI agent patches for \fc tasks. Specifically, the following examples are showcased:

\begin{itemize}
    \item Figure~\ref{fig:modin2} (\texttt{modin\_modin\_2}). \textbf{Failure mode: Incorrect triage; expert gained edge by identifying performance hotpath.} Modin has an expensive auto-switch backend logic that was being called even when all inputs shared the same backend. The agent was not able to identify the core issue, instead focusing on a caching issue that was not on the performance critical path. The human correctly identifies the issue and implemented a fix to the caching logic.
    \item Figure~\ref{fig:optuna6} (\texttt{optuna\_optuna\_4020}). \textbf{Failure mode: Correct triage; expert gained edge by \texttt{numpy} vectorization delegation.} Optuna's hypervolume computation used a naive recursive algorithm, when a faster $O(N^2)$ approach was possible. Both the human and the agent were able to identify and implement the algorithm. However, the human's solution used fully vectorized numpy operations, while the agent's solution used a Python-level sweep-line approach with \texttt{bisect}. This resulted in the human outperforming the agent despite both having the same asymptotic complexity.
    \item Figure~\ref{fig:optuna1} (\texttt{optuna\_optuna\_3647}). \textbf{Failure mode: Correct triage; expert implemented holistic full-module optimization.} Optuna's implementation for sorting non-dominated Pareto fronts used a naive algorithm that didn't scale well as number of trials increased. Both the human and the agent identified this issue; the agent's implementation utilized a Fenwick tree based algorithm which fixed a single hotpath (when inputs are 2D). However, the expert implementation implemented a holistic rewrite: it optimized the entire call chain to use vectorized numpy operations and merged separate pathways for 2D/N-D optimization, resulting in complementary improvements across the entire multi-objective optimization flow.
    \item Figure~\ref{fig:networkx4} (\texttt{networkx\_networkx\_7424}). The core issue was that NetworkX's BFS-based component discovery algorithm did not implement an early-termination optimization. Both the human and the agent fix this by implementing an early termination optimization. However, the agent outperforms the human by further optimizing the BFS implementation, achieving an additional $+0.0132$ advantage on top of the human's improvement.
    \item Figure~\ref{fig:pybamm1} (\texttt{pybamm\_team\_pybamm\_1}). A sensitivity computation in PyBaMM created a quadratic memory allocation bottleneck due to incremental concatenation without realizing that the full size was known in advance. Both the human and the agent identify the issue and collect all blocks first and concatenate once. The agent further optimizes the concatenation logic by consolidating multiple function calls into one and adding guards for empty inputs, resulting in a $+0.0167$ advantage.
    \item Figure~\ref{fig:shapely1} (\texttt{shapely\_shapely\_1982}). The \texttt{deprecate\_positional} decorator in Shapely called \texttt{inspect.signature} on every invocation, causing 300--1000\% slowdowns. The human and agent solution both converged on the same strategy: implementing a caching layer on the decorator. However, the agent implemented additional optimizations to skip the hot-path when no deprecated parameters existed, resulting in a $+0.0132$ advantage.
\end{itemize}

\clearpage

\noindent
\begin{minipage}[t]{0.485\linewidth}
\begin{fcHumanPatch}{16}
diff --git a/modin/core/storage_formats/pandas/query_compiler_caster.py
--- a/modin/core/storage_formats/pandas/query_compiler_caster.py
+++ b/modin/core/storage_formats/pandas/query_compiler_caster.py
@@ -1006,6 +1006,8 @@
         pin_target_backend = None
+        input_backends: set[str] = set()

         def register_query_compilers(arg):
+                input_backends.add(arg_backend)

@@ -1047,10 +1050,12 @@
-        inputs_pinned = (
-            len(input_query_compilers) < 2 and pin_target_backend is not None
-        )
-        if not AutoSwitchBackend.get() or inputs_pinned:
         # Skip casting when all inputs share a single backend and either
         # auto-switching is disabled or inputs are pinned
+        if len(input_backends) == 1 and (
+            not AutoSwitchBackend.get() or pin_target_backend is not None
+        ):
             f_to_apply = _get_extension_for_method(...)
             result = f_to_apply(*args, **kwargs)
-            if isinstance(result, QueryCompilerCaster) and inputs_pinned:
+            if (isinstance(result, QueryCompilerCaster)
+                    and pin_target_backend is not None):
                 result._set_backend_pinned(True, inplace=True)
             return result
\end{fcHumanPatch}
\end{minipage}
\hfill
\begin{minipage}[t]{0.485\linewidth}
\begin{fcAgentPatch}{16}
diff --git a/modin/core/storage_formats/pandas/query_compiler_caster.py
--- a/modin/core/storage_formats/pandas/query_compiler_caster.py
+++ b/modin/core/storage_formats/pandas/query_compiler_caster.py
@@ -137,23 +137,52 @@
-def _get_empty_qc_for_default_backend() -> BaseQueryCompiler:
-    """Get an empty query compiler for the default backend."""
+def _get_empty_qc_for_backend(backend: str) -> BaseQueryCompiler:
+    """Get an empty query compiler for the specified backend."""
     from modin.core.execution.dispatching.factories.dispatcher import FactoryDispatcher
-    return FactoryDispatcher.get_factory().io_cls.from_pandas(pandas.DataFrame())
+    original_backend = Backend.get()
+    try:
+        Backend.put(backend)
+        return FactoryDispatcher.get_factory().io_cls.from_pandas(pandas.DataFrame())
+    finally:
+        Backend.put(original_backend)

-_BACKEND_TO_EMPTY_QC: defaultdict[str, BaseQueryCompiler] = defaultdict(
-    _get_empty_qc_for_default_backend
-)
+_BACKEND_TO_EMPTY_QC: dict[str, BaseQueryCompiler] = {}

+def _get_cached_empty_qc(backend: str) -> BaseQueryCompiler:
+    if backend not in _BACKEND_TO_EMPTY_QC:
+        _BACKEND_TO_EMPTY_QC[backend] = _get_empty_qc_for_backend(backend)
+    return _BACKEND_TO_EMPTY_QC[backend]

@@ -1042,7 +1071,7 @@
-            input_qc_for_pre_op_switch = _BACKEND_TO_EMPTY_QC[input_backend]
+            input_qc_for_pre_op_switch = _get_cached_empty_qc(input_backend)
\end{fcAgentPatch}
\end{minipage}
\vspace{4pt}

\fcSummaryRow{16}
  {Disables \texttt{AutoSwitchBackend} by default and rewrites the casting-skip logic in \texttt{query\_compiler\_caster.py} to track the set of distinct input backends, skipping expensive query-compiler conversions when all inputs share a single backend. Updates four test files and adjusts metric assertions (not shown).}
  {Fixes a bug where the \texttt{defaultdict} factory ignores the requested backend when creating empty query compilers, replacing it with an explicit \texttt{\_get\_cached\_empty\_qc} function that temporarily switches \texttt{Backend.put()} to the correct backend. A correctness fix, but not on the performance-critical path.}

\fcTaskCaption{\texttt{modin\_project-modin\_7637}: Modin's \texttt{AutoSwitchBackend} feature, enabled by default, triggered an expensive type conversion even when all inputs shared the same backend. The agent solution (\texttt{openhands:claude-sonnet-4}) identified and fixed a real bug in the caching logic, but this was not on the performance-critical path, resulting in a $-0.1265$ advantage compared to the human expert's systemic fix that disabled \texttt{AutoSwitchBackend} by default and optimized the casting logic to track input backend diversity, skipping conversions when unnecessary.}{fig:modin2}

\noindent
\begin{minipage}[t]{0.485\linewidth}
\begin{fcHumanPatch}{13}
diff --git a/optuna/_hypervolume/wfg.py b/optuna/_hypervolume/wfg.py
--- a/optuna/_hypervolume/wfg.py
+++ b/optuna/_hypervolume/wfg.py
# New O(N^2) vectorized 3D hypervolume via coordinate compression
+def _compress_coordinate(coords: np.ndarray) -> tuple[np.ndarray, np.ndarray]:
+    sorted_indices = np.argsort(coords)
+    values = coords[sorted_indices]
+    r = np.zeros_like(sorted_indices)
+    r[sorted_indices] = np.arange(coords.shape[0], dtype=r.dtype)
+    return r, values

+def _compute_3d(sorted_pareto_sols: np.ndarray, reference_point: np.ndarray) -> float:
+    """O(N^2) 3D hypervolume via cumulative minimum on compressed coordinates."""
+    n = sorted_pareto_sols.shape[0]
+    x_vals = np.concatenate([sorted_pareto_sols[:, 0], reference_point[:1]])
+    y_ind, y_vals = _compress_coordinate(sorted_pareto_sols[:, 1])
+    y_vals = np.concatenate([y_vals, reference_point[1:2]])
+    z_max = reference_point[2]
+    zs = np.full((n, n), z_max)
+    for i in range(n):
+        zs[i, y_ind[i]] = sorted_pareto_sols[i, 2]
+    zs = np.minimum.accumulate(zs, axis=0)
+    zs = np.minimum.accumulate(zs, axis=1)
+    x_delta = x_vals[1:] - x_vals[:-1]
+    y_delta = y_vals[1:] - y_vals[:-1]
+    return np.sum((z_max - zs) * x_delta[:, np.newaxis] * y_delta[np.newaxis, :])

@@ -125,6 +178,13 @@ def compute_hypervolume(
     if reference_point.shape[0] == 2:
         hv = _compute_2d(sorted_pareto_sols, reference_point)
+    elif reference_point.shape[0] == 3:
+        hv = _compute_3d(sorted_pareto_sols, reference_point)
     else:
         hv = _compute_hv(sorted_pareto_sols, reference_point)
\end{fcHumanPatch}
\end{minipage}
\hfill
\begin{minipage}[t]{0.485\linewidth}
\begin{fcAgentPatch}{13}
diff --git a/optuna/_hypervolume/wfg.py b/optuna/_hypervolume/wfg.py
--- a/optuna/_hypervolume/wfg.py
+++ b/optuna/_hypervolume/wfg.py
# x-sweep with incremental 2D Pareto frontier via bisect
+def _compute_3d(sorted_pareto_sols, reference_point) -> float:
+    """Exact 3D hypervolume using x-sweep with incremental 2D frontier."""
+    import bisect
+    xs = pts[:, 0]
+    dx = np.maximum(xs_ext[1:] - xs_ext[:-1], 0.0)
+    y_list: list[float] = []
+    z_list: list[float] = []
+
+    def insert_frontier(y: float, z: float) -> None:
+        i = bisect.bisect_left(y_list, float(y))
+        if i > 0 and z >= z_list[i - 1]:
+            return  # dominated by left neighbor
         # ... (dominance-aware insertion: handle equal y,
         #      remove dominated points to the right)
+        y_list.insert(i, float(y))
+        z_list.insert(i, float(z))
+
+    for i in range(n):
+        insert_frontier(float(pts[i, 1]), float(pts[i, 2]))
+        if y_list:
+            yz = np.column_stack((np.asarray(y_list), np.asarray(z_list)))
+            areas[i] = _compute_2d(yz, ref_yz)
+    return float(np.dot(dx, areas))

@@ -126,7 +190,7 @@ def compute_hypervolume(
-        hv = _compute_hv(sorted_pareto_sols, reference_point)
+        hv = _compute_3d(...) if sorted_pareto_sols.shape[1] == 3 else _compute_hv(...)

\end{fcAgentPatch}
\end{minipage}
\vspace{4pt}

\fcSummaryRow{13}
  {Adds a specialized $O(N^2)$ \texttt{\_compute\_3d} function using a \texttt{\_compress\_coordinate} helper that maps $y$-coordinates to integer ranks via \texttt{np.argsort}, builds an $N \times N$ grid, and applies \texttt{np.minimum.accumulate} along both axes to compute dominated volume in fully vectorized numpy. Also adds a dedicated \texttt{elif} branch in \texttt{compute\_hypervolume} and parameterized tests (not shown).}
  {Adds a \texttt{\_compute\_3d} function using an $x$-sweep with incremental 2D Pareto frontier maintenance via \texttt{bisect} and Python lists. At each $x$-slice, the frontier is updated with dominance-aware insertion, then the 2D area is computed by delegating to \texttt{\_compute\_2d}. The dispatch in \texttt{compute\_hypervolume} is modified with an inline ternary for 3D inputs.}

\fcTaskCaption{\texttt{optuna\_optuna\_4020}: Optuna's \texttt{\_hypervolume.WFG} class used a naive recursive algorithm for hypervolume computation that had a $O(N^3)$ runtime for the common 3D case, when a $O(N^2)$ approach was possible. Both the human and the agent identified and implemented the faster algorithm. However, the human's solution used fully vectorized numpy operations, while the best agent (\texttt{terminus-2:gpt-5}) used a Python-level sweep-line approach with \texttt{bisect}. This resulted in the human outperforming the agent with a $-0.03964$ agent advantage despite both having the same asymptotic complexity.}{fig:optuna6}

\noindent
\begin{minipage}[t]{0.485\linewidth}
\begin{fcHumanPatch}{15}
diff --git a/optuna/study/_multi_objective.py b/optuna/study/_multi_objective.py
--- a/optuna/study/_multi_objective.py
+++ b/optuna/study/_multi_objective.py
@@ (selected excerpts)
-def _get_pareto_front_trials_2d(...):
-    ...  # Separate 2D implementation
-def _get_pareto_front_trials_nd(...):
-    ...  # Separate N-D implementation
-def _get_pareto_front_trials_by_trials(...):
-    if len(directions) == 2:
-        return _get_pareto_front_trials_2d(...)
-    return _get_pareto_front_trials_nd(...)
+def _get_pareto_front_trials_by_trials(...):
+    loss_values = np.asarray(...)
+    on_front = _is_pareto_front(loss_values,
+        assume_unique_lexsorted=False)
+    return [t for t, p in zip(trials, on_front) if p]

-def _fast_non_dominated_sort(
-    objective_values, *, penalty=None, n_below=None
+def _fast_non_domination_rank(
+    loss_values, *, penalty=None, n_below=None
 ) -> np.ndarray:
-    ...  # O(n^2) broadcast + defaultdict
+    ...  # Vectorized _calculate_nondomination_rank
+    ...  # + _is_pareto_front with lexsort
\end{fcHumanPatch}
\end{minipage}
\hfill
\begin{minipage}[t]{0.485\linewidth}
\begin{fcAgentPatch}{15}
diff --git a/optuna/study/_multi_objective.py b/optuna/study/_multi_objective.py
--- a/optuna/study/_multi_objective.py
+++ b/optuna/study/_multi_objective.py
@@ -189,42 +189,106 @@
 def _calculate_nondomination_rank(...):
     ...
     # Fast path for 2D objectives.
+    if objective_values.shape[1] == 2:
+        x = objective_values[:, 0]
+        y = objective_values[:, 1]
+        order = np.lexsort((y, x))
+        ys_unique = np.unique(y)
+        y_idx_all = np.searchsorted(ys_unique, y,
+            side='right')
+        m = len(ys_unique)
+        bit = np.zeros(m + 1, dtype=int)
+        def bit_query(i):  # Fenwick tree prefix max
+            ...
+        def bit_update(i, v):
+            ...
         # Process equal-x groups, BIT for rank
+        ...
+        return ranks, last_rank
+
     # Fallback: original O(n^2) broadcast for >=3D.
     domination_mat = np.all(...) & np.any(...)
\end{fcAgentPatch}
\end{minipage}
\vspace{4pt}

\fcSummaryRow{15}
  {Complete rewrite of \texttt{\_\allowbreak multi\_\allowbreak objective.py}. Renames \texttt{\_\allowbreak fast\_\allowbreak non\_\allowbreak dominated\_\allowbreak sort} to \texttt{\_\allowbreak fast\_\allowbreak non\_\allowbreak domination\_\allowbreak rank}, replaces the $O(n^2)$ broadcast-based algorithm with a vectorized \texttt{\_\allowbreak is\_\allowbreak pareto\_\allowbreak front} and \texttt{\_\allowbreak calculate\_\allowbreak nondomination\_\allowbreak rank} implementation, merges the separate 2D/N-D Pareto front functions, and updates all callers across the TPE sampler and NSGA-II selection strategy.}
  {Adds a specialized $O(n \log n)$ BIT (Fenwick tree) algorithm for 2D objectives in \texttt{\_\allowbreak calculate\_\allowbreak nondomination\_\allowbreak rank}, falling back to the original $O(n^2)$ broadcast for $\geq$3 objectives. While algorithmically superior for the 2D case, the agent only optimizes the inner ranking function without restructuring callers or the Pareto front computation.}

\fcTaskCaption{\texttt{optuna\_optuna\_3647}: The original implementation of Optuna's non-dominated sorting in multi-objective optimization cases emerged as a performance bottleneck when scaling to large number of trials ($\sim 10000$ trials). Both the best agent (\texttt{terminus-2:gpt-5}) and the human expert correctly identified the issue. The agent's solution focused on optimizing the inner ranking function with a specialized $O(n \log n)$ Fenwick-tree algorithm for 2D objectives, while retaining the original $O(n^2)$ broadcast-based approach for higher dimensions. In contrast, the human expert implemented a holistic rewrite of the entire call chain to use vectorized numpy operations and merged separate pathways for 2D/N-D optimization. This broader vectorized rewrite captured multi-objective speedups (5--6$\times$) that the agent's localized change missed, resulting in a $-0.03629$ agent advantage.}{fig:optuna1}

\noindent
\begin{minipage}[t]{0.485\linewidth}
\begin{fcHumanPatch}{18}
diff --git a/networkx/algorithms/components/connected.py
@@ -65,7 +65,7 @@ def connected_components(G):
     for v in G:
         if v not in seen:
-            c = _plain_bfs(G, n, v)
+            c = _plain_bfs(G, n - len(seen), v)
             seen.update(c)

diff --git a/networkx/algorithms/components/weakly_connected.py
@@ -62,7 +62,7 @@ def weakly_connected_components(G):
     for v in G:
         if v not in seen:
-            c = set(_plain_bfs(G, n, v))
+            c = set(_plain_bfs(G, n - len(seen), v))
             seen.update(c)
\end{fcHumanPatch}
\end{minipage}
\hfill
\begin{minipage}[t]{0.485\linewidth}
\begin{fcAgentPatch}{18}
diff --git a/networkx/algorithms/components/connected.py b/networkx/algorithms/components/connected.py
index ebe0d8c1..922d4904 100644
--- a/networkx/algorithms/components/connected.py
+++ b/networkx/algorithms/components/connected.py
@@ -62,12 +62,15 @@ def connected_components(G):
-    [-n-]{+remaining+} = len(G)
         ...
-            c = _plain_bfs(G, [-n-]{+remaining+}, v)
             seen.update(c)
+            remaining -= len(c)
             yield c
+            if remaining == 0:
+                break
@@ -203,14 +206,16 @@ def _plain_bfs(G, n, source):
+    seen_add = seen.add
     while nextlevel:
         thislevel = nextlevel
         nextlevel = []
+        append = nextlevel.append
             ...
                 if w not in seen:
-                    [-seen.add(w)-]{+seen_add(w)+}
-                    [-nextlevel.append(w)-]{+append(w)+}
diff --git a/networkx/algorithms/components/weakly_connected.py b/networkx/algorithms/components/weakly_connected.py
index ecfac50a..a89b7af8 100644
--- a/networkx/algorithms/components/weakly_connected.py
+++ b/networkx/algorithms/components/weakly_connected.py
@@ -59,12 +59,15 @@ def weakly_connected_components(G):
     # (same early-exit optimization as connected_components above)
@@ -166,32 +169,30 @@ def _plain_bfs(G, n, source):
     # (same local-variable caching as connected._plain_bfs above)
     # additionally, converted from generator (yield) to returning seen set:
-    yield source
+    ...
             if len(seen) == n:
-                return
+                return seen
+    return seen
\end{fcAgentPatch}
\end{minipage}
\vspace{4pt}

\fcSummaryRow{18}
  {Minimal single-line fix in both \texttt{connected\_components} and \texttt{weakly\_connected\_components}: passes \texttt{n\,-\,len(seen)} instead of \texttt{n} to \texttt{\_plain\_bfs}, tightening the BFS early-termination bound so it stops as soon as all remaining unseen nodes are found. No structural changes to the BFS itself.}
  {Multi-pronged optimization: tracks a \texttt{remaining} node count to break out of the component loop early, caches method lookups (\texttt{seen.add}, \texttt{nextlevel.append}) into local variables, and converts the weakly-connected \texttt{\_plain\_bfs} from a generator to a batch set return, eliminating per-node yield overhead.}

\fcTaskCaption{\texttt{networkx\_networkx\_7424}: NetworkX's \texttt{connected\_components} and \texttt{weakly\_connected\_components} passed the total graph node count \texttt{n} to \texttt{\_plain\_bfs} without accounting for already-discovered nodes, missing an early-termination optimization. For disconnected graphs with large components explored last, this caused dramatic slowdowns---up to 367$\times$ for adversarial cases with $n{=}1000$. Both the best agent (\texttt{openhands:gpt-5}) and the expert identified the core issue, and implemented the same early-termination optimization. However, the agent also implemented additional micro-optimizations that further reduced overhead, resulting in a $+0.0132$ advantage over the human's solution.}{fig:networkx4}

\noindent
\begin{minipage}[t]{0.485\linewidth}
\begin{fcHumanPatch}{22}
diff --git a/src/pybamm/solvers/processed_variable.py b/...
--- a/src/pybamm/solvers/processed_variable.py
+++ b/src/pybamm/solvers/processed_variable.py
@@ -443,16 +443,18 @@ class ProcessedVariable:
             dvar_dp_func = casadi.Function(
                 "dvar_dp", [t_casadi, y_casadi, p_casadi_stacked], [dvar_dp]
             )
-            for idx, t in enumerate(ts):
-                u = ys[:, idx]
-                next_dvar_dy_eval = dvar_dy_func(t, u, inputs_stacked)
-                next_dvar_dp_eval = dvar_dp_func(t, u, inputs_stacked)
-                if idx == 0:
-                    dvar_dy_eval = next_dvar_dy_eval
-                    dvar_dp_eval = next_dvar_dp_eval
-                else:
-                    dvar_dy_eval = casadi.diagcat(dvar_dy_eval, next_dvar_dy_eval)
-                    dvar_dp_eval = casadi.vertcat(dvar_dp_eval, next_dvar_dp_eval)
+            dvar_dy_eval = casadi.diagcat(
+                *[
+                    dvar_dy_func(t, ys[:, idx], inputs_stacked)
+                    for idx, t in enumerate(ts)
+                ]
+            )
+            dvar_dp_eval = casadi.vertcat(
+                *[
+                    dvar_dp_func(t, ys[:, idx], inputs_stacked)
+                    for idx, t in enumerate(ts)
+                ]
+            )

             # Compute sensitivity
             S_var = dvar_dy_eval @ dy_dp + dvar_dp_eval
\end{fcHumanPatch}
\end{minipage}
\hfill
\begin{minipage}[t]{0.485\linewidth}
\begin{fcAgentPatch}{22}
diff --git a/src/pybamm/solvers/processed_variable.py b/...
--- a/src/pybamm/solvers/processed_variable.py
+++ b/src/pybamm/solvers/processed_variable.py
@@ -436,29 +439,30 @@ class ProcessedVariable:
             dvar_dy = casadi.jacobian(var_casadi, y_casadi)
             dvar_dp = casadi.jacobian(var_casadi, p_casadi_stacked)

-            dvar_dy_func = casadi.Function(
-                "dvar_dy", [t_casadi, y_casadi, p_casadi_stacked], [dvar_dy]
-            )
-            dvar_dp_func = casadi.Function(
-                "dvar_dp", [t_casadi, y_casadi, p_casadi_stacked], [dvar_dp]
             # Single function returning both jacobians
+            grads_func = casadi.Function(
+                "pv_grads", [t_casadi, y_casadi, p_casadi_stacked],
+                [dvar_dy, dvar_dp]
             )
-            for idx, t in enumerate(ts):
+
+            dvar_dy_blocks = []
+            dvar_dp_blocks = []
+            for idx in range(ts.size):
+                t = ts[idx]
                 u = ys[:, idx]
-                next_dvar_dy_eval = dvar_dy_func(t, u, inputs_stacked)
-                next_dvar_dp_eval = dvar_dp_func(t, u, inputs_stacked)
-                if idx == 0:
-                    dvar_dy_eval = next_dvar_dy_eval
-                    dvar_dp_eval = next_dvar_dp_eval
-                else:
-                    dvar_dy_eval = casadi.diagcat(dvar_dy_eval, next_dvar_dy_eval)
-                    dvar_dp_eval = casadi.vertcat(dvar_dp_eval, next_dvar_dp_eval)
+                g_dy, g_dp = grads_func(t, u, inputs_stacked)
+                dvar_dy_blocks.append(g_dy)
+                dvar_dp_blocks.append(g_dp)
+
             # Concatenation in one shot
+            dvar_dy_eval = casadi.diagcat(*dvar_dy_blocks)
+            dvar_dp_eval = casadi.vertcat(*dvar_dp_blocks)

             # Compute sensitivity
             S_var = dvar_dy_eval @ dy_dp + dvar_dp_eval
\end{fcAgentPatch}
\end{minipage}
\vspace{4pt}

\fcSummaryRow{22}
  {Replaced the incremental per-timestep \texttt{casadi.\allowbreak diagcat}/\allowbreak \texttt{casadi.\allowbreak vertcat} loop with list comprehensions that build all Jacobian blocks first, then concatenate once via unpacking (\texttt{*blocks}). Also added a \texttt{CHANGELOG.md} entry (not shown).}
  {Consolidated the two separate \texttt{casadi.Function} objects (\texttt{dvar\_dy\_func}, \texttt{dvar\_dp\_func}) into a single \texttt{grads\_func} returning both Jacobians, reducing per-timestep function call overhead. Collects results in lists and concatenates once. Also adds guards for empty time series and empty result lists.}

\fcTaskCaption{\texttt{pybamm\_team-pybamm\_4735}: 
PyBaMM's \texttt{ProcessedVariable} sensitivity computation in \texttt{IDAKLUSolver} used an  incremental per-timestep concatenation operation, creating a quadratic memory allocation overhead. Both the best agent (\texttt{openhands:gpt-5}) and the expert identified that, instead of each loop iteration building a progressively larger matrix by concatenating to the existing result, it would be more efficient to first collect all blocks and then concatenate once at the end.  The agent added further micro-optimization: consolidating two accumulation function calls into one and added empty-input guards. This resulted in a $+0.0167$ agent advantage.}{fig:pybamm1}

\noindent
\begin{minipage}[t]{0.485\linewidth}
\begin{fcHumanPatch}{23}
diff --git a/shapely/decorators.py b/shapely/decorators.py
--- a/shapely/decorators.py
+++ b/shapely/decorators.py
-def deprecate_positional(should_be_kwargs, category=DeprecationWarning):
+def deprecate_positional(
+    should_be_kwargs: Iterable[str],
+    category: type[Warning] = DeprecationWarning,
+):
+    def decorator(func: Callable):
+        code = unwrap(func).__code__
+        pos_names = code.co_varnames[: code.co_argcount]
+        name_to_idx = {name: idx for idx, name in enumerate(pos_names)}
+        deprecate_positions = [
+            (name_to_idx[name], name) for name in should_be_kwargs
+            if name in name_to_idx]
+        if not deprecate_positions:
+            return func
+        warn_from = min(deprecate_positions)[0]
+
+        @lru_cache(10)
+        def make_msg(n_args: int):
+            used = [name for idx, name in deprecate_positions if idx < n_args]
             # ... (build warning message string)
+            return f"positional argument{plr} {args_txt} for `{func.__name__}` ..."

-    def decorator(func):
         @wraps(func)
         def wrapper(*args, **kwargs):
-            ret_value = func(*args, **kwargs)
-            sig = inspect.signature(func)
-            args_bind = sig.bind_partial(*args)
-            warn_args = [f"`{arg}`" for arg in args_bind.arguments.keys()
-                         if arg in should_be_kwargs]
-            if warn_args:
                 # ... (build warning message)
-                warnings.warn(msg, category=category, stacklevel=2)
-            return ret_value
+            result = func(*args, **kwargs)
+            n = len(args)
+            if n > warn_from:
+                warnings.warn(make_msg(n), category=category, stacklevel=2)
+            return result
         return wrapper
\end{fcHumanPatch}
\end{minipage}
\hfill
\begin{minipage}[t]{0.485\linewidth}
\begin{fcAgentPatch}{23}
diff --git a/shapely/decorators.py b/shapely/decorators.py
--- a/shapely/decorators.py
+++ b/shapely/decorators.py
     def decorator(func):
         # Cache signature inspection at decoration time
+        sig = inspect.signature(func)
+        param_names = list(sig.parameters.keys())
+        deprecated_positions = set()
+        for param_name in should_be_kwargs:
+            if param_name in param_names:
+                deprecated_positions.add(param_names.index(param_name))
+        if not deprecated_positions:
+            return func
+        max_deprecated_pos = max(deprecated_positions)

         @wraps(func)
         def wrapper(*args, **kwargs):
-            ret_value = func(*args, **kwargs)
-            sig = inspect.signature(func)
-            args_bind = sig.bind_partial(*args)
             # ... (per-call signature inspection)
             # Fast path: skip if not enough args
+            if len(args) <= max_deprecated_pos:
+                return func(*args, **kwargs)
             # Only check deprecated positions
+            warn_positions = [pos for pos in deprecated_positions if pos < len(args)]
+            if warn_positions:
+                args_bind = sig.bind_partial(*args)
                 # ... (build and emit warning)
+            return func(*args, **kwargs)
         return wrapper
\end{fcAgentPatch}
\end{minipage}
\vspace{4pt}

\fcSummaryRow{23}
  {Completely rewrote the \texttt{deprecate\_positional} decorator: replaced \texttt{inspect.signature} with \texttt{inspect.unwrap} and direct \texttt{\_\_code\_\_} introspection at decoration time, added an \texttt{lru\_cache}-backed \texttt{make\_msg} helper to avoid rebuilding warning strings, and included type annotations and a comprehensive 138-line test suite.}
  {Cached \texttt{inspect.signature} at decoration time and pre-computed deprecated parameter positions as a set. Added an early-return fast path when no deprecated parameters exist and a second fast path skipping checking when argument count is below the threshold.}

\fcTaskCaption{\texttt{shapely\_shapely\_1982}: The \texttt{deprecate\_positional} decorator in Shapely called \texttt{inspect.signature} and \texttt{sig.bind\_partial} on every decorated function invocation, causing a 300--1000\% performance regression. Users reported significant Polygon creation slowdowns. The best agent (\texttt{terminus-2:claude-sonnet-4}) and the human expert converged on nearly identical core strategies. Both implemented a caching layer to move signature inspection from call time to decoration time. The agent added additional micro-optimizations to skip checks when no deprecated parameters exist or when the argument count is below the threshold. This resulted in a $+0.0131$ advantage over the human's solution.}{fig:shapely1}

\begin{figure}
    \centering
    \includegraphics[width=\linewidth]{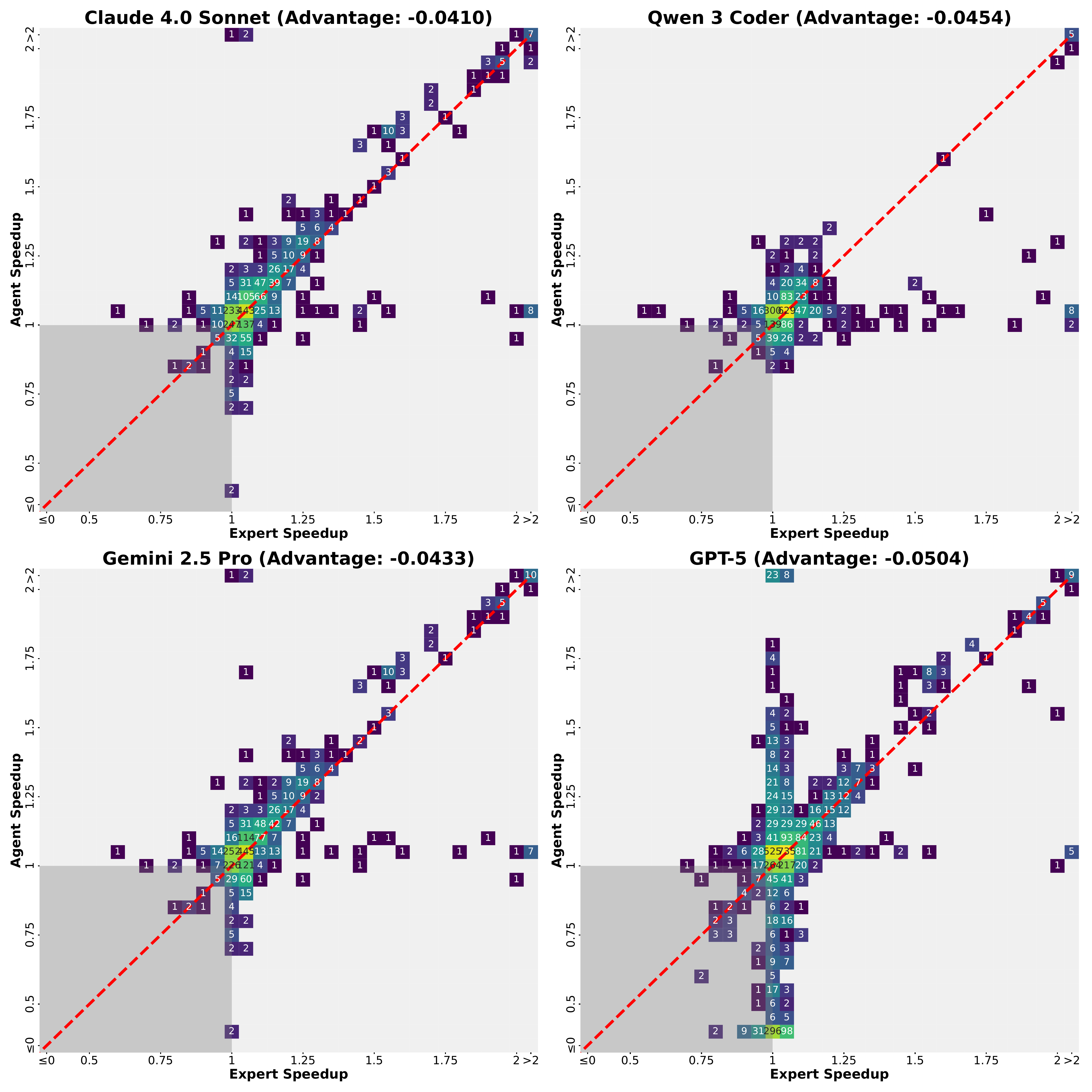}
    \caption{
    Visualization of advantage for Terminus 2 Agents. Refer to Figure \ref{fig:advantage-example} for an explanation of each region. Each square represents the number of workloads in that region (within $0.5$ units). A speedup of $1.0$ indicates no deviation from baseline performance. The red dotted line represents equal advantage.
    This visualization is helpful to gauge the \textit{holistic} behavior of models across the entire workload distribution. For instance, Claude 4.0 Sonnet (Top Left) achieves a better overall advantage than GPT-5 (Bottom Right) by making measured and surgical optimizations that align with the equal-advantage line, whereas optimizations proposed by GPT-5 are more volatile, with more workloads experiencing performance degradations and effectively bringing the overall advantage down.
    }
    \label{fig:advantage-terminus}
\end{figure}

\end{document}